\documentclass[12pt]{article}

\usepackage{graphicx}

\title{Observations and perspectives on the variation of biodiversity}
\author{Dirson Jian Li \footnote{E-Mail: dirson@mail.xjtu.edu.cn}\\ \normalsize{\it \small Department of Applied Physics, School of Science, Xi'an Jiaotong University, Xi'an 710049, China} }

\date{}
\begin{document}
\maketitle
\sloppy
\baselineskip18pt
\setlength{\parskip}{12pt}

{\bf \begin{center}{Abstract}\end{center}
Based on statistical analysis of the complete genome sequences, a remote relationship has been observed between the evolution of the genetic code and the three domain tree of life. The existence of such a remote relationship need to be explained. The unity of the living system throughout the history of life relies on the common features of life: the homochirality, the genetic code and the universal genome format. The universal genome format has been observed in the genomic codon distributions as a common feature of life at the sequence level. A main aim of this article is to reconstruct and to explain the Phanerozoic biodiversity curve. It has been observed that the exponential growth rate of the Phanerozoic biodiversity curve is about equal to the exponential growth rate of genome size evolution. Hence it is strongly indicated that the expansion of genomes causes the exponential trend of the Phanerozoic biodiversity curve, where the conservative property during the evolution of life is guaranteed by the universal genome format at the sequence level. In addition, a consensus curve based on the climatic and eustatic data is obtained to explain the fluctuations of the Phanerozoic biodiversity curve. Thus, the reconstructed biodiversity curve based on genomic, climatic and eustatic data agrees with Sepkoski's curve based on fossil data. The five mass extinctions can be discerned in this reconstructed biodiversity curve, which indicates a tectonic cause of the mass extinctions. And the declining origination rate and extinction rate throughout the Phanerozoic eon might be due to the growth trend in genome size evolution.}

\noindent {\bf Keywords:} Phanerozoic biodiversity curve $\vert$ tectonic cause of mass extinctions $\vert$ genome size evolution $\vert$ universal genome format $\vert$ declining origination and extinction rates $\vert$ adaptation strategy

\section{Introduction}

Some interesting relationships have been observed between biological data in traditionally separate fields. In the previous articles, the concurrent origin of the genetic code and the homochirality might be explained in the triplex picture (Li 2018-I), and the three-domain tree of life might be reconstructed in consideration of the evolution of the genetic code in the triplex2duplex picture (Li 2018-II). The present article is an exploratory attempt to explain the trend and fluctuations in the Phanerozoic biodiversity curve. Considering the hypothetical nature of this study, we should take critical look at the interpretations herein. Here, a statistical picture of genome size evolution in the universal genome format is proposed to explain the reasonable existence of the remote relationship between the genetic code and the tree of life throughout the entire history of life, according to the universal genome format and the adaptation strategy of life to the changing environments on this planet. It should be emphasised that the universal genome format at the sequence level plays an essential role in the interactions among species at the species level. And it has been realised that the homochirality of life at the molecular level also plays an essential role in the robustness of the living system at the species level so as to survive the five mass extinctions in the Phanerozoic eon. 

The Phanerozoic biodiversity curve can be reconstructed with both biological and geological considerations. The trend of the reconstructed Phanerozoic biodiversity curve is based on genome size evolution, while the fluctuations of the reconstructed Phanerozoic biodiversity curve is based on climatic and eustatic data. The reconstructed Phanerozoic biodiversity curve agrees with Sepkoski's biodiversity curve based on fossil records, where the exponential growth rate in the biodiversity curve may be determined by the exponential growth rate in genome size evolution. According to such a reconstruct the Phanerozoic biodiversity curve, both the declining origination rate and the declining extinction rate for the Phanerozoic biodiversity curve can be explained, which are due to the exponential growth rate in the biodiversity curve. All the five mass extinctions have been discerned in the reconstructed biodiversity curve, which indicates that the mass extinctions might be explained by interactions of earth spheres, without having to resort to extraterrestrial triggers. Actually, the mass extinctions are not harmful to the long-term growth of biodiversity. The accidental events can hardly change the trend in genome size evolution at the sequence level and can hardly wipe out the living systems from the earth at the species level, which consists of a huge amount of homochiral materials. 

\subsection{Problems}

What is the driving force in the evolution of life? Although an initial driving force is proposed at the molecular level in the first part of this series (Li 2018-I) by considering the relative stabilities of triplex base pairs, the driving force throughout the entire history of life need to be clarified. A remote relationship between the prebiotic sequence evolution and the diversification of life has been observed at the sequence level, where the unity of life during evolution need to be explained. 

The interaction between biosphere and the other earth's spheres such as lithosphere, hydrosphere, atmosphere need to be explained. There are different results on the Phanerozoic climate changes in literatures; a consensus climate changes need to be obtained. There are different results on the sea level fluctuations in literatures; a consensus sea level fluctuations need to be obtained. The log-normal distribution of genome sizes need to be explained. The exponential growth trend in genome size evolution need to be explained. And the agreement between the trend in genome size evolution and the trend in Phanerozoic biodiversity curve need to be explained. 

The Phanerozoic biodiversity curve need to be reconstructed, where the trend and fluctuations of the Phanerozoic biodiversity curve need to be explained based on biological and geological data. The triggers of the five mass extinctions need to be explained. And the declining trends for both extinction rates and origination rates need to be explained. And the adaptation strategy of life during evolution need to be explained. 

\subsection{Data and observations}

The genome sizes of animals are from the Animal Genome Size Database (Gregory et al. 2007), and the genome sizes of plants are from the Plant DNA C-values Database (Plant DNA C-values database, release 5.0, Dec. 2010). In this paper, we generally follow the taxonomy in the Animal Genome Size Database and the taxonomy in the Plant DNA C-values Database. The distribution of genome sizes in certain taxa is a useful method in the statistical analysis of genome sizes, where $7$ Eukaryotic taxa, $19$ animal taxa and $53$ Angiosperm taxa are considered respectively. Concretely speaking, the $7$ Eukaryotic taxa are as follows: Diploblastica, Protostomia, Deuterostomia, Bryophyte, Pteridophyte, Gymnosperm and Angiosperm. The $19$ animal taxa are as follows: Ctenophores, Sponges, Cnidarians, Anthropod, Molluscs, miscellaneous Invertebrates, Nematodes, Flatworms, Annelid, Myriapods, Rotifers, Tardigrades, Amphibian, Bird, Fish, Mammal, Reptile, Echinoderm, and Chordates. Among the $53$ Angiosperm taxa, the $34$ Dicotyledoneae taxa are as follows: Lentibulariaceae, Cruciferae, Rutaceae, Oxalidaceae, Crassulaceae, Rosaceae, Boraginaceae, Labiatae, Vitaceae, Cucurbitaceae, Onagraceae, Leguminosae, Myrtaceae, Polygonaceae, Euphorbiaceae, Convolvulaceae, Chenopodiaceae, Plantaginaceae, Rubiaceae, Caryophyllaceae, Amaranthaceae, Malvaceae, Umbelliferae, Solanaceae, Papaveraceae, Compositae, Cactaceae, Passifloraceae, Orobanchaceae, Cistaceae, Asteraceae, Ranunculaceae, Loranthaceae, and Paeoniaceae, and the $19$ Monocotyledoneae taxa are as follows: Cyperaceae, Juncaceae, Bromeliaceae, Zingiberaceae, Iridaceae, Orchidaceae, Araceae, Gramineae, Palmae, Asparagaceae, Agavaceae, Hyacinthaceae, Commelinaceae, Amaryllidaceae, Xanthorrhoeaceae, Asphodelaceae, Alliaceae, Liliaceae, and Aloaceae. 

The origin times for the $7$ taxa are as follows respectively: Diploblastica, $560$ Million years ago ($Ma$); Protostomia, $542\ Ma$ (Ediacaran-Cambrian); Deuterostomia, $525\ Ma$; Bryophyte, $488.3\ Ma$ (Cambrian-Ordovician); Pteridophyte, $416.0\ Ma$ (Silurian-Devonian); Gymnosperm, $359.2\ Ma$(Devonian-Carboniferous); Angiosperm, $145.5\ Ma$ (Jurassic-Cretaceous) (Willis and McElwain 2014; Gibling and Davies 2012; Couvreur et al. 2011). 

Sepkoski's biodiversity curve based on fossil records refers to (Bambach et al. 2004). The data on the climate change in the Phanerozoic eon refer to (Boucot and Gray 2001; Boucot et al. 2009; Berner 1998; Berner and Kothavala 2001; Berner et al. 2003; Raymo 1991; Veizer et al. 1999; Veizer et al. 1999; Veizer et al. 2000). The data on the sea level change in the Phanerozoic eon refer to (Haq et al 1987; Haq and Schutter 2008; Hallam 1992). 

\subsection{Main results}

According to the statistical analysis of genome sizes, it is shown that the genome sizes of species in a taxon abide by the logarithmic normal distribution. There is a rough linear relationship between the means of the logarithmic genome sizes of species in the $7$ Eukaryotic taxa and the corresponding origin times of the $7$ Eukaryotic taxa (Fig 1a). In general, the later the origin time is, the greater the mean of the logarithmic genome sizes is. So, the genome size evolution can be regarded as a log-normal stochastic process (Fig 2), which results in the exponential growth trend of genome size evolution. 

I found that the exponential growth rate in genome size evolution is approximately equal to the exponential growth rate of the Phanerozoic biodiversity curve (Fig 1a). The logarithmic normal distribution of genome sizes might be explained by genome duplication. In the process of genome duplication, the genome sizes can vary in a wide range, but the features of genomic codon distributions and the universal genome format are conservative (Fig 4d in Li 2018-II). So in general, the genome sizes vary greatly in a certain taxon whose genomic codon distribution has yet invariant features. Roughly speaking, the greater the average genome size is, the greater the number of species in a taxon is. So, the exponential growth trend in the Phanerozoic biodiversity curve originated from the exponential growth trend in genome size evolution. 

A consensus climatic curve has been obtained by averaging the four climatic curves in literatures (Fig 3b), which are based on the Berner model, the climate sensitive deposition, the strontium isotope ($Sr$) and the oxygen isotope ($O$), respectively. A consensus eustatic curve has been obtained by averaging two eustatic curves in literatures (Haq et al 1987; Haq and Schutter 2008; Hallam 1992) (Fig 3c). Hence a climato-eustatic curve has be obtained by averaging the consensus climatic curve and the consensus eustatic curve, considering that the higher the temperature was, the lower the biodiversity was and that the sea regression often leaded to extinction. I found that the oscillations of the climato-eustatic curve coincides with the oscillations of the the Phanerozoic biodiversity curve (Fig 1b), where not only the five mass extinctions but also most minor extinctions can be discerned in the climato-eustatic curve.

Thus a Phanerozoic biodiversity curve has been reconstructed based on both the exponential growth trend in genome size evolution and the oscillations in the climato-eustatic curve (Fig 1c). The reconstruction is based on genomic, climatic and eustatic data (Fig 4), while Sepkoski's biodiversity curve is based on fossil records. Amazingly, the reconstructed Phanerozoic biodiversity curve agrees quite well with Sepkoski's biodiversity curve. The reconstruction of Sepkoski's biodiversity curve based on molecular evolution and interactions of the earth's spheres shows that the five mass extinctions were directly caused by both climatic and eustatic sea level changes. It is also shown that the declining trends in both extinction rates and origination rates originated from a dilution effect due to rapid genome size increment (Fig 1d); namely the ratios of extinct or novel species decreased when the biodiversity increased exponentially. Life adapts excellently to the earth's changing environment via sacrifice of a large number of redundant species. 

\section{The picture at species level}
\subsection{Genome evolution}

In the second part of this series (Li 2018-II), it was shown that the additivity of genomic codon distribution brings about the universal genome format and the conservative features of genomic codon distribution. And it was revealed that the universal genome format of genomic codon distributions in observations results from the evolution from triplex DNA to duplex DNA. The universal genome format is an invariant property in genome size evolution, which shows that there is an inherent rule in genome evolution. 

Genome size can be regarded as an ultra-coarse-grained description of a genome. Consequently, genome size evolution can be regarded as an overall description of genome evolution. The driving force in the evolution of life comes from either the genome size evolution ab intra or the earth's changing environment ab extra. It is shown that the genome size evolution via large scale duplication can be described by a log-normal stochastic process, by which the features of genome size distributions become reluctant to change. 

Biodiversity originated from the diverse assembly of genomes in the triplex2duplex picture. The genome evolution at sequence level brought forth numerous redundant species, which were potentially able to adapt to any possible environments. In the case of the earth's environment, only a small part of the species had been selected by the earth's environment. The so called adaptation just means the fact that there exists a certain group of species who can survive the special environment on a certain planet. 

\subsection{Adaptation}

Life is a long-lasting phenomenon resulted from multi-scale interactions in spacetime. The adaptation of life to the earth's changing environments is ensured by the common features of life: the homochirality, the genetic code and the universal genome format. These unchanging common features guarantee that life is a nonrandom phenomenon, whereas randomness prevails in both the microcosmic physical world and the macroscopic physical world. These conservative common features bridge the gap between the informative molecules and the changing environments. The interactions between the informative molecules and the environments cause that the microcosmic world acquires the capacity to remould the macroscopic world. If randomness also prevails in the phenomenon of life, life cannot change the earth into a habitable planet. 

Based on the sequence evolution in the universal genome format, numerous redundant potential genomes can be continuously generated, which is independent of the earth's changing environments. So the steady increasing trend in biodiversity throughout the history of life at the species level is due to the evolution of life via duplications at the sequence level. The climate changes and the sea level fluctuations can directly choose certain ``adaptable'' species and discard the other ``unadaptable'' species. If the evolution of life can repeat itself once more, the history of life on the earth will not be too strange for us. But if there is another habitable planet, a different group of species would be chosen as the corresponding adaptable species, although the sequences still evolve in the same universal genome format. Although there were many extinctions events, the increasing trend in the Phanerozoic biodiversity curve did not changed, which is due to the exponential trend of the underlying genome size evolution. 

\section{Trend of Phanerozoic biodiversity}
\subsection{Genome size evolution}

Genome size evaluates the total amount of DNA contained within the genome of a species. Genome sizes vary greatly among species. The genome size evolution is a fundamental and complicated problem in molecular evolution, which is concerned with the macroevolution of life (Gregory 2004; Palazzo and Gregory 2014; Lynch 2007). Several models were proposed to explain the genome size evolution from different perspectives (Petrov 2002; Karev et al. 2003; Lygeros et al. 2008). But the mechanism for genome size evolution is not clear, and the distributional pattern of genome sizes remains unknown (Gregory 2005). Genome size evolution is a complex process, which is driven by either the evolutionary factors or the ecological factors. Numerous small- and large-scale duplication events have played significant roles in genome size evolution (Gregory 2005). Owing to the following intricate multi-factorial drivers, the genome size evolution can be taken as a stochastic process through the long evolutionary history. Gene family expansion is an ongoing phenomenon, and large gene families are a feature of metazoan genomes. Polyploidy plays an especially important role in plant genome evolution. Polyploidy is also a widespread phenomenon in the animal kingdom. Transposable elements often contribute significantly to genome size changes. The mechanisms for amplification and spread of transposable elements are transposition, horizontal transfer, and sexual reproduction. Loss of transposable elements can also occur. Although the overall trend of genome size evolution has been in the direction of increase, the change of genome size is not a uni-directional process. In many cases, the genome size will shrink back very quickly after whole genome duplication due to selective constrains. There are also some groups of animals whose genome underwent extensive gene loss. Hence the distribution of genome sizes of the closely-related species in a taxon may fit certain statistical distribution. The statistical features of genome sizes reveal insight into their evolutionary history.

There are some statistical features of genome sizes of species among taxa. The genome size distributions can be obtained respectively for the $7$ Eukaryotic taxa, the $19$ animal taxa, and the $53$ Angiosperm taxa. The genome size distributions for these taxa approximately fit log-normal distributions, respectively. The genome size distribution for animal and the genome size distribution for plant can be obtained, respectively, both of which fit log-normal distribution (Fig 6a). Owing to the additivity of normal distribution, the genome size distribution for both animal and plant also fits log-normal distribution (Fig 6a). Genome size layout for a certain taxon is to describe the pattern of the genome size distributions of the next-lower-rank taxa, which is defined by the correlationship between the genome size variation ranges of the next-lower-rank taxa and the corresponding standard deviations of logarithmic genome sizes (Fig 6b, 6c). The genome size layout for a taxon indicates its evolutionary history. It is observed that the smaller the standard deviation of logarithmic genome sizes is, the smaller (or greater) the logarithmic mean of genome sizes for Dicotyledoneae (or Monocotyledoneae) taxa is (Fig 6c). The genome size layout for animal is roughly $\Lambda$-shaped (Fig 6b) and the genome size layout for Angiosperm is V-shaped, where most genome size ranges for Dicotyledoneae taxa slants to the small genome size side, while most genome size ranges for Monocotyledoneae taxa slants to the large genome size side (Fig 6c). 

\subsection{A stochastic process model}

The genome size evolution follows a log-normal stochastic process. A log-normal stochastic process model is developed to simulate the statistical features of genome sizes. The log-normal distributions of genome sizes and the different types of genome size layouts are simulated by the model (Fig 7), by which the statistical features of genome sizes are explained. The statistical features of genome sizes for animal and plant in simulations (Fig 8) agree with the observations (Fig 6). 

Duplications contribute much to the genome size evolution (Gregory 2005), whose contribution can be quantitatively estimated by a genome size growth rate. From a statistical perspective, the trend in genome size evolution follows an exponential growth trend. The change of genome size $\delta S$ is proportional to the genome size $S$ itself: $\delta S = S\ \lambda \delta t$, where $t$ is time and $\lambda$ is the genome size growth rate that is independent of genome size $S$. So the trend of genome size evolution obeys the exponential growth trend: $S(t) = S_0 \exp(\lambda t)$. The exponential growth trend in genome size evolution was suggested by studying the relationship between the origin time of taxa and the estimated average genome sizes (Sharov 2007); and it was also obtained in a formula to estimate the genome sizes (Li and Zhang 2010). Here, I obtained the exponential growth trend in genome size evolution based on the relationship between the logarithmic mean of genome sizes of species in $7$ taxa and the corresponding origin time of the $7$ taxa (Fig 1a). 

The genome size evolution can be regarded as a stochastic process. When considering the stochastic impact on genome size growth, we can obtain $\delta S = S \lambda \delta t + S \lambda \delta n$, where $n$ is a standard normal random variable. So the genome size evolution obeys the stochastic equation $S(t) = S \exp(\lambda t +\lambda n)$. I propose a log-normal stochastic process model (Ghanem 1999) to simulate both the observation of the log-normal distributions of genome sizes and the observation of the exponential growth rate of genome size evolution. In the model, the genome size $S$ increases step by step at certain probability $P$ from the initial value $S(1)=1$ base pair at step $1$ to the final value $S(N)$ base pairs at step $N$, where $N$ is the maximum number of steps. From step $n-1$ to step $n$ ($n=2, 3, ..., N$), the genome size doubles at probability $P$, or remains the same at probability $1-P$. The net result leads to a certain expected growth rate, which ensures that the genome size tends to increase exponentially. This is a stochastic process, where the probability $P$ brings about the log-normal distribution of genome size $S(N)$. The exponential growth rate in genome size evolution is constant in the simulation, when the probability $P$ keeps constant. The whole process from step $1$ to step $N$ can be divided into $M$ stages. $M$ steps $n=t*(N/M), t=1, 2, ..., M$ are chosen in the process, where $M$ is the number of stages and $t$ can be interpreted as the time in genome size evolution. Thus $M$ genome size distributions $S(t*(N/M))$ are obtained at the time $t=1, 2, ..., M$ respectively. In the simulation, these genome size distributions $S(t*(N/M))$ are calculated based on the results of genome sizes at time $t$ by running the process from $1$ to $n(t)$ for sufficiently many times. The values of genome sizes changes randomly due to the probability $P$, but the simulated genome size distributions always follow log-normal distribution. The logarithmic means of genome sizes for the simulated taxa increase with respect to time $t$ due to the increasing numbers of steps $n(t)$ (Fig 7b). These obtained $M$ genome size distributions $S(t*(N/M))$ can be interpreted as the genome size distributions of one taxon at different stages ($t=1, 2, ..., M$) in the evolution (Fig 7a).

Based on the log-normal stochastic process model, a number of genome size distributions can be obtained to simulate the statistical features of genome sizes for a taxon (Fig 7a). All the genome size distributions by the simulation fit log-normal distributions (Fig 7a). The exponential growth rates in genome size evolution are common in the simulation due to the same probability $P$. The logarithmic mean of genome sizes increases approximately linearly with respect to time $t$ (Fig 7b), which is equivalent to that the genome size increases exponentially. So the observation of the exponential trend in genome size evolution (Fig 1a) might be explained by the log-normal stochastic process model (Fig 7b). The statistic features of genome sizes in observation (Fig 6b, 6c) can be explained by developing the stochastic process model from single-taxon model to multiple-taxa model. In the multiple-taxa model, the probabilities $P$ is assigned with multiple constant values so as to simulate the genome size distributions for multiple taxa. The $\Lambda$-shaped and V-shaped genome size layouts in observation can be simulated by the multiple-taxa model if choosing different genome size growth rates for different stages in genome size evolution. The difference between the simulation of $\Lambda$-shaped layout and the simulation of V-shaped genome size layout results from the different assignment schemes for the probabilities $P$ between the ancestor taxa and the descendant taxa. 

The genome size layout for animal is about $\Lambda$-shaped in observation (Fig 6b), which can be simulated by the model (Fig 8a, 8b). The genome size layout for Angiosperm is about V-shaped in observation (Fig 6c), which can be simulated by the model (Fig 8c, 8d). The simulation indicates a relationship between the genome size layout and the phylogeny of life. The different evolutionary histories for animal and Angiosperm lead to their different genome size layouts. Almost all phyla of animal originated nearly together at the Cambrian explosion (Conway-Morris 1993, Conway-Morris 2003, Valentine 2001, Shu 2008). It is generally believed that animal is monophyletic (Wainright et al. 1993, Cavalier-Smith et al. 1996). But the classes in Angiosperms did not originate together (Willis and McElwain 2014). The two Angiosperm branches, the Monocotyledons and Dicotyledons, may originate independently. So the Angiosperms would be polyphyletic (Stuessy 2004). The simulation of different types of genome size layouts helps to understand the different origins for animal and Angiosperm. A $\Lambda$-shaped genome size layout for animal indicates a common ancestor for animals (Fig 8a, 8b), which support its monophyletic origin. And a V-shaped genome size layout for Angiosperm can be simulated by different ancestors for Dicotyledons and Monocotyledons (Fig 8c, 8d). The simulation results indicate that Angiosperm is not monophyletic. The agreement between the observations and simulations shows that the genome size evolution can be described by a log-normal stochastic process. 

\section{Driving force in the evolution} 

Evolution has taken place and is continuing to occur at both the species level and the sequence level. History archives of biodiversity on the earth have been stored not only in the fossil records at the species level (Sepkoski 2002; Bambach et al. 2004; Rohde and Muller 2005) but also in the genomes of contemporary species at the sequence level. A Phanerozoic biodiversity curve can be plotted based on the fossil records
(Bambach et al. 2004) (Fig 1c, 2h). The growth trend of this curve is exponential (Bambach et al. 2004) (Fig
1c, 2h), and the deviations from the exponential trend represent extinctions and originations (Bambach et al. 2004) (Fig 1b, 2h). Environmental changes substantially influence the evolution of life at the species level, but do not directly impact on the molecular evolution. The genome size is an ultra-coarse-grained representation of a species at the sequence level. So, the evolution of life can be outlined by the evolution of genome size. And the genome size increased mainly via large scale duplications. Considering the relative invariance of the genomic codon distributions in the duplications (Fig 4d In the second part of this series (Li 2018-II), the features of species at the sequence level are conservative throughout genome size evolution. So, it is relatively independent between the evolution at the species level and the evolution at the sequence level. In this paper, a split and reconstruction scheme is proposed to explain the Phanerozoic biodiversity curve (Fig 1c, 2h, 4): the exponential growth trend of the Phanerozoic biodiversity curve is postulated to be driven by the genome size evolution (Fig 1a), while the extinctions and originations in the fluctuations of the Phanerozoic biodiversity curve is due to the Phanerozoic eustatic and climatic changes (Fig 1b). 

It is necessary to shed light on the genome size evolution. It will be shown that the growth trend of genome size evolution is also exponential in general (Sharov 2007; Li and Zhang 2010). The genome size distributions of species in certain taxa generally satisfy logarithmic normal distributions, according to the statistical analysis of genome sizes in the databases Animal Genome Size Database (Gregory et al. 2007) and Plant DNA C-values Database (Plant DNA C-values database) (Fig 2a). The genome size distributions for the $7$ higher taxa of animals and plants, namely, Diploblastica, Protostomia, Deuterostomia, Bryophyte, Pteridophyte, Gymnosperm and Angiosperm, generally satisfy logarithmic normal distributions, respectively. The genome size distribution for all the species in the two databases satisfies logarithmic normal distribution very well, due to the additivity of normal distributions (Fig 2a).

The logarithmic normal distributions of genome sizes for taxa can be simulated by the log-normal stochastic process model (Fig 2b). The genome sizes obtained at a certain time in the simulation represent the genome sizes of species in a taxon at that time, which satisfy logarithmic normal distribution. The simulation indicates that, along the direction of time, the logarithmic mean of genome sizes for a taxon $G_{logMean}$ increases, and the logarithmic standard deviation of genome sizes for a taxon $G_{logSD}$ also increases (Fig 2b). According to the simulation, the logarithmic mean of genome sizes for the ancestor of a taxon at the origin time $G_{orilogMean}$ can be estimated by the genome sizes at present. Details are as follows. When genome size evolves in an exponential growth trend, $G_{orilogMean}$ at the origin time is always less than $G_{logMean}$ at present for a certain taxon. The greater $G_{logSD}$ is (e.g. $G_{logSD}^B > G_{logSD}^A$ in Fig 2b), the earlier the origin time is (e.g. $t_{origin}^B < t_{origin}^A$ in Fig 2b), and thereby the less $G_{orilogMean}$ is (e.g. $G_{orilogMean}^B < G_{orilogMean}^A$ in Fig 2b). Thus, the genome size for ancestor at the origin time $G_{orilogMean}$ is obtained by calculating the genome sizes at present: $G_{logMean}$ minus $G_{logSD}$ times an undetermined factor $\chi$ (Fig 2b, 2g).

Based on the genome sizes of contemporary species in the databases, the logarithmic means of genome sizes $G_{logMean}$ and the logarithmic standard deviations of genome sizes $G_{logSD}$ are obtained for the $7$ higher taxa of animals and plants, respectively (Fig 2g). The genome sizes for the ancestors of the $7$ higher taxa at their origin times $G_{orilogMean}$ are less than the corresponding $G_{logMean}$ at present, whose differences can be estimated by the corresponding $G_{logSD}$ times an undetermined factor $\chi$, respectively (Fig 2g). The trend of the genome size evolution cannot be worked out just only based on the genome sizes of contemporary species. Furthermore, a chronological reference has to be introduced here. The origin times for taxa can be estimated according to the fossil records. In this paper, the origin times for the $7$ higher taxa are assumed as follows respectively: Diploblastica, $560\ Ma$; Protostomia, $542\ Ma$; Deuterostomia, $525\ Ma$; Bryophyte, $488.3\ Ma$; Pteridophyte, $416.0\ Ma$; Gymnosperm, $359.2\ Ma$; Angiosperm, $145.5\ Ma$. The main results in this paper cannot be weakened by disagreements among the above origin times in literatures or by expansion of the genome size databases. A rough exponential growth trend of genome size evolution is observed in the plot whose abscissa and ordinate represent the origin time and $G_{logMean}$, respectively (Fig 2g), where the differences between the genome sizes at the origin time and that at present are not considered yet. A more reasonable plot is introduced by taking $G_{orilogMean}$ rather than $G_{logMean}$ as the ordinate (Fig 2g). In the calculation, $G_{orilogMean}$ are obtained respectively for the $7$ higher taxa by (Fig 2g) $$G_{orilogMean} = G_{logMean} - \chi \cdot G_{logSD}.$$ According to the regression analysis, a regression line approximates the relationship between the origin time and $G_{orilogMean}$ (Fig 2g). Let the intercept of the regression line be the logarithmic mean of genome sizes for all the species in the databases, then the above undetermined factor is $\chi = 1.57$ (Fig 2g). The slope of the regression line represents the exponential growth speed of genome size evolution, which doubles for each $177.8\ Ma$ (Fig 2g, 1a). It does not violate the common sense that the genome size at the time of the origin of life about $3800\ Ma$ is about several hundreds base pairs, according to the regression line.

There are more evidences to support the method for calculating the ancestors' genome size based on the data of contemporary species. The values of $G_{logMean}$, $G_{logSD}$ and $G_{orilogMean}$ are obtained for the $19$ animal taxa or for the $53$ Angiosperm taxa, respectively (Fig 2c). The numbers of species in these taxa are great enough for statistical analysis. The chronological order for the $19$ animal taxa is obtained by comparing $G_{orilogMean}$ (Fig 2e). The order of the three superphyla is suggested to be: $1$, Diploblastica; $2$, Protostomia; $3$, Deuterostomia, which can be inferred from the chronological order of the $19$ animal taxa and the superphylum affiliation of the $19$ taxa (Fig 2e). This result supports the three-stage pattern in Metazoan origination based on fossil records (Conway-Morris 1993, 1989; Budd and Jensen 2000; Valentine 2001; Shu 2005, 2008; Shu et al. 2004; Shu et al. 2009). The chronological order for the $53$ Angiosperm taxa is also obtained by comparing $G_{orilogMean}$ (Fig 2f), which confirms the common sense that dicotyledoneae originated earlier than monocotyledoneae. Let the abscissa and ordinate represent respectively $G_{orilogMean}$ and $G_{logMean}$ for the $7$ higher taxa and $19+53$ taxa, where the variation ranges of logarithm of genome sizes $G_{logMean} \pm G_{logSD}$ were also indicated on the ordinate. The upper triangular distribution of the genome size variations thus obtained (Fig 2d) reveals intrinsic features of the genome size evolution. The greater $G_{orilogMean}$ is, the greater $G_{logMean}$ is, and the less $G_{logSD}$ is (Fig 2d). The upper vertex of the triangle indicates approximately the upper limit of genome sizes of contemporary species. The upper triangular distribution (Fig 2d) can be explained by the simulation of genome size evolution (Fig 2b). The simulation shows that $G_{orilogMean}$ and $G_{logMean}$ increase along the evolutionary direction; the earlier the time of origination is, the greater $G_{logSD}$ is (Fig 2b, 2d).

The exponential growth trend of Phanerozoic biodiversity curve might be explained by the exponential growth trend of genome size evolution. Bambach et al. obtained the Phanerozoic biodiversity curve based on fossil records, which satisfies the exponential growth trend (Fig 1c). The number of genera doubles for each $172.7\ Ma$ in the Phanerozoic eon (Fig 1a). It has been obtained that the genome size doubles for each $177.8\ Ma$ in the Phanerozoic eon (Fig 1a). The exponential growth rate for the number of genera agrees with the exponential growth rate for the genome size. So it is conjectured that the genome size evolution at the sequence level drives the evolution of life at the species level. 

Note that Bambach et al.'s curve is chosen as the Phanerozoic biodiversity curve in this paper. It is based on the following considerations. Based on fossil records, Sepkoski obtained the Phanerozoic biodiversity curve at family level for the marine life first (Sepkoski 1981), then obtained the curve at genus level (Sepkoski 1997). Sepkoski did not include single fossil records in order to reduce the impact on the curve at genus level from the samplings and time intervals. In order to reduce such an impact furthermore, Bambach et al. chose only boundary-crossing taxa in construction of the Phanerozoic biodiversity curve at genus level (Bambach et al. 2004). Rohde and Muller obtained a curve for all genera and another curve for well-resolved genera (removing genera whose time intervals are epoch or period, or single observed genera) respectively based on SepkoskiÔs Compendium, which can roughly be taken as the upper limit and lower limit of the Phanerozoic curve (Fig 1c). These Phanerozoic biodiversity curves are roughly in common in their variations, from which the five mass extinctions O-S, F-F, P-Tr, Tr-J, K-Pg can be discerned. There are several models to explain the growth trend of the Phanerozoic biodiversity curve. Sepkoski and other co-workers tried to explain the growth trend by $2$-dimensional Logistic model (Sepkoski 1978, 1979, 1984). Benton et al. explained it by exponential growth model (Hewzulla et al. 1999). The Bambach et al.'s curve is lower in the Paleozoic era, and obviously higher in the Cenozoic era than Sepkoski's curve. It is more obvious that the Bambach et al.'s curve is in exponential growth trend than Sepkoski's curve.

The exponential growth trend and the net fluctuations of the Bambach et al.'s biodiversity curve can be explained at both the molecular level and the species level respectively by a split and reconstruction scheme (Fig 4). In this scheme, the exponential growth trend of biodiversity can be attributed to the driving force from the genome size evolution, considering the approximate equality between the exponential growth rate of number of genera and the exponential growth rate of genome size (Fig 1a). The homochirality of life, the genetic code and the universal genome format are invariant during genome evolution. Expansion of genome sizes at sequence level can benefit the emergence of new species. According to the split and reconstruction scheme, the extinctions and originations by the deviations from the exponential growth trend of the biodiversity curve is interpreted as a biodiversity net fluctuation curve. A raw biodiversity net fluctuation curve is obtained by subtracting the growth trend from the logarithmic Phanerozoic biodiversity curve (Fig 4), and the biodiversity net fluctuation curve (Fig 1b, 3e) is defined by dividing the corresponding standard deviation. The biodiversity net fluctuation curve might be explained by the Phanerozoic eustatic curve and Phanerozoic climatic curve in next section (Fig 1b).

\section{Variation of Phanerozoic biodiversity}

According to the split and reconstruction scheme for the Phanerozoic biodiversity curve, the growth trend of biodiversity can be attributed to the genome size evolution at the sequence level, and the net fluctuations deviated from the growth trend can be attributed to the couplings among earth's spheres. The history of biodiversity in the Phanerozoic eon can be outlined by the biodiversity net fluctuation curve (Fig 1b): the curve reached the high point after the Cambrian radiation and the Ordovician radiation; then the curve reached the low point after a triple plunge in the following three O-S, F-F and P-Tr mass extinctions; and after P-Tr mass extinction, the curve climbed higher and higher except for mass extinctions and recoveries around Tr-J and K-Pg. The biodiversity net fluctuation curve can be reconstructed through a climato-eustatic curve based on climatic and eustatic data (Fig 1b), which is free from the fossil records. Accordingly, it is advised that the tectonic movement account for the five mass extinctions.

Numerous explanations on the cause of mass extinctions have been proposed, such as climate change, marine transgression and regression, celestial collision, large igneous province, water anoxia, ocean acidification, release of methane hydrates and so on. According to a thorough study of control factors on mass extinctions, it is almost impossible to explain all the five mass extinctions by a single control factor. Although the O-S mass extinction may be attributed to the glacier age, the other mass extinctions almost happened in high temperature periods. The other glacier ages in the Phanerozoic eon had always avoided the mass extinctions. Although the mass extinctions tend to happen with marine regressions, it is commonly believed that no marine regression happened in the end-Changhsingian, the greatest mass extinction (Wignall and Hallam 1992, 1993; Wignall and Twitchett 1996; Yin and Tong 1998; Erwin et al. 2002). Large igneous province may play roles only in the last three mass extinctions (Renne and Basu 1991; Campbell et al. 1992). The K-Pg mass extinction was most likely to be caused by celestial collision, but there are still significant doubts on this explanation (Alvarez et al. 1980; Keller 2004; Stewart and Allen 2002). According to the high resolution study of the mass extinctions based on fossil records, more understandings on the complexity of mass extinctions arise. For example, the P-Tr mass extinction can be divided into two phases: End-Guadalupian and end-Changhsingian (Jin 1991, 1993; Jin et al. 2000; Stanley and Yang 1994; Shen et al. 2011); furthermore the end-Changhsingian stage can be subdivided into main stage (B line) and final stage (C line) (Xie et al. 2005; Chen and Benton 2012), and the course of extinction may happened very rapidly (Bowring et al. 1999; Eshet et al. 1995; Rampino et al. 2000; Ward et al. 2000; Twichett et al. 2001). Such a complex pattern of the mass extinction challenges any explanations on this problem. it is also almost impossible to attribute a mass extinction to a single control factor. Based on the above discussions, it is wise to consider a multifactorial explanation for the mass extinctions. In consideration of the interactions among the lithosphere, hydrosphere, atmosphere and biosphere, the tectonic movement of the aggregation and dispersal of Pangaea influenced the biodiversity net fluctuation curve at several time scales through the impacts of hydrosphere and atmosphere (Fig 1b).

Before explaining the mass extinctions based on the tectonic movement, two preparations are required: first, to construct a Phanerozoic climatic curve; second, to construct a Phanerozoic eustatic curve.

\subsection{Consensus climatic curve}

The Phanerozoic climatic curve is based on the following four results (Fig 3b): ($i$) the climatic gradient curve based on climate indicators in the Phanerozoic eon (Boucot and Gray 2001; Boucot et al. 2009) (Fig 3a); ($ii$) the atmospheric $CO_2$ fluctuations in the Berner's models that emphasise the weathering role of tracheophytes (Berner 1998; Berner and Kothavala 2001; Berner et al. 2003); ($iii$) the marine carbonate $^{87}Sr/^{86}Sr$ record in the Phanerozoic eon in Raymo's model (Raymo 1991; Veizer et al. 1999). The more positive $^{87}Sr/^{86}Sr$ values, assumed to indicate enhanced $CO_2$ levels and globally colder climates; more negative $^{87}Sr/^{86}Sr$ values, indicating enhanced sea-floor speeding and hydrothermal activity, are associated with increased $CO_2$ levels; ($iv$) the marine $\delta ^{18}O$ record in the Phanerozoic eon based on thermodynamic isotope effect (Veizer et al. 1999; Veizer et al. 2000). The above four independent results are quite different to one another (Fig 3b). Considering the complexity of the problem on the Phanerozoic climate change yet, all the four methods are reasonable in principle on their grounds respectively. A consensus Phanerozoic climatic curve is obtained based on the above four independent results after nondimensionalization at equal weights. Namely, the Phanerozoic climatic curve is defined as the average of nondimensionalized climate indicator curve (Boucot and Gray 2001; Boucot et al. 2009), Berner's $CO_2$ curve (Berner 1998), $^{87}Sr/^{86}Sr$ curve (Raymo 1991; Veizer et al. 1999) and $\delta ^{18}O$ curve (Veizer et al. 1999; Veizer et al. 2000) (Fig 3b).

Nondimensionalization is a practical and reasonable method to obtain the consensus Phanerozoic climatic curve. As the temperature in a certain time is concerned, when the four results agree to one other, the common temperatures are adopted; when the four results disagree, the majority consensus temperatures are adopted (Fig 3b). According to the Phanerozoic climatic curve, the temperature was high from Cambrian to Middle Ordovician (Fig 3b); the temperature was low during O-S (Fig 3b); the temperature increases from Silurian to Devonian, and it reach high during F-F (Fig 3b); the temperature is extreme low during D-C (Fig 3b); the temperature rebounded slightly during Missisippian, and the temperature became lower again during Pennsylvanian, and the temperature decreased to extreme low during early Permian (Fig 3b); the temperature increase in Permian, and especially the temperature increase rapidly during the Lopingian, and reach an extreme high peak during P-Tr (Fig 3b); the temperature is high from Lower Triassic to Lower Jurassic, and it reached to extreme high during Lower Jurassic (Fig 3b); the temperature became low from Middle Jurassic to Lower Cretaceous, where it is coldest in the Mesozoic era during J-K, which was not low enough to form glaciation (Fig 3b); the temperature is high again from Upper Cretaceous to Paleocene, and the detailed high temperature during the Paleocene-Eocene event is observed (Fig 3b); the temperature decreased from Eocene to present, especially reached to the minimum in the Phanerozoic eon during Quaternary (Fig 3b). Four glacier ages are observed in the Phanerozoic eon (Fig 3b): Hirnantian glacier age, Tournaisian glacier age, early Permian glacier age and Quaternary glacier age, especially bipolar ice sheet at present. The Phanerozoic climatic curve also indicates the extreme low temperature during J-K (Fig 3b), which corresponds to possible continental glaciers at high latitudes in the southern hemisphere (Stoller and Schrag 1996). In short, the fine-resolution Phanerozoic climatic curve agrees with the common sense on climate change at the tectonic time scale.

The Phanerozoic climatic curve is taken as a standard curve to evaluate the four independent methods. The climate indicator curve is the best result among the four. It accurately reflects the four high temperature periods and four low temperature periods, and the varying amplitudes of the curve are relatively modest and reasonable. But it overestimated the temperature during Pennsylvanian (Fig 3b). The $\delta ^{18}O$ curve is also reasonable. It also reflects the four high temperature periods and four low temperature periods. But it overestimated the temperature during Guadalupian, and underestimated the temperature during Jurassic (Fig 3b). The $^{87}Sr/^{86}Sr$ curve is reasonable to some extent. It reflects the high temperature during Mesozoic and low temperature during Carboniferous and Quaternary. But it underestimated the temperature during Cambrian, and overestimated the temperature during O-S and J-K (Fig 3b). The Berner's $CO_2$ curve is relatively rough. It reflects the high temperature during Mesozoic, But it overestimated the temperature from Cambrian to Devonian (Fig 3b).

As the climate changes at the tectonic time scale is concerned, the Berner's $CO_2$ curve indicates relatively long term climate changes; the climate indicator curve indicates relatively mid term climate changes; and the $^{87}Sr/^{86}Sr$ and $\delta ^{18}O$ curves indicate relatively short term climate changes (Fig 3b). The scale of the variation of the Phanerozoic climatic curve is relatively accurate because this consensus climatic curve is obtained by combining all the different time scales of the above four climatic curves. Four high temperature periods Cambrian, Devonian, Triassic, K-Pg and four low temperature periods O-S, Carboniferous, J-K, Quaternary appeared alternatively in the Phanerozoic climatic curve, which generally reflect four periods in the climate changes in the Phanerozoic eon (Fig 3b). Such a time scale of the four periods of climate changes is about equal to the average interval of mass extinctions, and is also equal to the time scale of the four periods of the variation of the relative abundance of microbial carbonates in reef during Phanerozoic (Riding and Liang 2004; Riding 2005, 2006; Kiessling et al. 1999). The agreement of the time scales indicates the relationship between the climate change and the fluctuations of biodiversity in the Phanerozoic eon.

\subsection{Consensus eustatic curve}

The Phanerozoic eustatic curve is based on the following two results (Fig 3c): (i) Haq sea level curve (Haq et al 1987; Haq and Schutter 2008), (ii) Hallam sea level curve (Hallam 1992). In 1970s, Seismic stratigraphy developed a method called sequence stratigraphy to plot the sea level fluctuations. Vail in the Exxon Production Research Company obtained a sea level curve based on this method, which concerns controversial due to business confidential data. Haq et al. replotted a sea level curve for the mesozoic era and cenozoic era also based on this method (Haq et al 1987), and a sea level curve for the paleozoic era (Haq and Schutter 2008). The Haq sea level curve here is obtained by combining these two curves (Fig 3c). Hallam obtained a sea level curve by investigating large amounts of data on sea level variations (Hallam 1992) (Fig 3c). Generally speaking, it is not quite different between the Haq sea level curve and the Hallam sea level curve. Therefore, a Phanerozoic eustatic curve can be obtained by averaging the Haq sea level curve and the Hallam sea level curve after nondimensionalization at equal weights (Fig 3c).

\subsection{Climato-eustatic curve}

The sea level fluctuations and the climate change have an important impact on the fluctuations of biodiversity. With respect to the impact of the sea level fluctuations, this paper follows the traditional view that sea level fluctuations play positive contributions to the fluctuations of biodiversity. With respect to the impact of the climate change, this paper believes that climate change plays negative contributions to the fluctuations of biodiversity. The climate gradient curve is opposite to the climate curve (Fig 3b). The higher the temperature is, the lower the climate gradient is, and consequently the less the global climate zones are; while the lower the temperature is, the higher the climate gradient is, and consequently the more the global climate zones are. When the climate gradient is high, there were tropical, arid, temperate and frigid zones among the global climate zones, which boosts the biodiversity. Actually, all organisms must die at ultrahigh temperature, which is also a common feature of life. So, this paper believes that the biodiversity increases with the climate gradient. This point of view agrees with the following observations. The biodiversity rebounded during the low temperature period Carboniferous (Fig 3e), and increased steadily during the ice age during Quaternary (Fig 3e); and the four mass extinctions F-F, P-Tr, Tr-J, K-Pg appeared in high temperature periods (Fig 3e). Although the O-S mass extinction occurred in the ice age (Fig 3e), this case is quite different from other mass extinctions. In the first mass extinction during O-S, mainly the Cambrian fauna, which originated in the high temperature period, cannot survive from the O-S ice age. Mass extinctions no longer occurred thereafter in the ice ages for those descendants who have successfully survived from the O-S ice age (Fig 3e).

Based on the above analysis, the eustatic curve and the climatic gradient curve is positively related to the fluctuations of biodiversity in the Phanerozoic eon. Accordingly, a climato-euctatic curve is defined by the average of the nondimensionaliezed eustatic curve and the nondimensionaliezed climatic grandaunt curve, which indicates the impact of environment on biodiversity (Fig 3d). It should be emphasised that the climato-euctatic curve is by no means based on fossil records, which is only based on climatic and eustatic data. It is obvious that the climato-euctatic curve agrees with the biodiversity net fluctuation curve (Fig 1b). Judging from the overall features, the outline of the history of the Phanerozoic biodiversity can be reconstructed in the climato-eustatic curve, as far as the Cambrian radiation and the Ordovician radiation, the Paleozoic diversity plateau, the P-Tr mass extinction, and the Mesozoic and Cenozoic radiation are concerned (Fig 1b). Judging from the mass extinctions, the five mass extinctions O-S, F-F, P-Tr, Tr-J, K-Pg can be discerned from the five steep descents in the climato-eustatic curve respectively (Fig 1b, 3g). As to K-Pg especially, there is obviously a pair of descent and ascent around K-Pg in the climato-eustatic curve (Fig 1b), so celestial explanations are actually not necessary here. Judging from the minor extinctions, most of the minor extinctions can be discerned in the climato-eustatic curve in details, as far as the following are concerned: $3$ extinctions during Cambrian, Cm-O extinction, S-D extinction, $2$ extinctions during Carboniferous, $1$ extinction during Permian, $1$ extinction during Triassic, $3$ extinctions during Jurassic, $2$ extinctions during Cretaceous, $2$ extinctions during Paleogene, $1$ extinction during Neogene and the extinction at present (Fig 1b, 3g). Judging from the increment of the biodiversity, the originations and radiations can be discerned from the ascents in the climato-eustatic curve, as far as the following are concerned: the Cambrian radiation, the Ordovician radiation, the Silurian radiation, the Lower Devonian diversification, the Carboniferous rebound, the Triassic rebound, the Lower Jurassic rebound, the Middle Cretaceous diversification, the Paleogene rebound (Fig 1b, 3g). The derivative of the climato-eustatic curve indicates the variation rates (Fig 1b, 3g). The extinctions are discerned by negative derivative, and the rebounds and radiations by positive derivative. It is obvious to discern the five mass extinctions, most of minor extinctions, and rebounds and radiations in the derivative curve of the climato-eustatic curve, just like in the climato-eustatic curve (Fig 1b, 3g).

\section{Cause of mass extinctions}

The interactions among the lithosphere, hydrosphere, atmosphere, biosphere are illustrated by the eustatic curve, climatic curve, biodiversity net fluctuation curve and the aggregation and dispersal of Pangaea (Fig 3e, 3f). There were about two long periods in the eustatic curve, which were divided by the event of the most aggregation of Pangaea (Fig 3e). The first period comprises the grand marine transgression from the Cambrian to the Ordovician and the grand marine regression from the Silurian to the Permian, and the second period comprises the grand marine transgression during Mesozoic and the grand regression during Cenozoic (Fig 3e). These observations reveal the tectonic cause of the grand marine transgressions and regressions in the Phanerozoic eon, considering the aggregation and dispersal of Pangaea. The coupling between the hydrosphere and the atmosphere can be comprehended by comparing the eustatic curve and the climatic curve. The global temperature was high during both the grand marine transgressions (Fig 3e). And all the four ice ages in the Phanerozoic occurred during the grand marine regressions (Fig 3e). Generally speaking, the climate gradient curve varies roughly synchronously with the eustatic curve from the Cambrian to the Lower Cretaceous along with the aggregation of Pangaea, except for the regression around the Tournaisian ice ages (Fig 3e). And the climate gradient curve varies roughly oppositely with the eustatic curve thereafter from the Upper Cretaceous to the present along with the dispersal of Pangaea (Fig 3e); namely the climate gradient curve increases as the eustatic curve decreases. These observations reveal the tectonic cause of the climate change at tectonic time scale in the Phanerozoic eon, considering the impact on the climate from the grand transgressions and regressions. The biodiversity responses almost instantly to both the eustatic curve and the climatic curve, because the life spans of individuals of any species are much less than the tectonic time scale. Hence, the biodiversity net fluctuation curve agrees with the climato-eustatic curve very well (Fig 1b). In conclusion, the tectonic movement has substantially influenced the biodiversity through the interactions among the earth's spheres (Fig 3e, 3f).

The interactions among the earth's spheres can explain the outline of the biodiversity fluctuations in the Phanerozoic eon. During Cambrian and Ordovician, both the sea level and the climate gradient increased, so the biodiversity increased (Fig 3e). From Silurian to Permian, both the sea level and the climate gradient decreased, so the biodiversity decreased (Fig 3e), except for the biodiversity plateau during Carboniferous due to the opposite increasing climate gradient and decreasing sea level (Fig 3e). From Triassic to Lower Cretaceous, both the sea level and the climate gradient increased, so the biodiversity increased (Fig 3e). From Upper Cretaceous to present, the sea level decreased and the climate gradient increased rapidly, so the biodiversity increased (Fig 3e).

The interactions among the earth's spheres can explain the five mass extinctions within the same theoretical framework, though their causes are different in detail. The mass extinctions O-S, K-Pg occurred around the peaks in the eustatic curve respectively, and the P-Tr mass extinction around the lowest point in the eustatic curve. There were about two periods in the eustatic curve and about four periods in the climatic curve in the Phanerozoic eon. So the period of climate change was about a half of the period of sea level fluctuations, as far as their changes at the tectonic time scale are concerned (Fig 3e). By comparison, the F-F mass extinction occurred in the low level of climate gradient. There was a triple plunge from the high level of biodiversity during the end-Ordovician to the extreme low level of biodiversity during Triassic in the biodiversity net fluctuation curve, which comprised a series of mass extinctions O-S, F-F, P-Tr (Fig 1b). It is interesting that a relatively less great extinction may occur after each of the triple plunge mass extinctions respectively: namely the S-D extinction after the O-S mass extinction, the Carboniferous extinction after the F-F mass extinction, and the Tr-J mass extinction after the P-Tr mass extinction (Fig 1b). Especially during the end of Permian, the aggregation of the Pangaea pull down both the eustatic curve and the climate gradient curve, and hence pull down the biodiversity curve (Fig 3e). After a long continual descent in the biodiversity net fluctuation curve and after the two mass extinctions O-S, F-F, the ecosystem was more fragile at the end of Permian than in the previous period when biodiversity was at the high level, therefore the P-Tr mass extinction occurred as the most severe extinction event in the Phanerozoic eon (Fig 1b). The P-Tr mass extinction can be divided into the first phase at the end of Guadalupian and the second phase at the end of Changhsingian (Fig 1b). The extinction of Fusulinina mainly occurred at the end of Guadalupian; while the extinction of Endothyrina mainly occurred at the end of Changhsingian. Note that the number of Endothyrina did not decrease obviously at the first phase. Different extinction patterns in the above two phases indicate that the triggers for the two phases are different to each other. Marine regression occurred at the end of the Guadalupian, whereas it is indicated that no marine regression occurred at the end of the Changhsingian. The second phase of the end-Changhsingian extinction may be caused by the declining climate gradient (or increasing temperature). All the four independent methods in the above (climatic-sensitive sediments, Berner's $CO_2$, $^{87}Sr/^{86}Sr$ and $\delta ^{18}O$) predict the rapidly increasing temperature at the end of the Changhsingian (Fig 3b). Either Siberia trap or the methane hydrates can raise the temperature at the end of the Changhsingian, which may trigger the P-Tr mass extinction. Four mass extinctions F-F, P-Tr, Tr-J, K-Pg and the Cm-O extinction occurred during the low level of climate gradient (Fig 3e); as a special case, the O-S mass extinction occurred during the high level of climate gradient (Fig 3e). Taken together, these observations show that low climate gradient can contribute to the mass extinctions substantially. Generally speaking, the mass extinctions might occur coincidentally when brought together two or more factors among the low climate gradient, marine regression, and the like (Fig 3e).

The interactions among the earth's spheres can explain the minor extinctions, radiations and rebounds. The minor descents in the biodiversity net fluctuation curve relate to mid-term or short-term changes in sea level and climate at the tectonic time scale, which correspond approximately to the minor extinctions respectively (Benton and Harper 1997; Sepkoski 1998) (Fig 1b, 3g). It is shown that most minor extinctions in the Paleozoic era corresponded to low climate gradient, and most minor extinctions during the Mesozoic and Cenozoic corresponded to marine regressions (Fig 1b, 3g). After mass extinctions, both the climate gradient and sea level tended to increase; henceforth the radiations and rebounds occurred. Especially for the case of the P-Tr mass extinction, the climate gradient remained low throughout the Lower Triassic (Fig 1b, 3e), so it took a long time for the recovery from the P-Tr mass extinction.

\section{On the Phanerozoic biodiversity curve}
\subsection{Reconstruction of the Phanerozoic biodiversity curve}

Now, let's reconstruct the Phanerozoic biodiversity curve. Previously, the exponential growth trend of the Phanerozoic biodiversity curve might have been explained by the exponential growth trend of the genome size evolution (Fig 1a), and the deviation from the exponential growth trend of the Phanerozoic biodiversity curve might have been explained by the climato-eustatic curve (Fig 1b). By combining the exponential growth trend of the genome size evolution and the climato-eustatic curve together, the biodiversity curve based on fossil records can be reconstructed based on the genomic, climatic and eustatic data (Fig 1c). The technique details are shown in Fig 4. The reconstructed biodiversity net fluctuation curve is obtained based on the climato-eustatic curve times a weight, which is equal to the standard deviation of the biodiversity net fluctuation curve (Fig 4). And the slope and the intercept of the line that corresponds to the exponential growth trend of the reconstructed biodiversity is obtained as follows. Its slope is defined by the slope of the line that corresponds to the exponential growth trend of genome size evolution (Fig 4), and its intercept is defined by the intercept of the line that corresponds to the exponential growth trend of the biodiversity curve (Fig 4). The reconstructed biodiversity curve thus obtained agrees nicely with the Bambach et al.'s biodiversity curve within the error range of fossil records (Fig 1c).

\subsection{Declining extinction and origination rates}

The above split and reconstruction scheme of the Phanerozoic biodiversity curve can explain not only the exponential growth trend and the detailed fluctuations, but also the roughly synchronously declining origination rate and extinction rate throughout the Phanerozoic eon (Flessa and Jablonski 1985; Van Valen 1985; Sepkoski 1991; Gilinsky 1994; Alroy 1998) (Fig 1d, 5a, 5b, 5c). Ramp and Sepkoski first found the declining extinction rate throughout the Phanerozoic eon (Raup and Sepkoski 1982; Newman and Eble 1999). And Sepkoski also found the declining origination rate throughout the Phanerozoic eon (Sepkoski 1998). The numbers of origination genera and extinction genera are marked in the Bambach et al.'s biodiversity curve. The origination rate and extinction rate are obtained from the numbers of origination or extinction genera divided by the corresponding time intervals (Fig 5c). The difference between the origination rate and extinction rate is the variation rate of number of genera (Fig 5c). It is shown that both the origination rate and extinction rate decline roughly synchronously throughout the Phanerozoic eon (Jablonski 2008) (Fig 5c).

The rough synchronicity between the origination rate and the extinction rate can be explained according to the split and reconstruction scheme. The Phanerozoic biodiversity curve can be divided into two independent parts: an exponential growth trend and a biodiversity net fluctuation curve (Fig 2h, 4). Notice that there is no declining trend in the climato-eustatic curve, the expectation value of which is constant (Fig 1b, 5a); whereas the biodiversity curve was continuously boosted upwards by the exponential growth trend of genome size evolution (Fig 1a). The expectation value of the difference between the origination rate and extinction rate corresponds to the exponential growth trend of biodiversity boosted by the genome size evolution (Fig 5a, 5b). If too large or too small origination rate brings the biodiversity to deviate from the exponential growth trend, the extinction rate has to response by increasing or decreasing so as to maintain the constant exponential growth trend of the biodiversity (Fig 5b, 5c). The rough synchronicity between the origination rate and the extinction rate shows that the growth trend of biodiversity at the species level was constrained strictly by the exponential growth trend of genome size evolution at the sequence level.

The declining trends of the origination rate and the extinction rate might be explained according to the split and reconstruction scheme. It is necessary to assume that there is an upper limit of ability to create new species in the living system, which is due to the constraint of the earth's environment at the species level, the limited mutation rate at the sequence level and the physical constraint of the size of cellular membrane (Fig 5a). Although there is no declining trend for the variation rate of the biodiversity net fluctuation curve (Fig 3g, 5a), the trend of the variation rate of the Phanerozoic biodiversity curve tends to decline in general, because the exponential growth trend in genome size evolution is added in the denominator in the calculations (Fig 1d, 5b, 5c). The declining variation rate throughout the Phanerozoic eon shows that the genome size evolution played a primary role in the growth of biodiversity. The exponential growth trend of biodiversity driven by the genome size evolution can gradually reduce the impact of sea level fluctuations and climate changes on the biodiversity fluctuations (Fig 1d, 5b). The rough synchronicity between the origination rate and the extinction rate can explain both the declining origination rate and the declining extinction rate throughout the Phanerozoic eon (Fig 5b, 5c).

\subsection{Strategy of the evolution of life}

The lithosphere consists of huge mass of solid matter, the hydrosphere huge mass of liquid matter and the atmosphere huge mass of gaseous matter. But the matter in biosphere is much less than the matter in other spheres on the earth's surface. Actually, the biosphere consists of quite a lot of homochiral carbonaceous matter. The origin of homochirality is so difficult that it only happened at the beginning of life; the ability to produce a homochiral system has to be passed down from generation to generation throughout the evolution of life. The key in the reproductive process of life would be regarded as passing the general ability of producing homochial carbonaceous matter by diverse species, rather than be trivially regarded as passing the diverse genetic materials to the respective offsprings. It seems that what the biosphere has been doing is to continuously increase the production of homochiral carbonaceous matter, which becomes strategic material reserves for the robust living system to survive the changing environments. 

There exists a growing trend for the complexity of life throughout the evolution of life. But there are no obvious increasing or decreasing trends in the complexity of lithosphere, hydrosphere or atmosphere. The driving force to boost complexity of life might be attributed to the sequence evolution that distinguishes biosphere from other spheres on the earth's surface. According to statistical analysis of genome sizes of contemporary species, there is a rough correlation between the average complexity and the average genome sizes in taxa, which might be invalid in the case of individual species (Gregory 2005). For example, the average genome size for Diploblostica is less than the average genome size for more complex Protostomia, and the latter is still less than the average genome size for still more complex Deuterostomia. So the growing trend in genome size evolution may indicate the growing trend in complexity of life (Sharov 2007; Li and Zhang 2010). There are diverse species within biosphere, and the biodiversity can be boosted by the increasing amounts of genome sequences. A universal genome format has been observed in the genomic codon distributions as a common feature of life (Li 2018-II). Such a common feature of life at the sequence level guarantees that the sequence evolution of life in a statistical manner keeps away from random sequences. 

The living system differs from non-living system in that the latter tends to abide by the ergodic theory in statistical physics while the former tends to judge upon the information of bounded rationality stored in their genetic materials; and the living system is not only able to store valuable information but also able to develop itself. The biodiversity evolves independently at the sequence level and at the species level. This ensures the adaptation of life to the earth's changing environment. The homochirality is an essential feature of life. At the beginning of life, the homochirality, the genetic code, and the universal genome format had been established. The earth's changing environment influenced the living system only at the species level, which cannot directly threaten the evolution of chiral macromolecules. The strategy of adaptation is that the living system can generate numerous redundant species based on the independent evolution of chiral macromolecules; thus the living system sacrificed majority of extinct species in exchange for the minority of outstanding species that survived from the earth's changing environments. By virtue of this strategy of adaptation, the chiral living system may adapt to a wide range of environmental changes on planets. 

\section{Conclusion and discussion}

This study is an exploratory attempt to explain the Phanerozoic biodiversity curve. The Phanerozoic biodiversity curve based on fossil record has been reconstructed based on genomic data, climatic data and eustatic data. In this article, the trend in genome size evolution throughout the history of life might be explained based on genome size data. And a consensus climatic curve and a consensus eustatic curve have been obtained based on the respective results in literatures. A Phanerozoic biodiversity curve was reconstructed by combining the trend of genome size evolution and the fluctuations in the consensus climatic and eustatic curves, which agrees with the Phanerozoic biodiversity curve based on fossil records. And the declining extinction and origination trends during Phanerozoic eon might be explained according to the impact of the evolution of life at sequence level on the evolution at the species level. 

\section*{Acknowledgements} My warm thanks to Jinyi Li for valuable discussions. I wish to thank the contributors of the biological and geological data used in this study. Supported by the Fundamental Research Funds for the Central Universities.

\clearpage

\begin{figure}
  \centering
  {\small \bf a} \includegraphics[width=16cm]{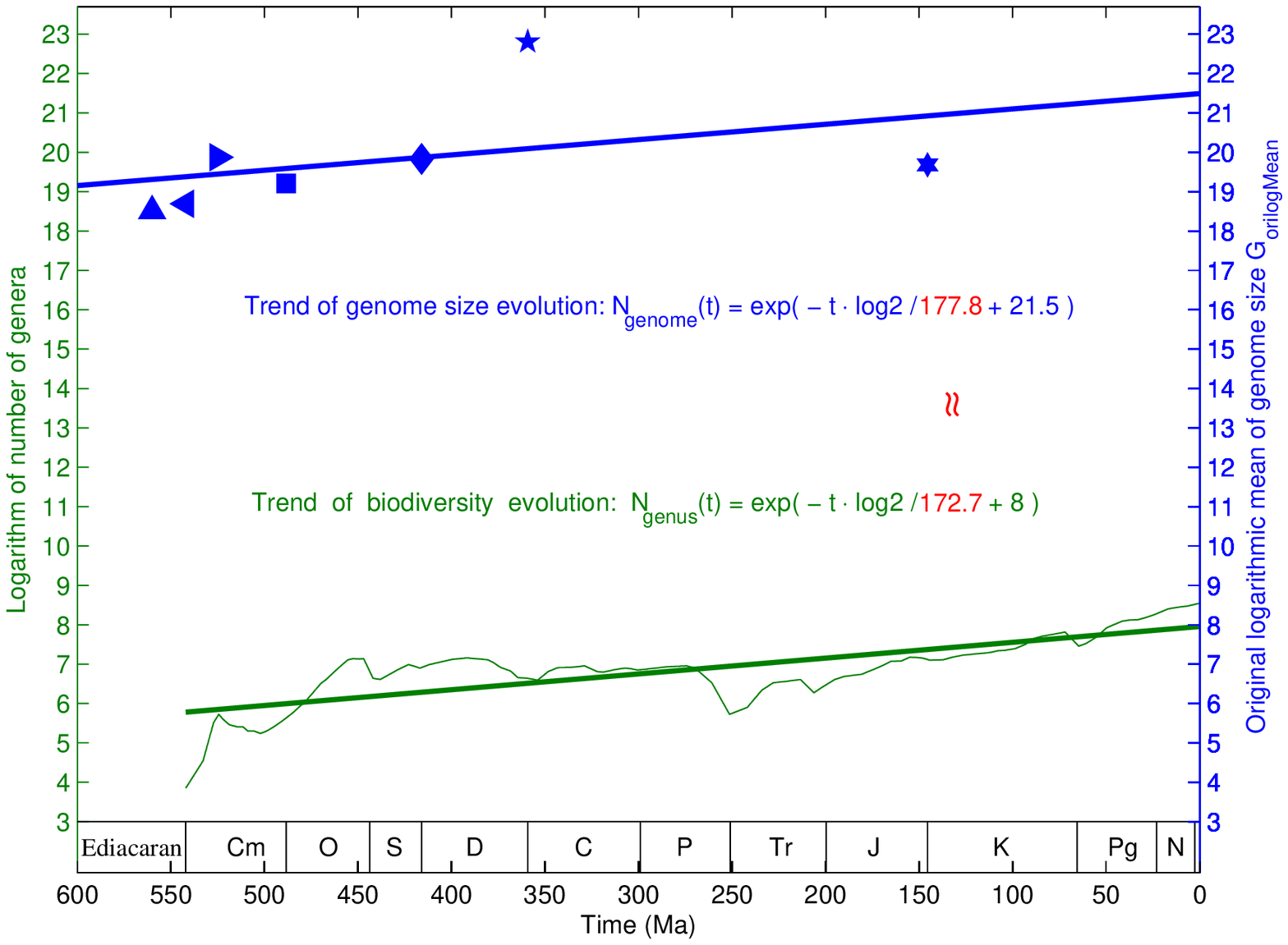}\\ \vspace{0.5cm}
  {\small \bf b} \includegraphics[width=16cm]{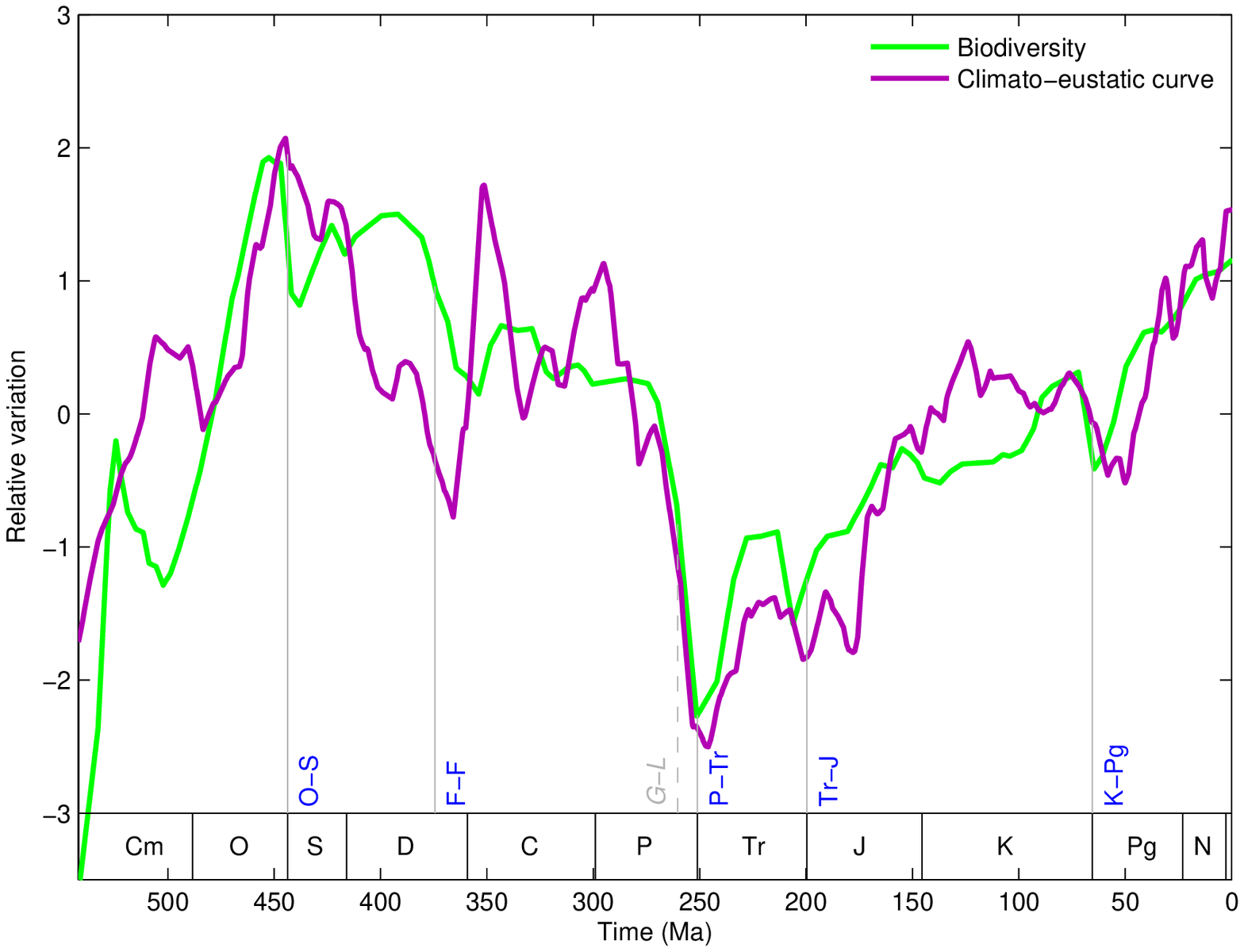}
\end{figure}

\begin{figure}
  \centering
  {\small \bf c} \includegraphics[width=16cm]{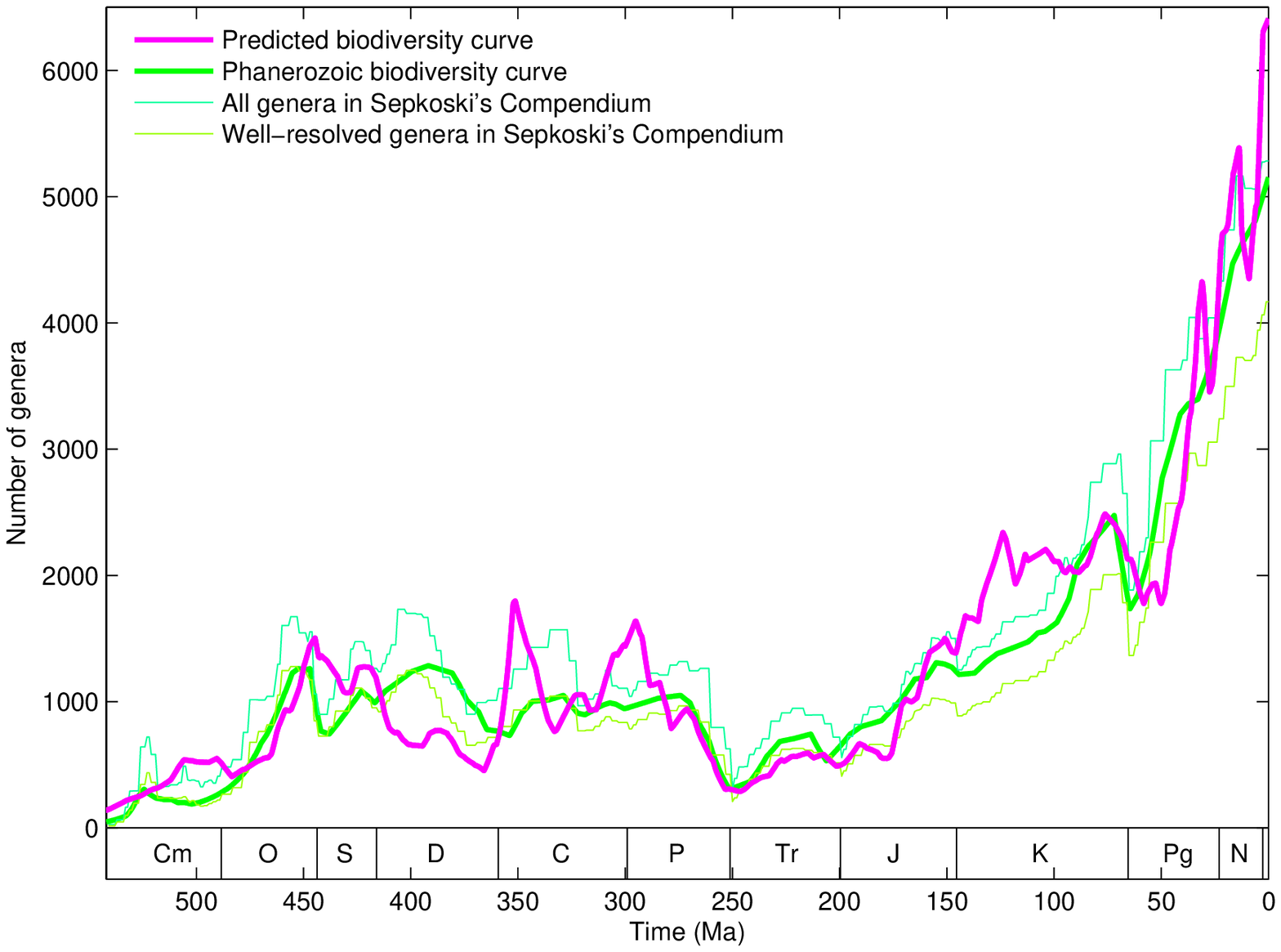}\\ \vspace{0.5cm}
  {\small \bf d} \includegraphics[width=16cm]{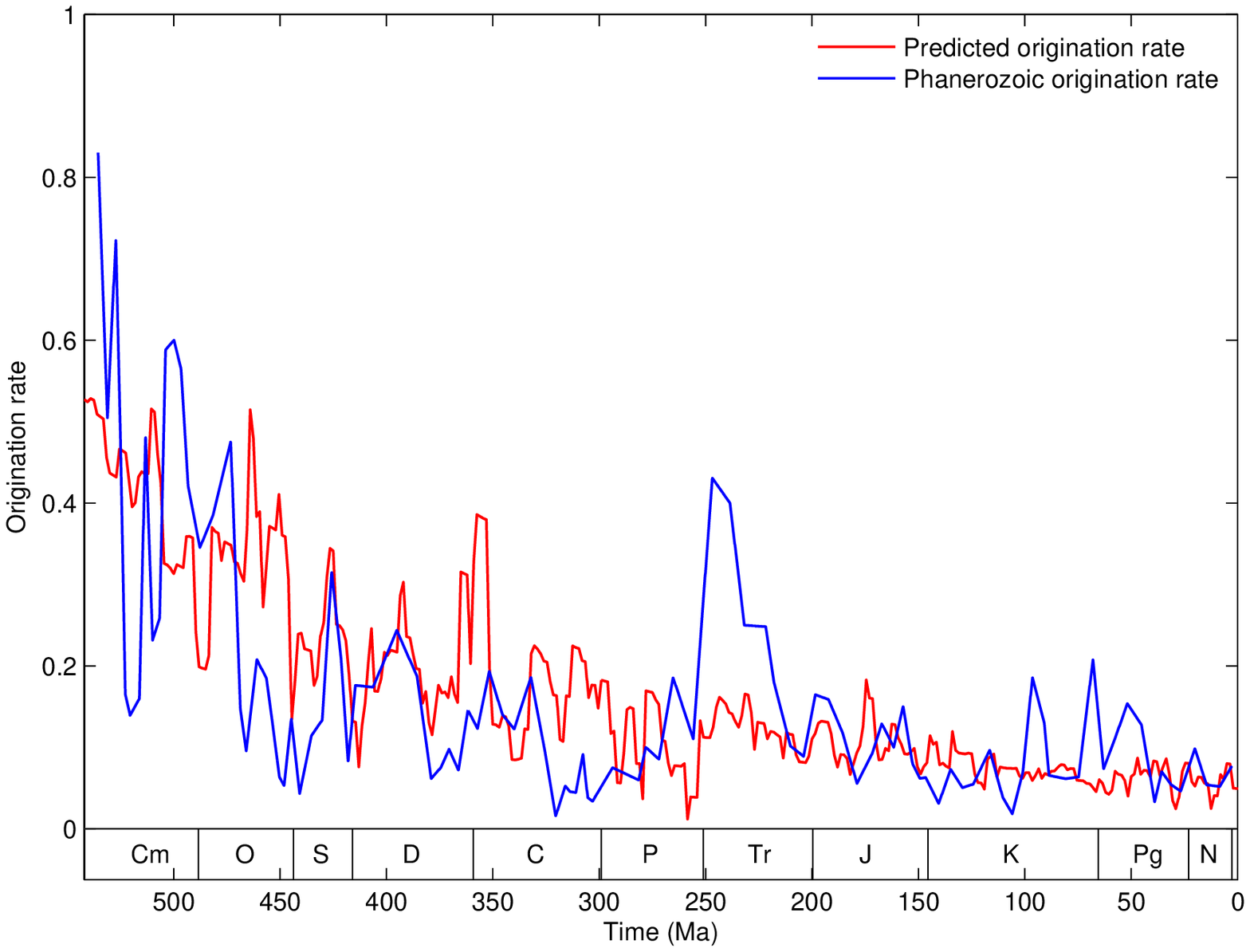}
\end{figure}

\begin{figure}
  \centering
  \caption{Explanation of the Phanerozoic biodiversity curve. {\bf a} Explanation of the trend of the Phanerozoic biodiversity curve by the trend of genome size evolution. The exponential growth rate of the Phanerozoic biodiversity curve is about equal to the exponential growth rate of genome size evolution (indicated in red). Refer to Fig 2g and Fig 2h to see the exponential trend of the genome size evolution and the exponential trend of the Phanerozoic biodiversity, respectively. {\bf b} Explanation of the fluctuations of the Phanerozoic biodiversity curve based on climate change and sea level fluctuations, which indicates the tectonic cause of the mass extinctions. Refer to Fig 3d and Fig 2h to see the climato-eustatic curve and the biodiversity net fluctuation curve, respectively. {\bf c} Reconstruction of the Phanerozoic biodiversity curve based on genomic, climatic and eustatic data, whose result agrees with the biodiversity curve based on fossil records. The Bambach et al.'s Phanerozoic biodiversity curve (Fig 2h) based on fossil records is in green, and the predicted biodiversity curve in pink is based on the climato-eustatic curve in Fig 3d and the exponential trend of genome size evolution in Fig 2g. Refer to Fig 4 for details. {\bf d} Explanation of the declining origination rate throughout the Phanerozoic eon. Refer to Fig 5c to see the declining origination rate and extinction rate based on fossil records; and refer to Fig 5b to see the explanation of these declining rates.}
\end{figure}

\clearpage \begin{figure}
 \centering
 {\small \bf a} \includegraphics[width=8cm]{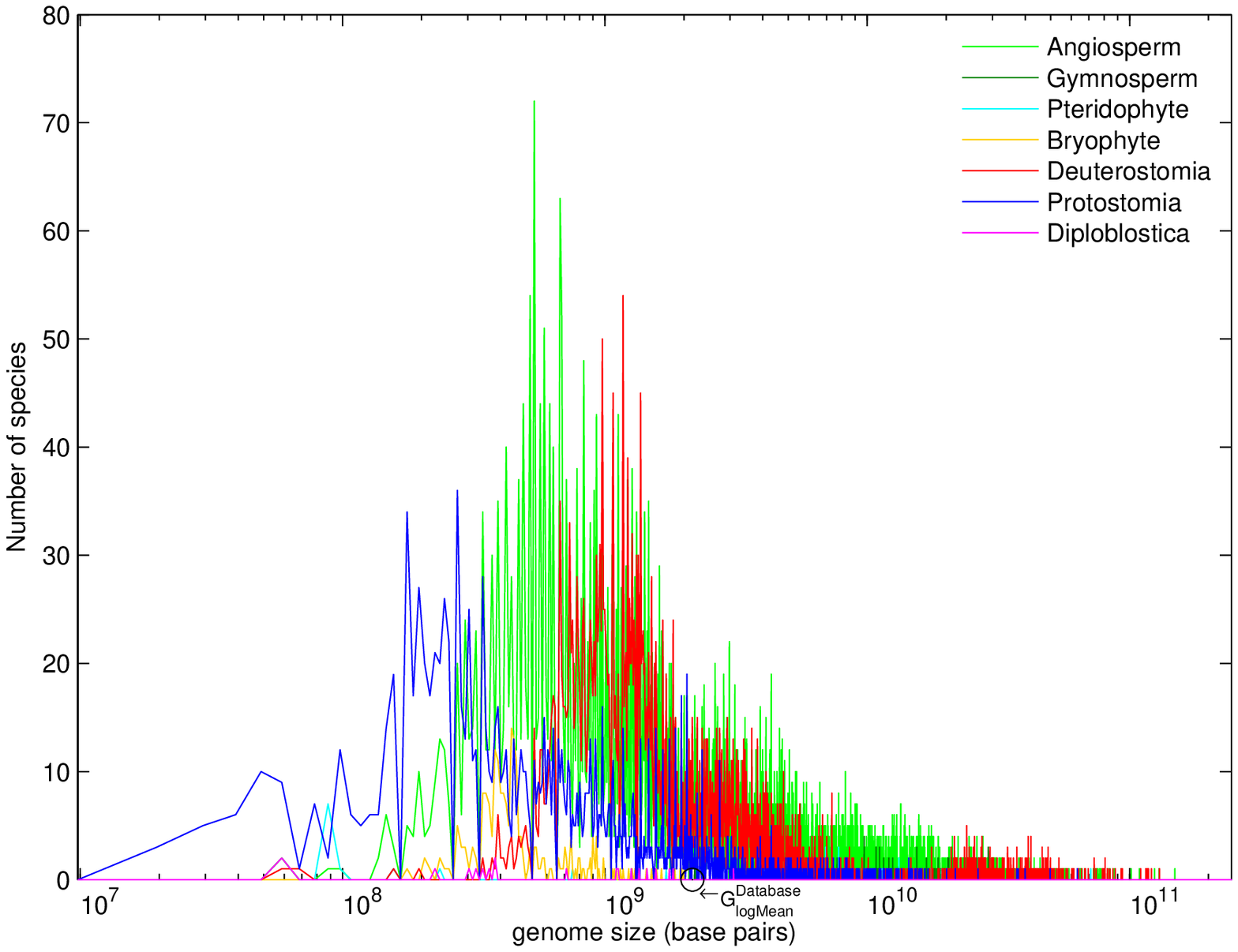}
 {\small \bf b} \includegraphics[width=8cm]{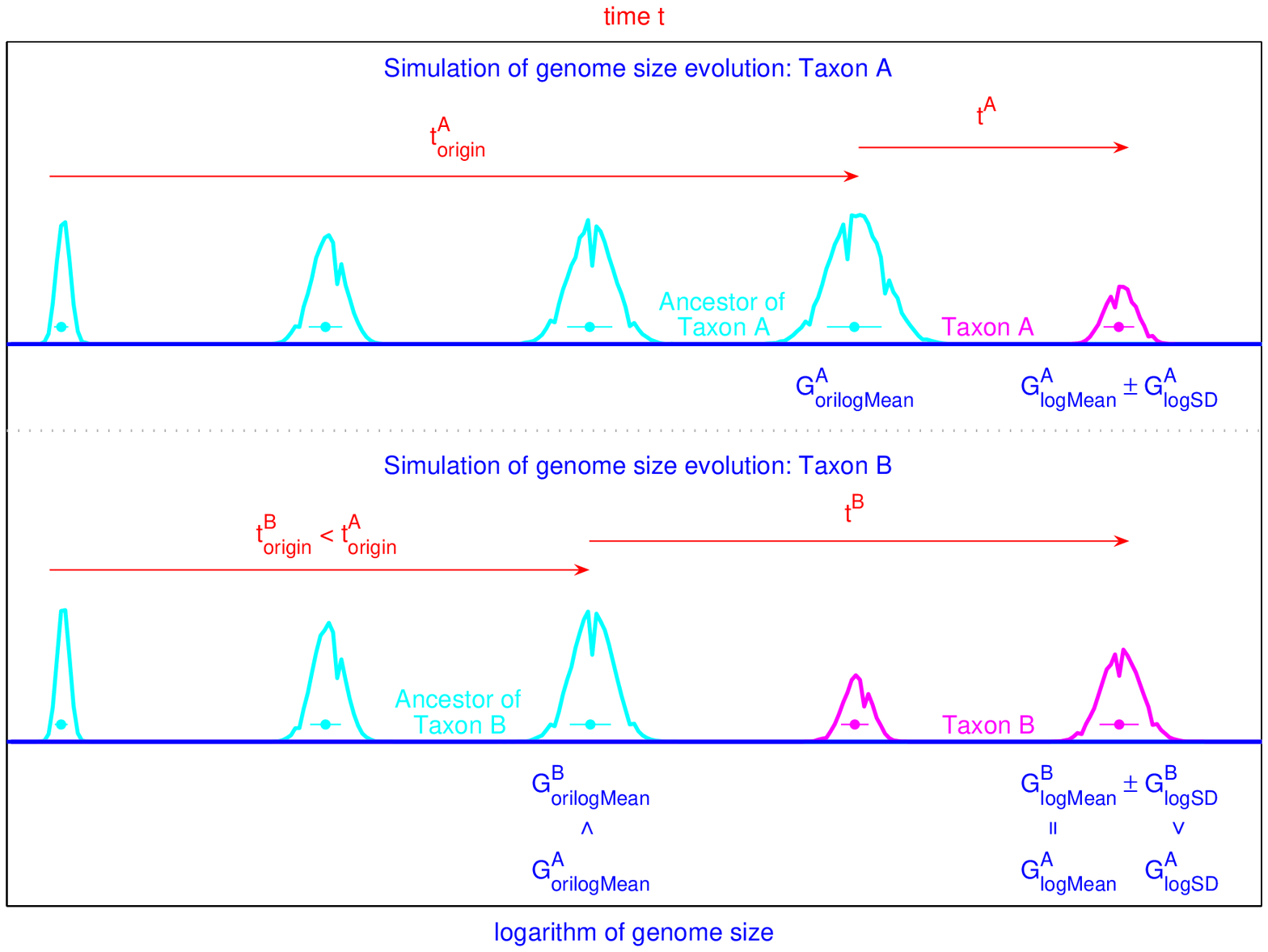}\\
 {\small \bf c} \includegraphics[width=8cm]{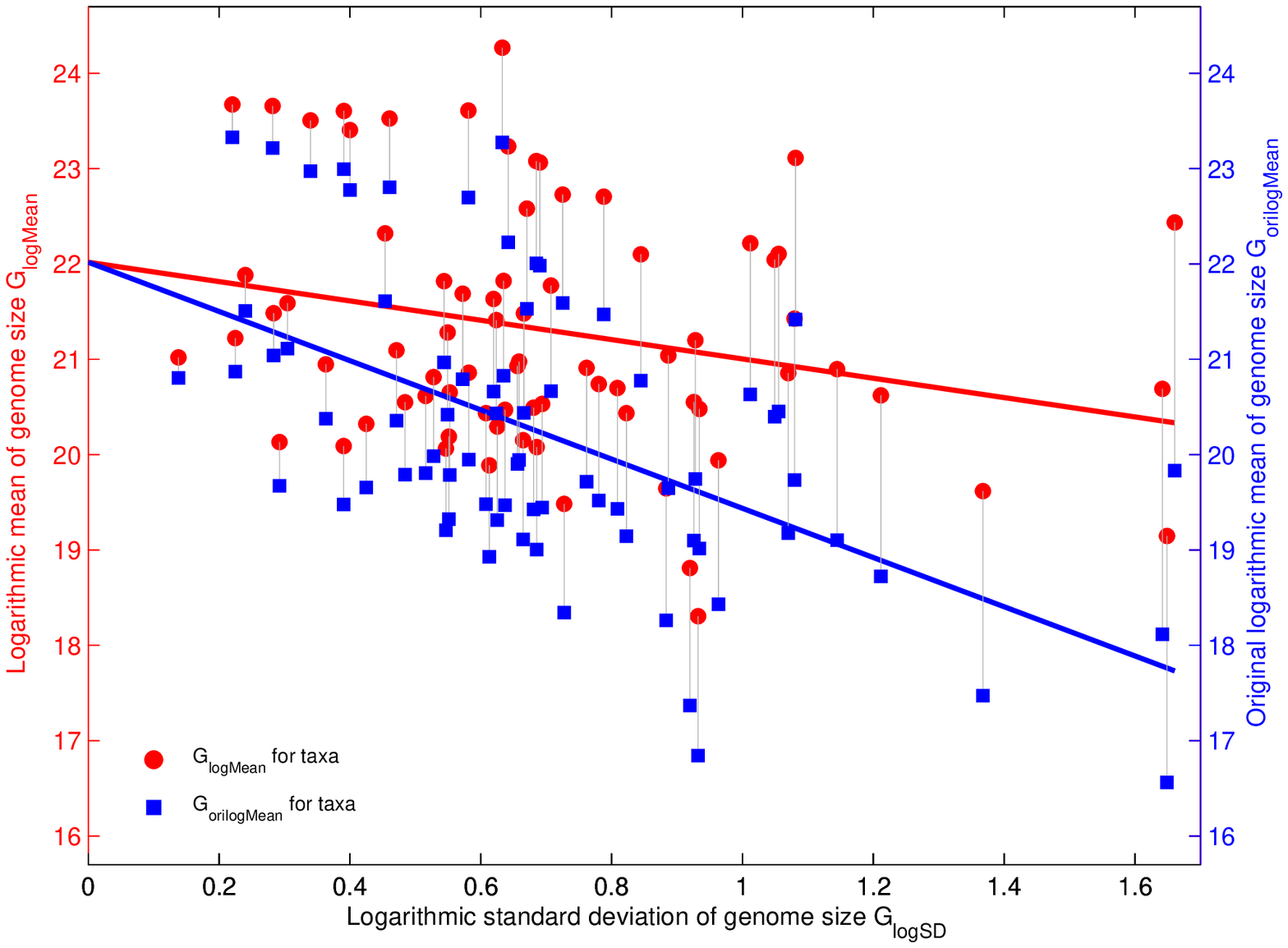}
 {\small \bf d} \includegraphics[width=8cm]{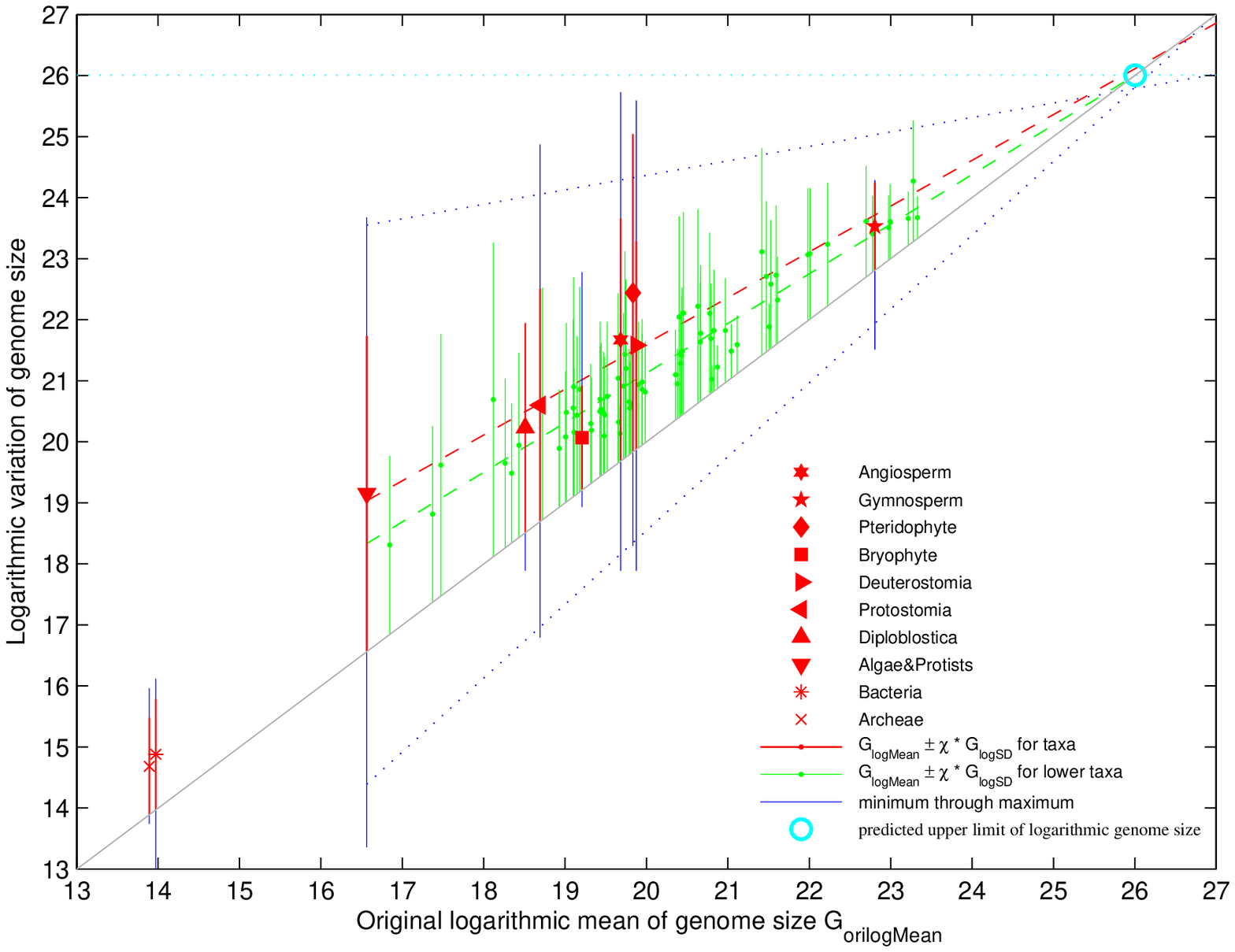}\\
 {\small \bf e} \includegraphics[width=8cm]{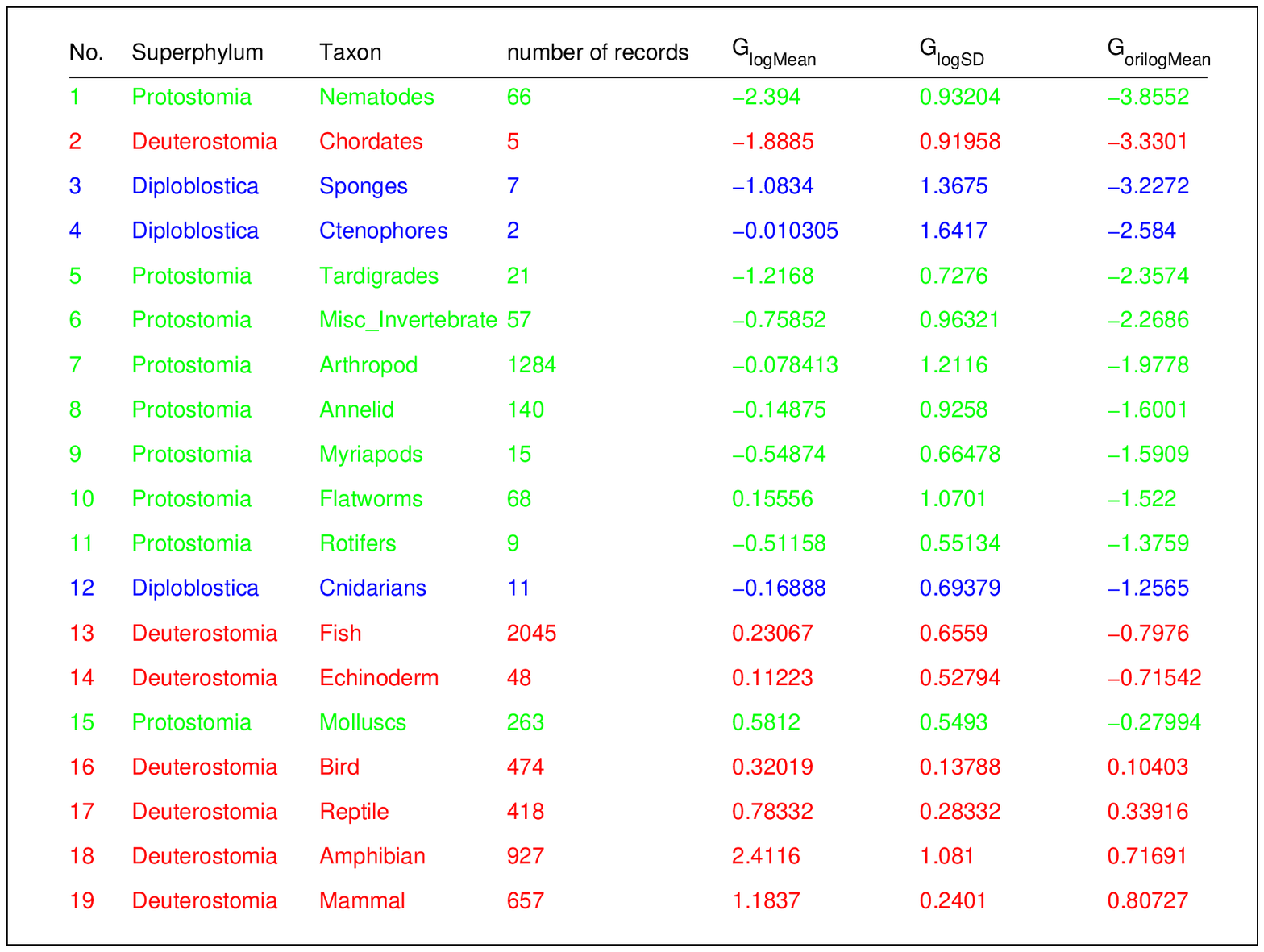}
 {\small \bf f} \includegraphics[width=8cm]{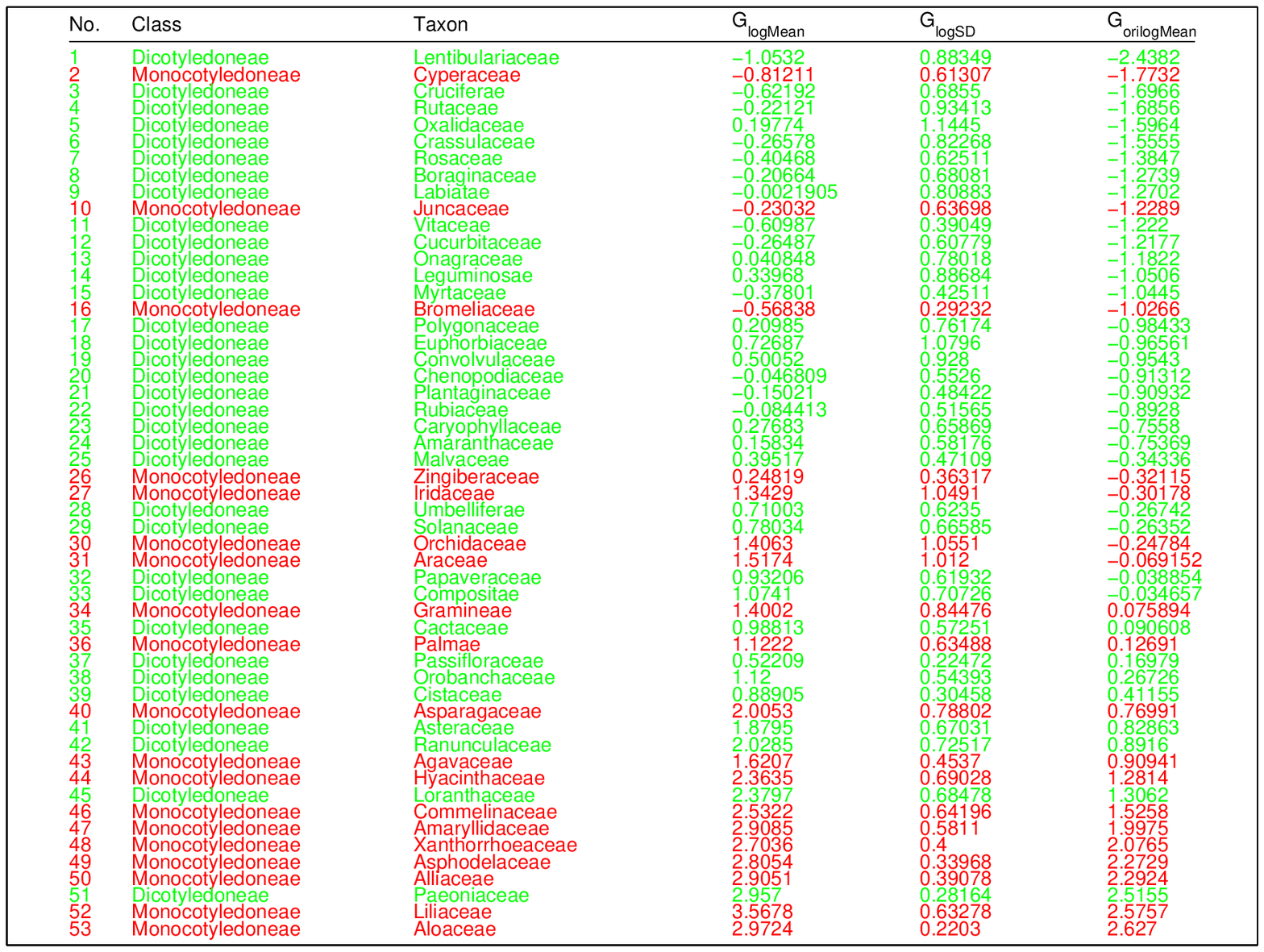}\\
 {\small \bf g} \includegraphics[width=8cm]{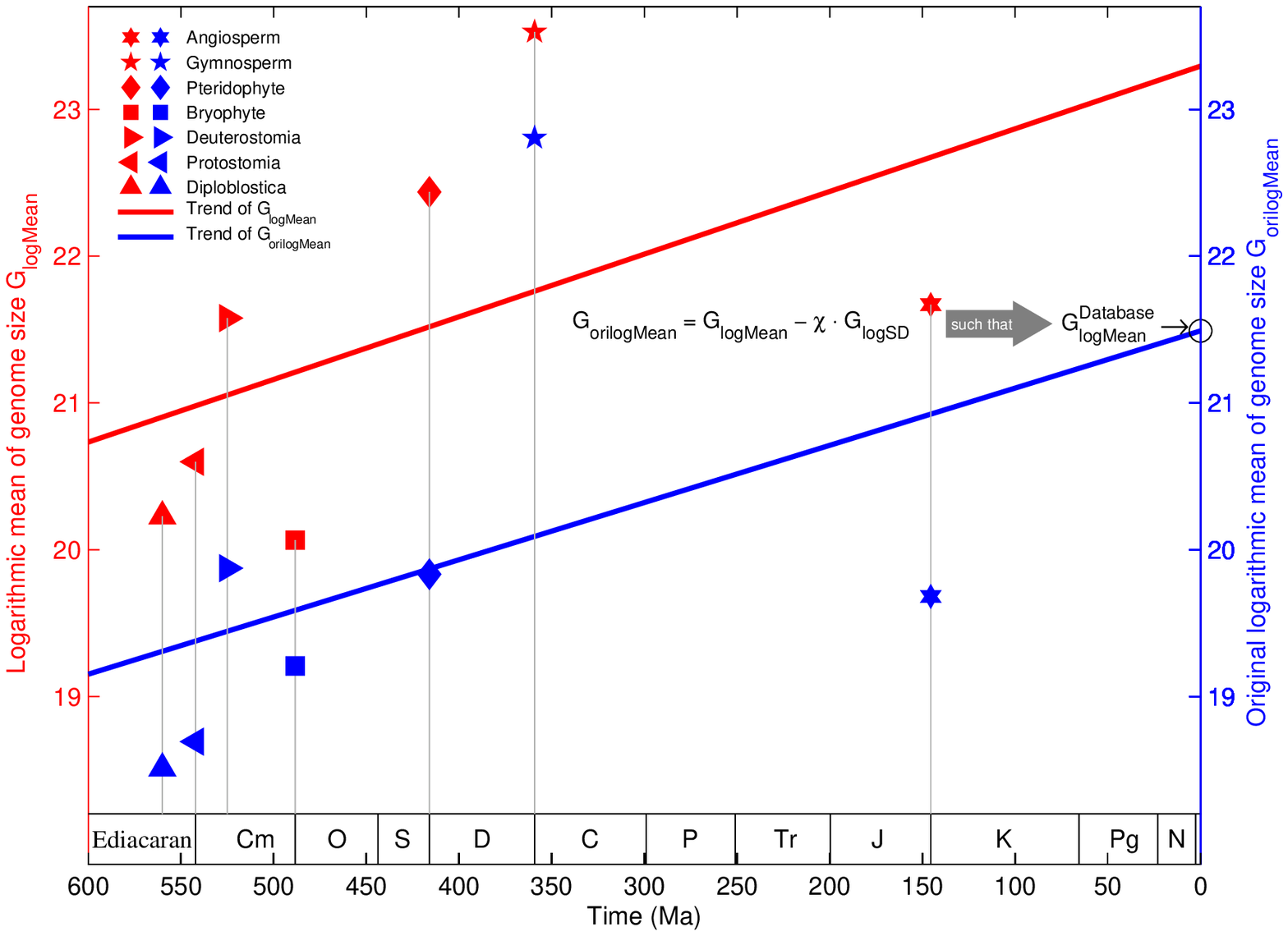}
 {\small \bf h} \includegraphics[width=8cm]{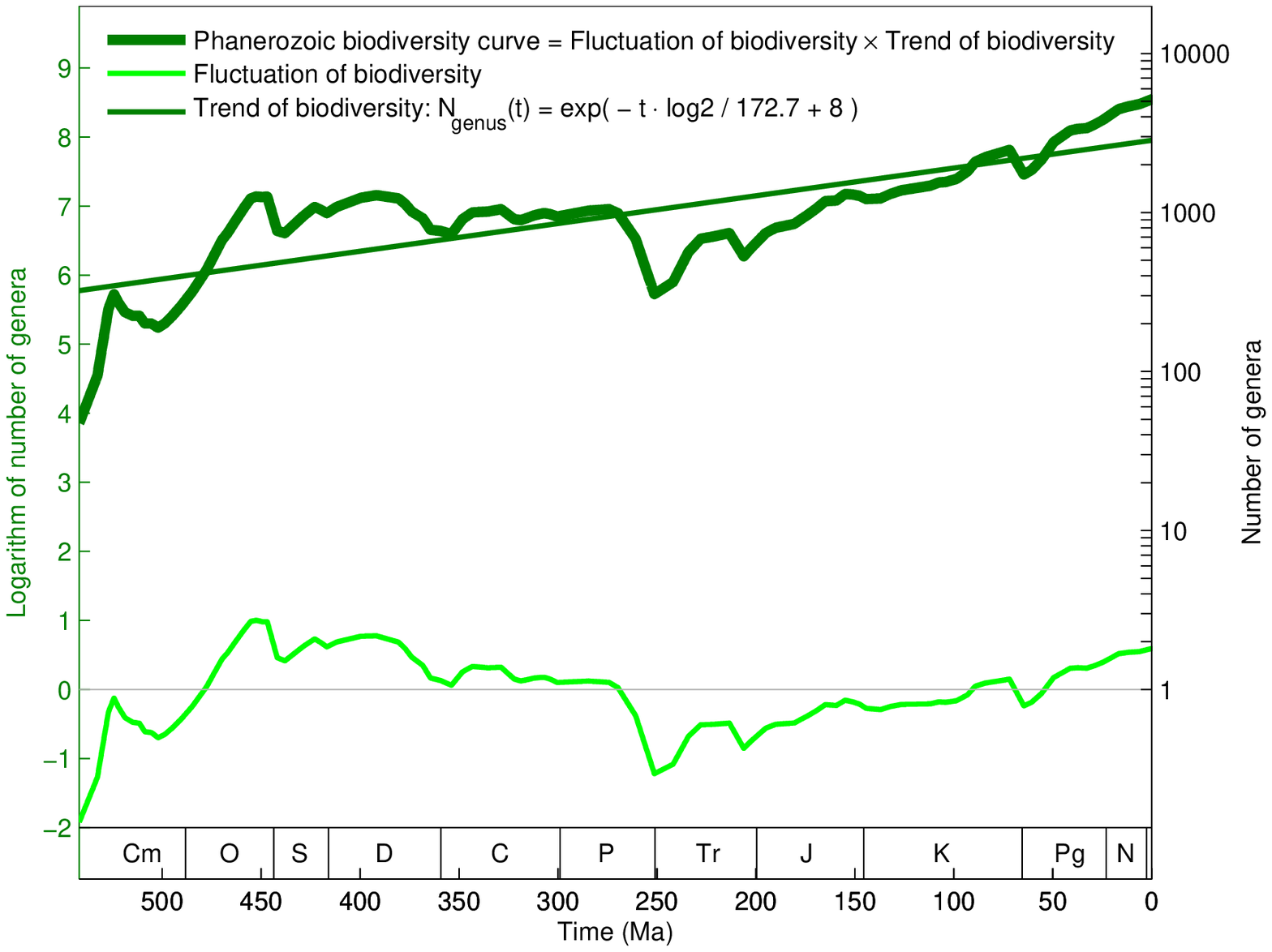}
\end{figure}

\clearpage \begin{figure}
 \centering
 \caption{}
\end{figure}

{Detailed explanation of the trend of the Phanerozoic biodiversity curve. {\bf a} Logarithmic normal distribution of genome sizes for taxa. The genome size distribution for all the species in the two databases satisfies logarithmic normal distribution very well, due to the additivity of normal distributions. The logarithmic mean of genome sizes for all the species in the two databases is marked in the present figure, which represents the intercept of the trend of genome size evolution in Fig 2g. {\bf b} Simulation of the evolution of genome size. The genome size doubles in the stochastic process model, which results in the logarithmic genome size distributions. The invariance of the genomic codon distributions after genome duplication is considered in the model (Fig 4d in Li 2018-II). Two taxa $A$ and $B$ are compared in the simulation. The greater the logarithmic standard deviation of genome sizes in a taxon is ($G^{B}_{logSD}$ $>$ $G^{A}_{logSD}$), the earlier the taxon originated ($t^{B}_{origin}$ $<$$t^{A}_{origin}$). Besides, the less the original logarithmic mean of genome size is ($G^{B}_{orilogMean}$ $<$ $G^{A}_{orilogMean}$), the earlier the taxon originated ($t^{B}_{origin}$ $<$$t^{A}_{origin}$). {\bf c} The greater the logarithmic standard deviation of genome sizes $G_{logSD}$ is, the less the original logarithmic mean of genome sizes $G_{orilogMean}$ is, and meanwhile also the less the logarithmic mean of genome sizes $G_{logMean}$ is. These statistical analyses of genome sizes for the taxa are obtained based on the databases for animals and plants. {\bf d} The less the original logarithmic mean of genome sizes $G_{orilogMean}$ is, the less the logarithmic mean of genome sizes $G_{logMean}$ is (red), and meanwhile the greater the logarithmic standard deviation of genome sizes $G_{logSD}$ is (green or blue). The original logarithmic mean of genome sizes $G_{orilogMean}$ indicate the origin time of taxon (Fig 2b). The convergent point (namely the upper vertex of the triangle area in the present figure) indicates the upper limit of logarithmic genome sizes at present. {\bf e} The three animal superphyla Diploblostica, Protostomia and Deuterostomia have been roughly distinguished based on the order of the original logarithmic means of genome sizes $G_{orilogMean}$ for the animal taxa (Fig 2d). This supports that the order of $G_{orilogMean}$ indicate the evolutionary chronology. {\bf f} The two Angiosperm classes Dicotyledoneae and Monocotyledoneae have been roughly distinguished based on the order of the original logarithmic means of genome sizes $G_{orilogMean}$ for the Angiosperm taxa (Fig 2d). This also supports that the order of $G_{orilogMean}$ indicate the evolutionary chronology. {\bf g} The exponential trend of genome size evolution based on the relation between the origin time of taxa and their original logarithmic means of genome sizes. The red regression line of $G_{logMean}$ against the origin time should be shifted downwards to the blue regression line of $G_{orilogMean}$ against the origin time, by considering the inverse relationship between $G_{logSD}$ and the origin time (Fig 2b). The parameter $\chi$ in the definition of $G_{orilogMean}$ is determined by letting the intercept of the blue regression line be the average logarithmic genome size $G^{Database}_{logMean}$ (Fig 2a). {\bf h} The exponential trend of Bambach et al.'s Phanerozoic biodiversity curve. The raw biodiversity net fluctuation curve (thin light green) is obtained by subtracting the linear trend of logarithmic biodiversity curve (thin dark green) from the logarithmic biodiversity curve (thick dark green).}

\clearpage \begin{figure}
 \centering
 {\small \bf a} \includegraphics[width=8cm]{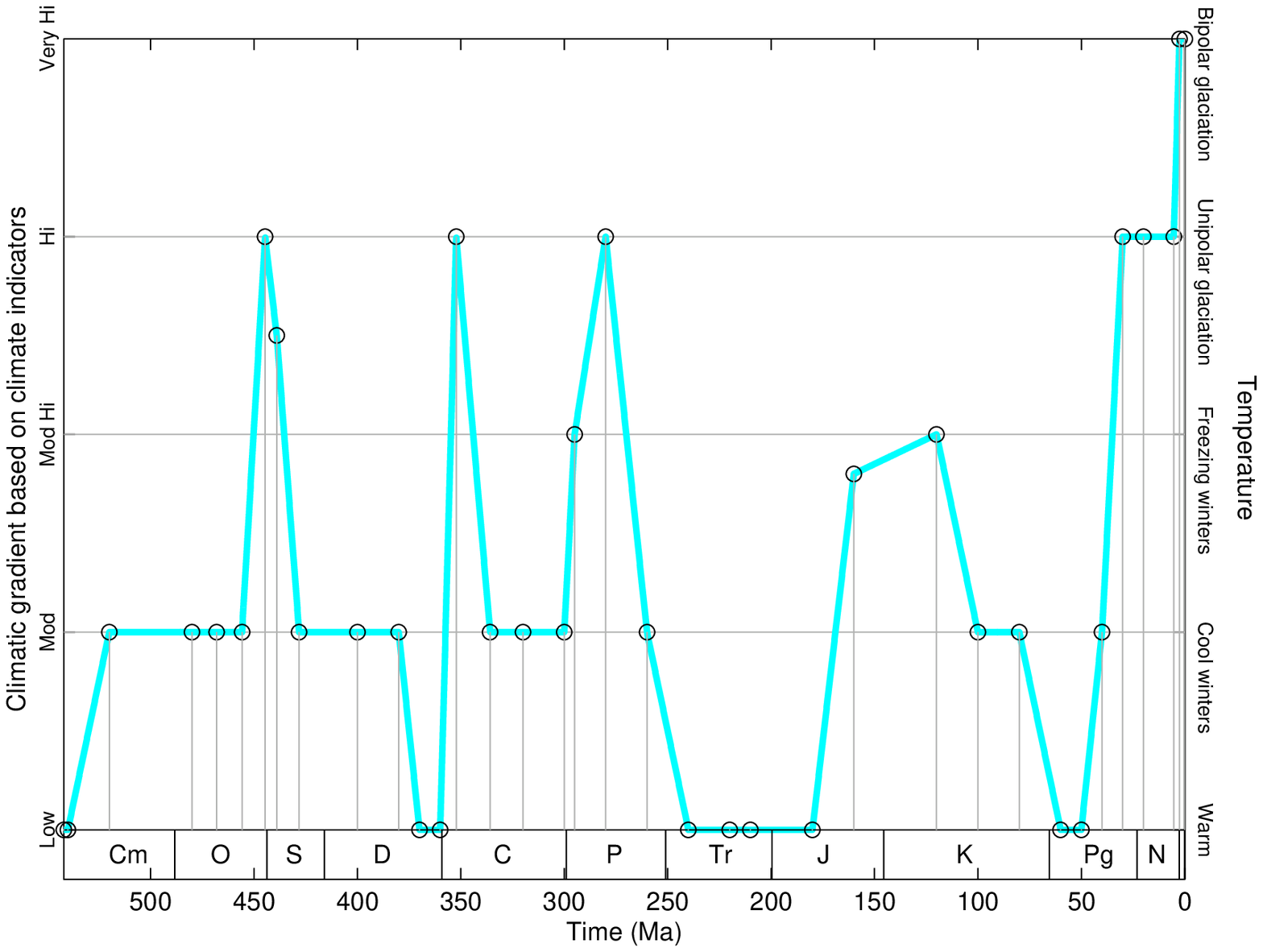}
 {\small \bf b} \includegraphics[width=8cm]{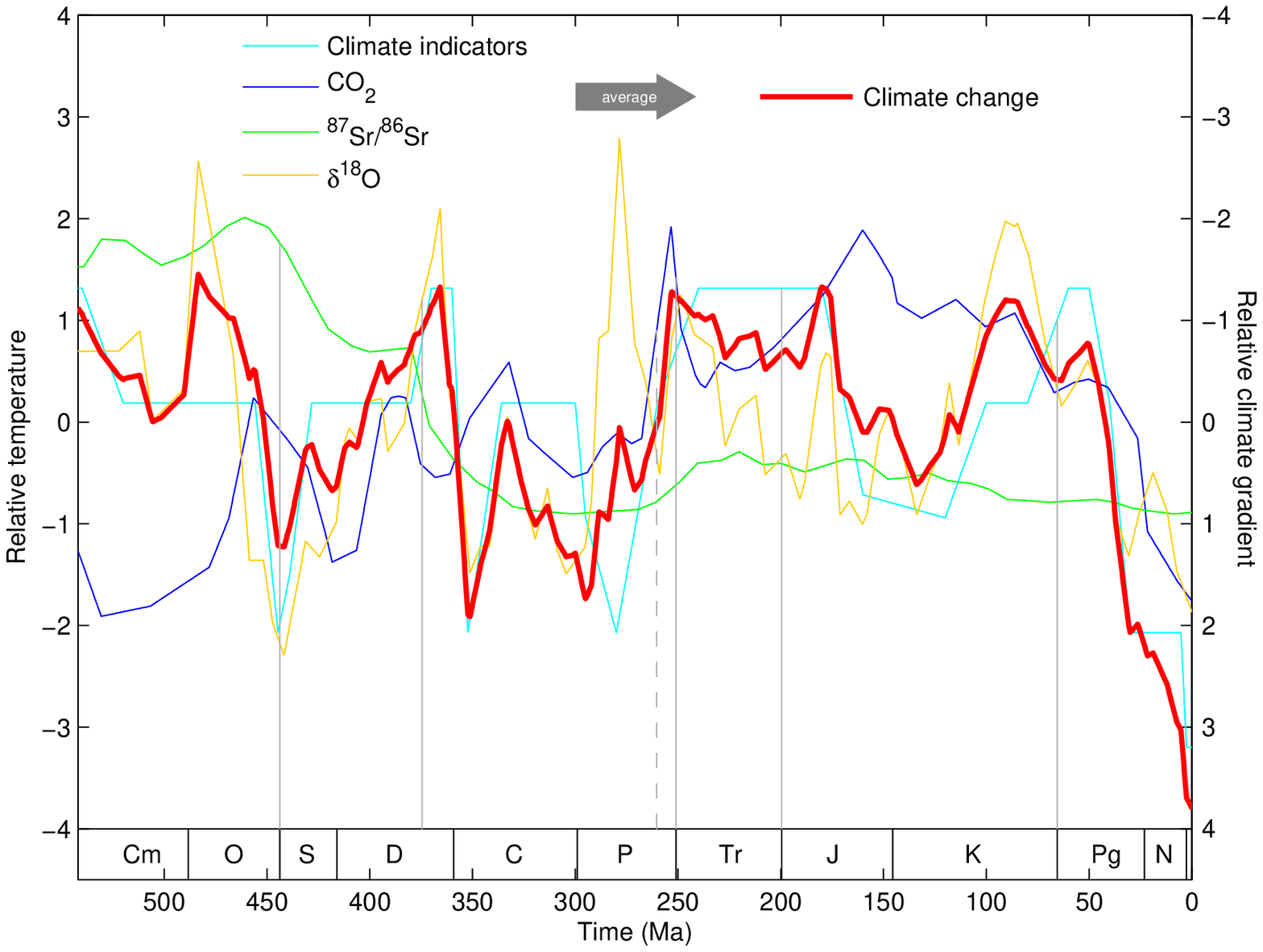}\\
 {\small \bf c} \includegraphics[width=8cm]{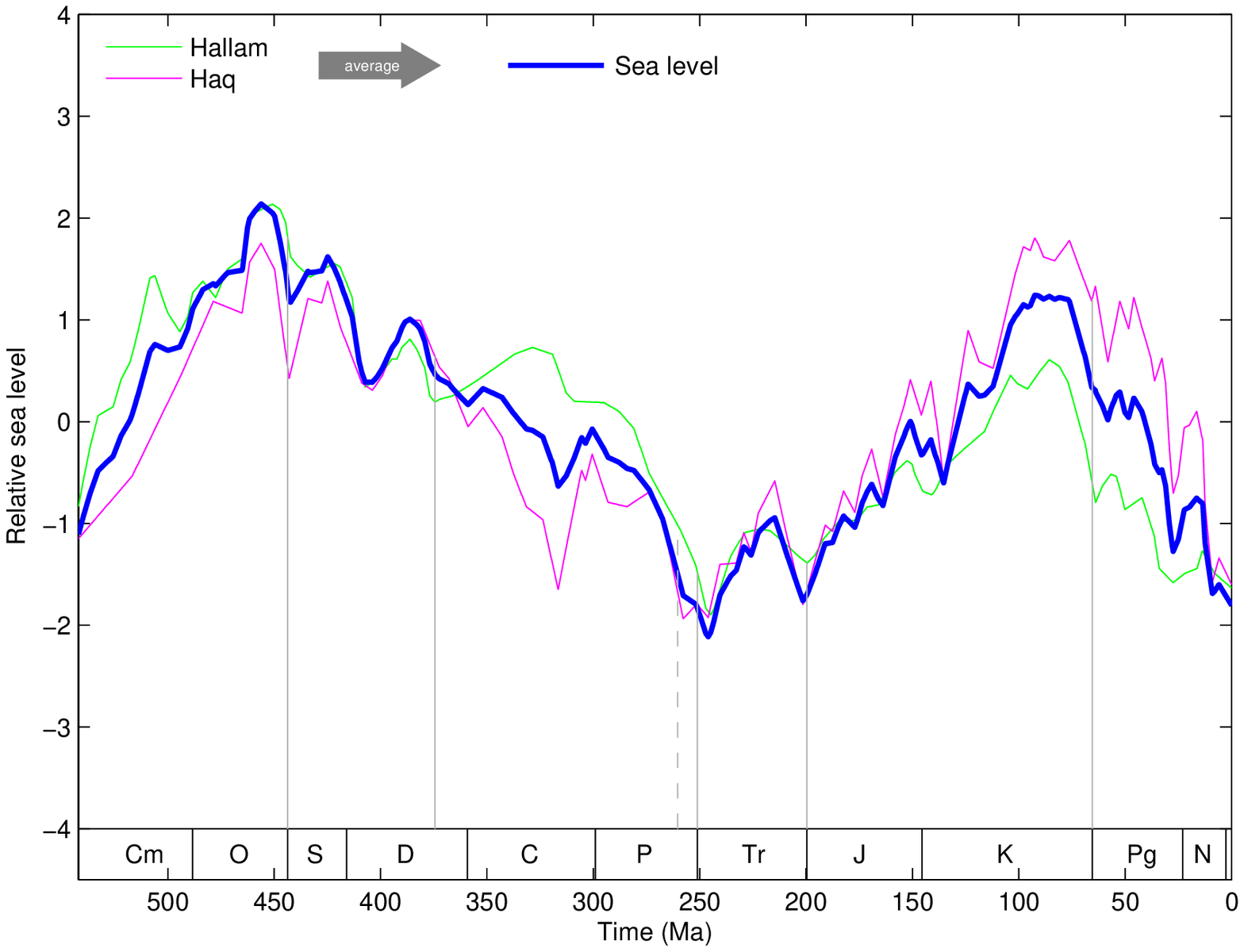}
 {\small \bf d} \includegraphics[width=8cm]{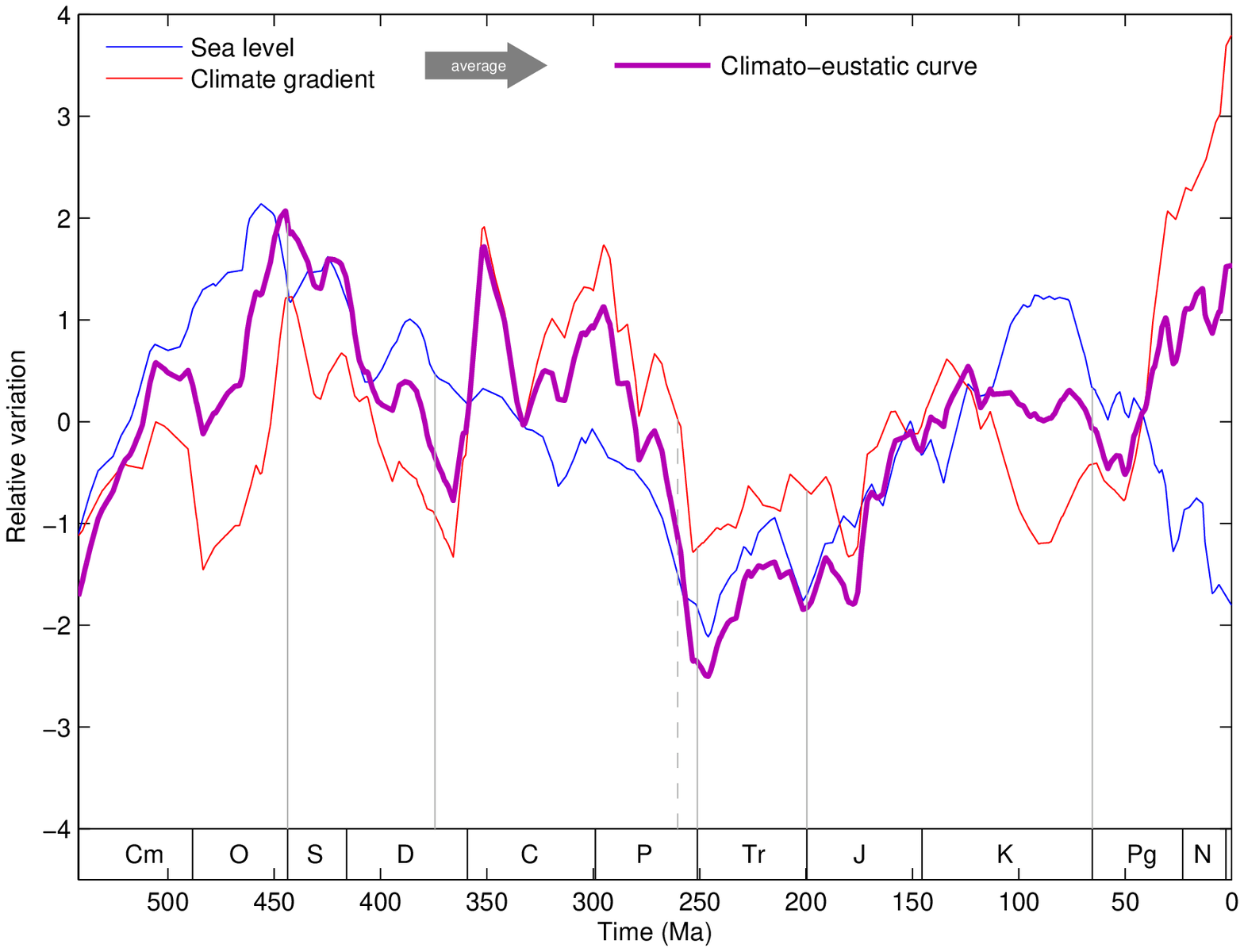}\\
 {\small \bf e} \includegraphics[width=16cm]{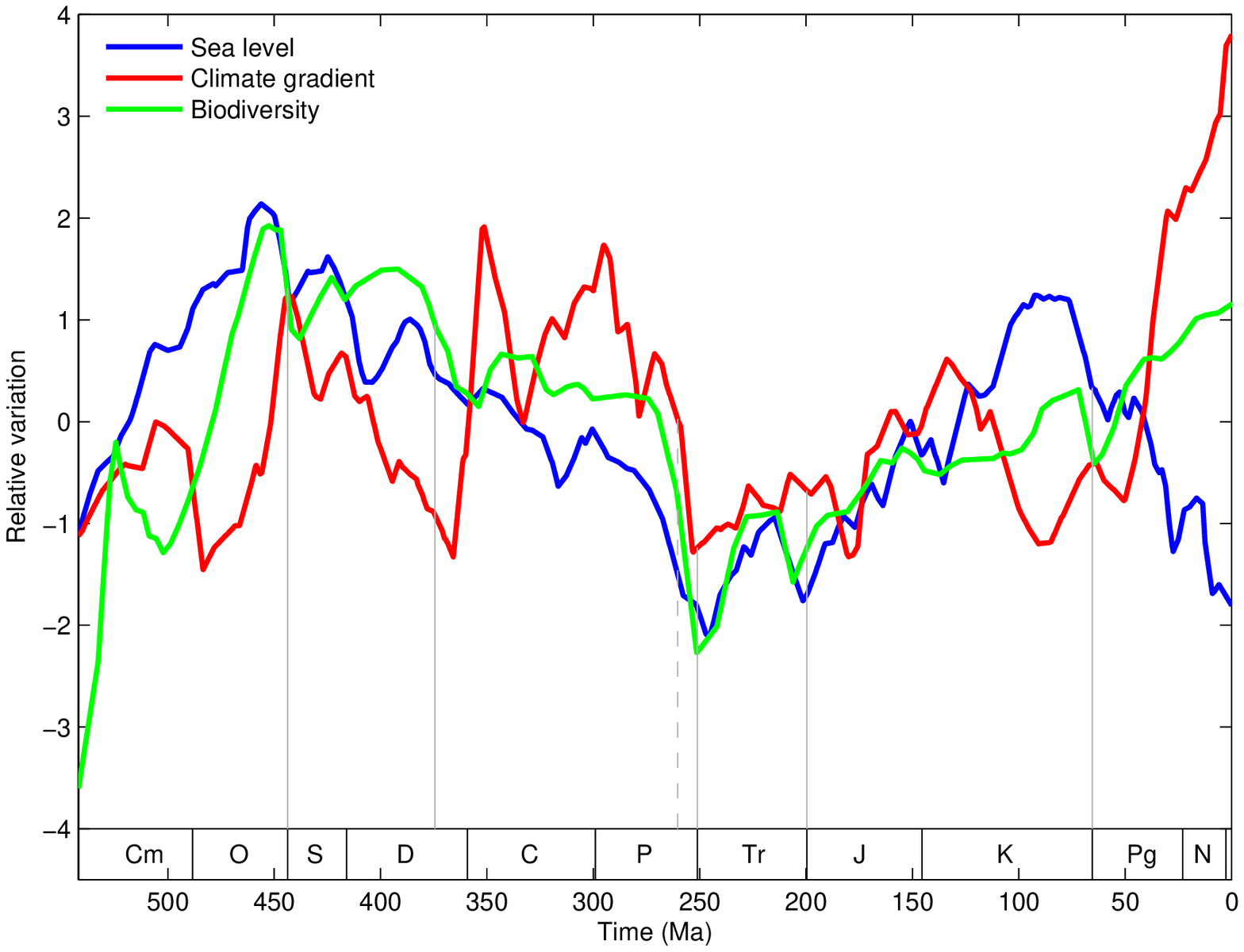}
\end{figure}

\begin{figure}
 \centering
 {\small \bf f} \includegraphics[width=10.20cm]{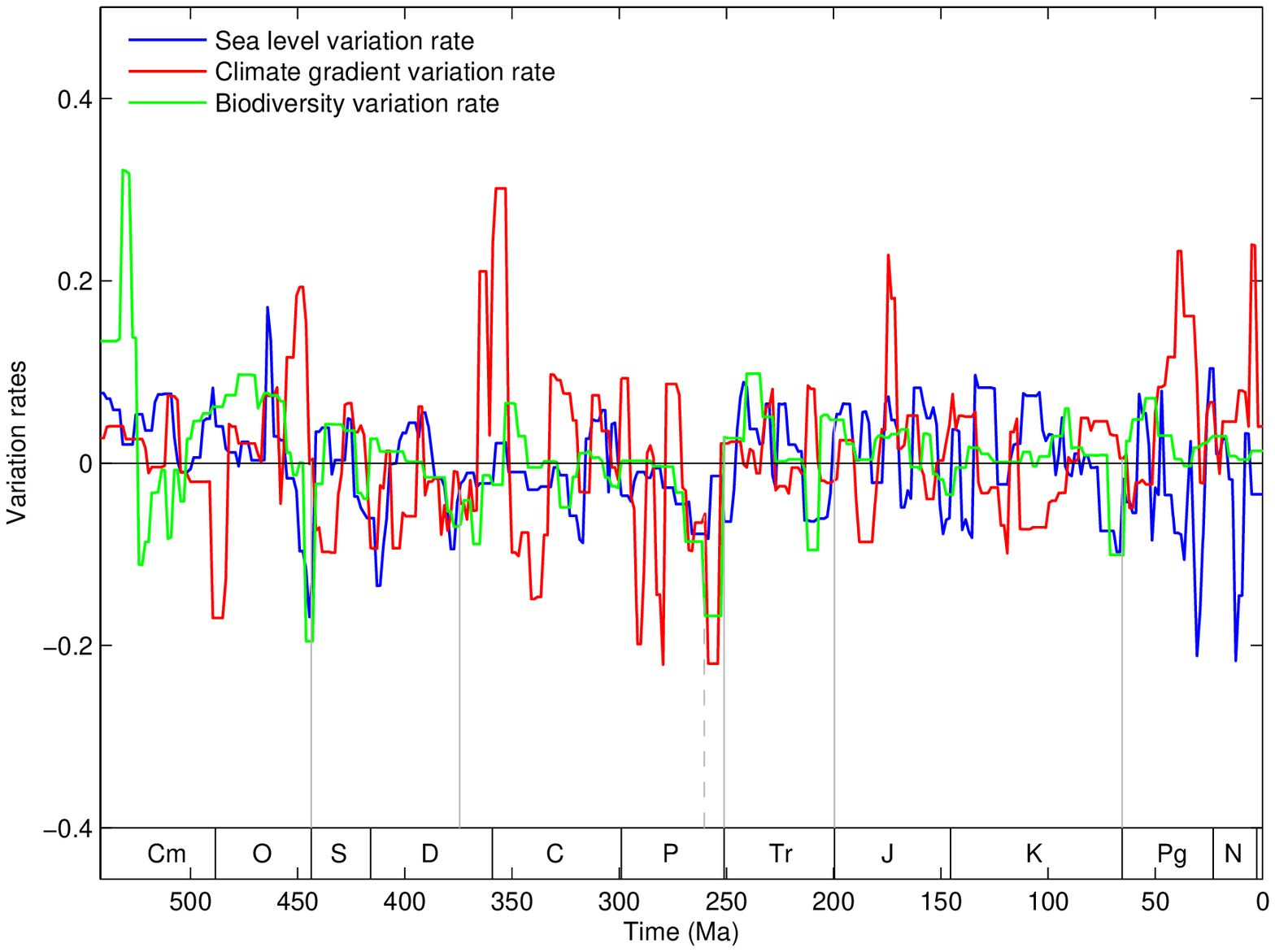}\\
 {\small \bf g} \includegraphics[width=10.20cm]{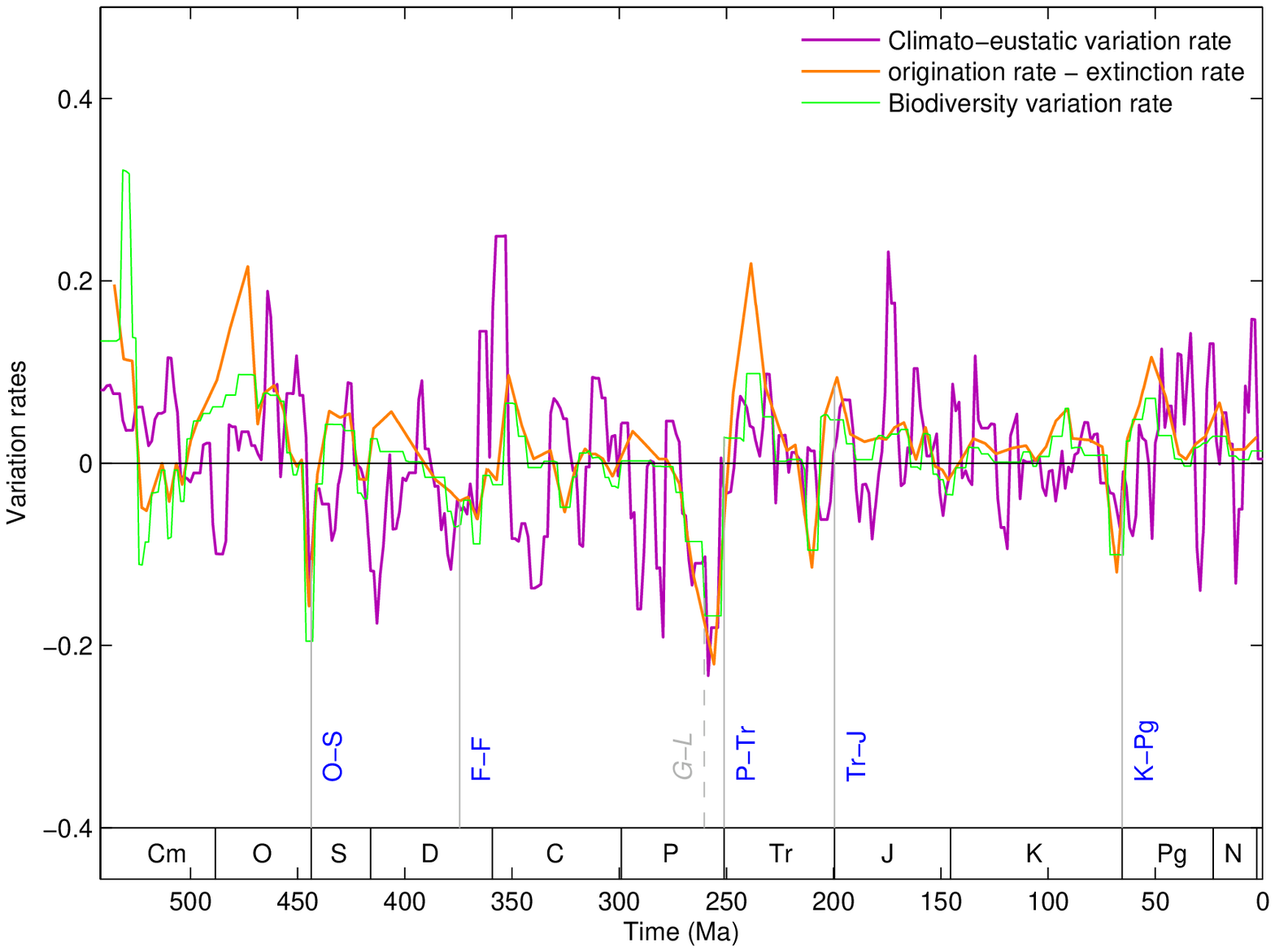}\\
 \caption{Detailed explanation of the fluctuations of the Phanerozoic biodiversity curve. {\bf a} Climatic gradient curve in the Phanerozoic eon based on climate indicators. The climatic gradient curve is opposite to the climate curve. {\bf b} The Phanerozoic climate curve is obtained by averaging the four independent results (climate indicators, Berner's $CO_2$, marine $^{87}Sr/^{86}Sr$ and marine $\delta ^{18}O$). {\bf c} The Phanerozoic eustatic curve is obtained by averaging the Hallam's and Haq's results on sea level fluctuations. {\bf d} The climato-eustatic curve is obtained by averaging the Phanerozoic climate gradient curve (Fig 3b) and the Phanerozoic eustatic curve (Fig 3c). {\bf e} Relationships among the eustatic curve (Fig 3c), the climate gradient curve (Fig 3b) and the biodiversity net fluctuation curve (Fig 2h) in the Phanerozoic eon. {\bf f} Variation rate of the eustatic curve, variation rate of the climate gradient curve and variation rate of the biodiversity net fluctuation curve in the Phanerozoic eon (the derivative curves of the three curves in Fig 3e). {\bf g} Variation rate of the climato-eustatic curve and variation rate of the biodiversity net fluctuation curve in the Phanerozoic eon (the derivative curves of the two curves in Fig 1b). The five mass extinctions (including two phases in the P-Tr extinction) are marked in the present figure (Fig 1b), and the minor extinctions can also be observed in the variation rate of the climato-eustatic curve. In addition, the difference between origination rate and extinction rate is plotted (Fig 5c).}
\end{figure}

\clearpage \begin{figure}
 \centering
 \includegraphics[width=18.3cm]{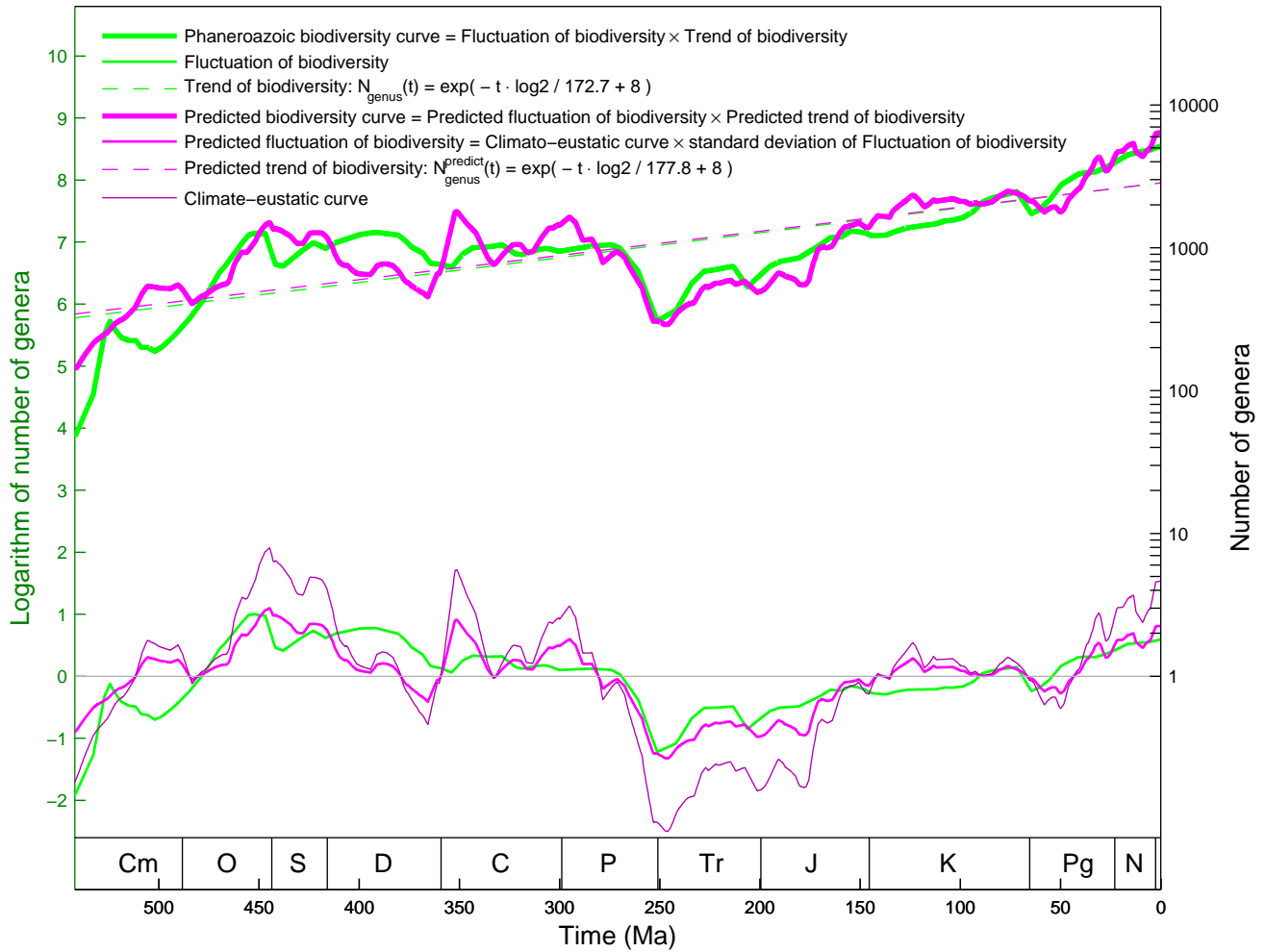}
 \caption{Detailed explanation of the reconstruction of the Phanerozoic biodiversity curve based on genomic, climatic and eustatic data (Fig 1a, 1b, 1c). The exponential growth trend of the predicted biodiversity curve (obtained based on the exponential trend in genome size evolution in Fig 2g) corresponds to the exponential growth trend in the Phanerozoic biodiversity curve (obtained based on the fossil records in Fig 2h); The fluctuations of the predicted biodiversity curve (obtained based on the climato-eustatic curve in Fig 3d) corresponds to the biodiversity net fluctuation curve (obtained based on the fossil records in Fig 2h).}
\end{figure}

\begin{figure}
 \centering
 {\small \bf a} \includegraphics[width=9.5cm]{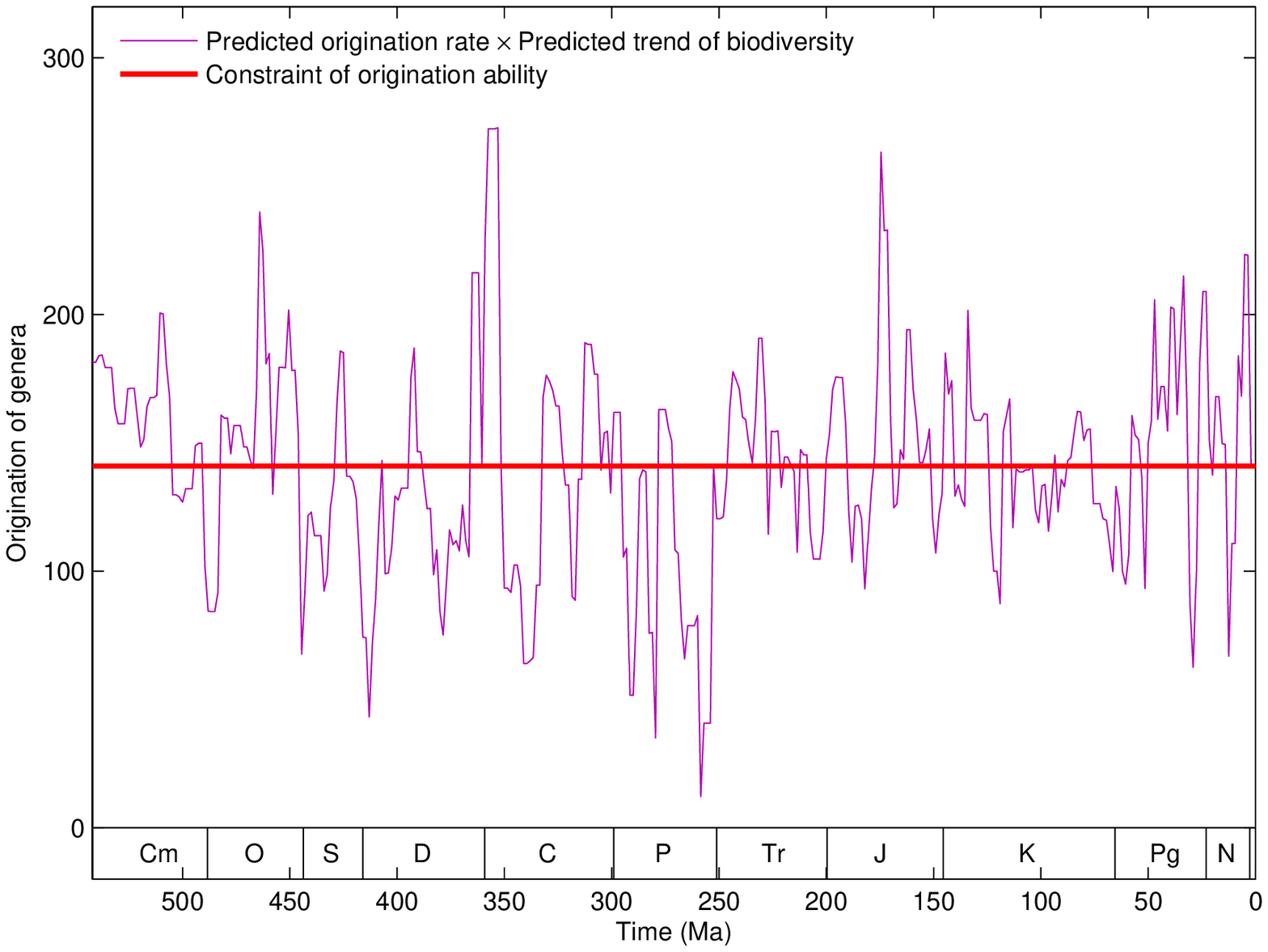}\\
 {\small \bf b} \includegraphics[width=9.5cm]{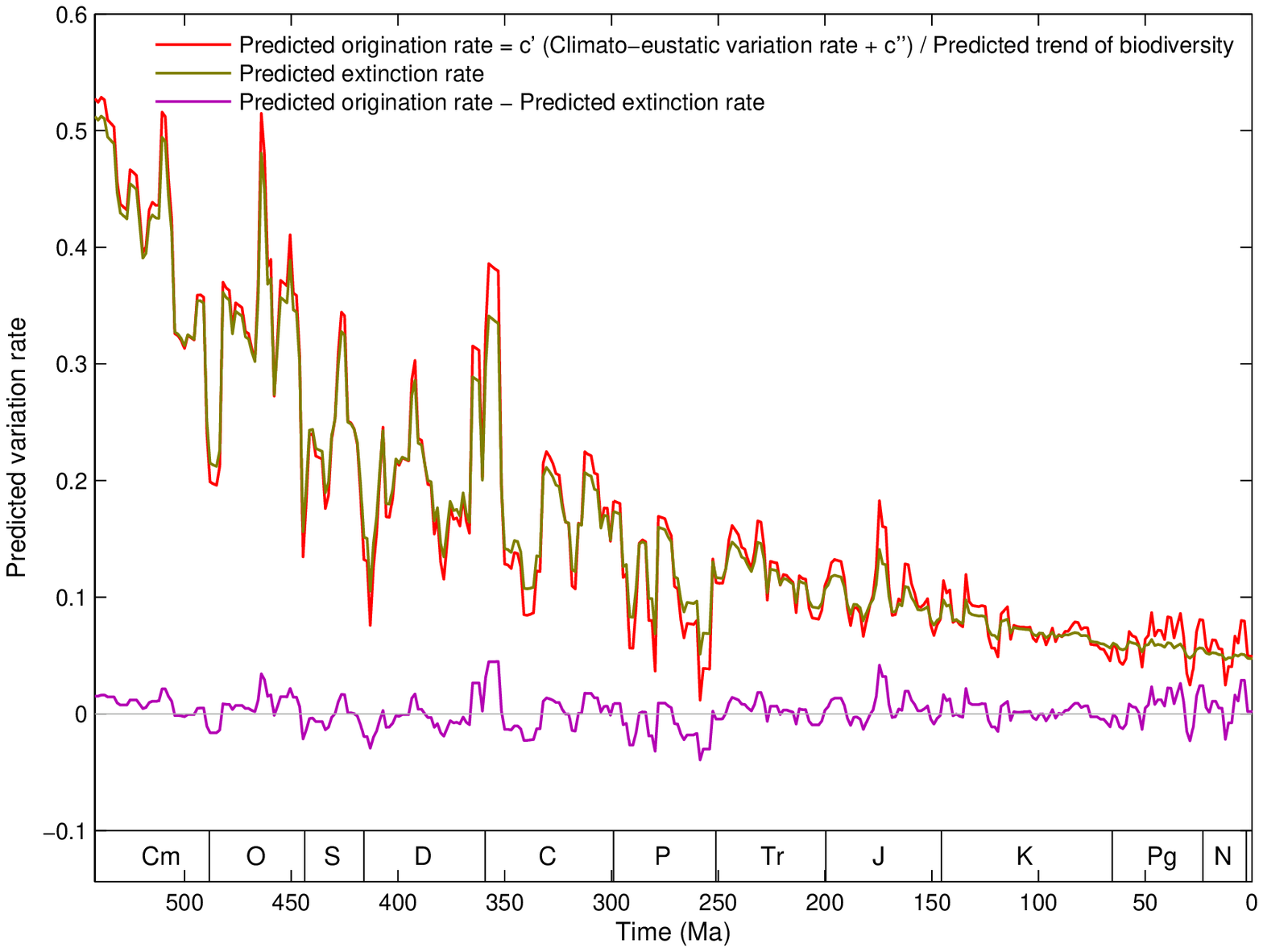}\\
 {\small \bf c} \includegraphics[width=9.5cm]{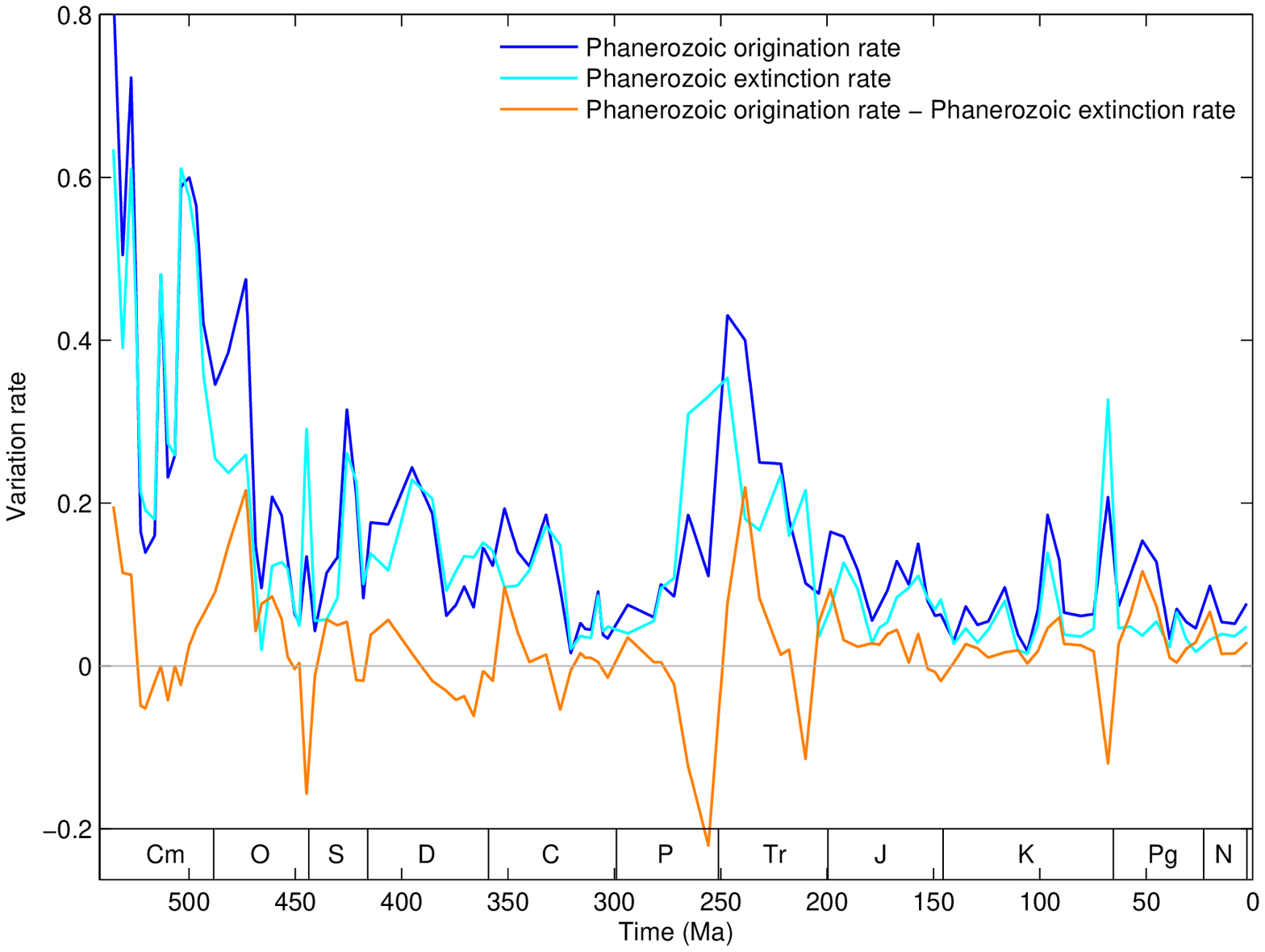}
 \caption{Detailed explanation of the declining origination rate and extinction rate throughout the Phanerozoic eon. {\bf a} Constraint of origination ability of life on earth, which results in the declining origination rate through the Phanerozoic eon (Fig 1d). {\bf b} Explanation of the declining origination rate and extinction rate in the Phanerozoic eon, according to the reconstruction of the Phanerozoic biodiversity curve (Fig 4), which agrees with the results based on fossil records (Fig 5c). {\bf c} The declining origination rate and extinction rate in the Phanerozoic eon based on fossil records.}
\end{figure}

\clearpage
\begin{figure}
  \centering
  {\small \bf a} \includegraphics[width=9.5cm]{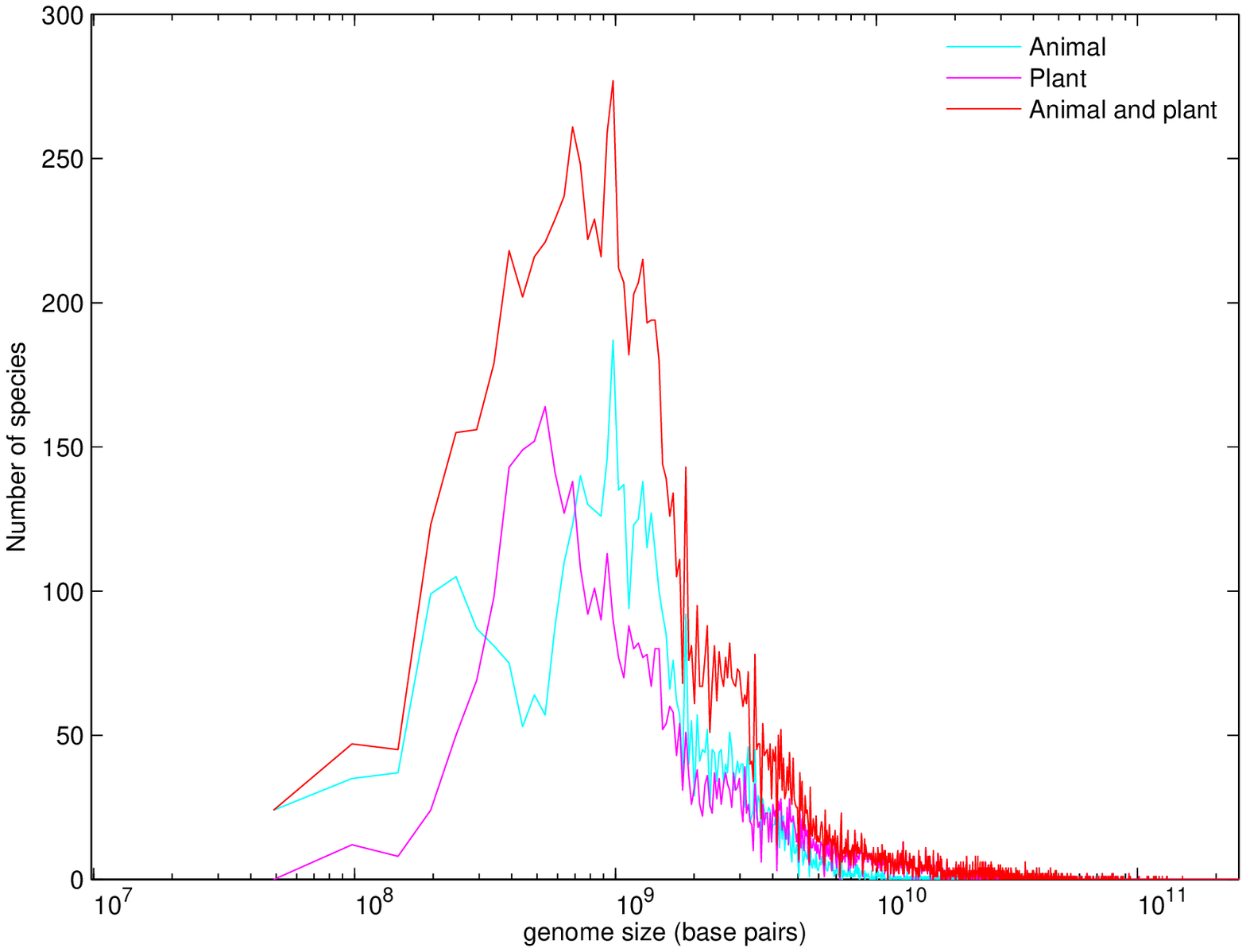}\\
  {\small \bf b} \includegraphics[width=9.5cm]{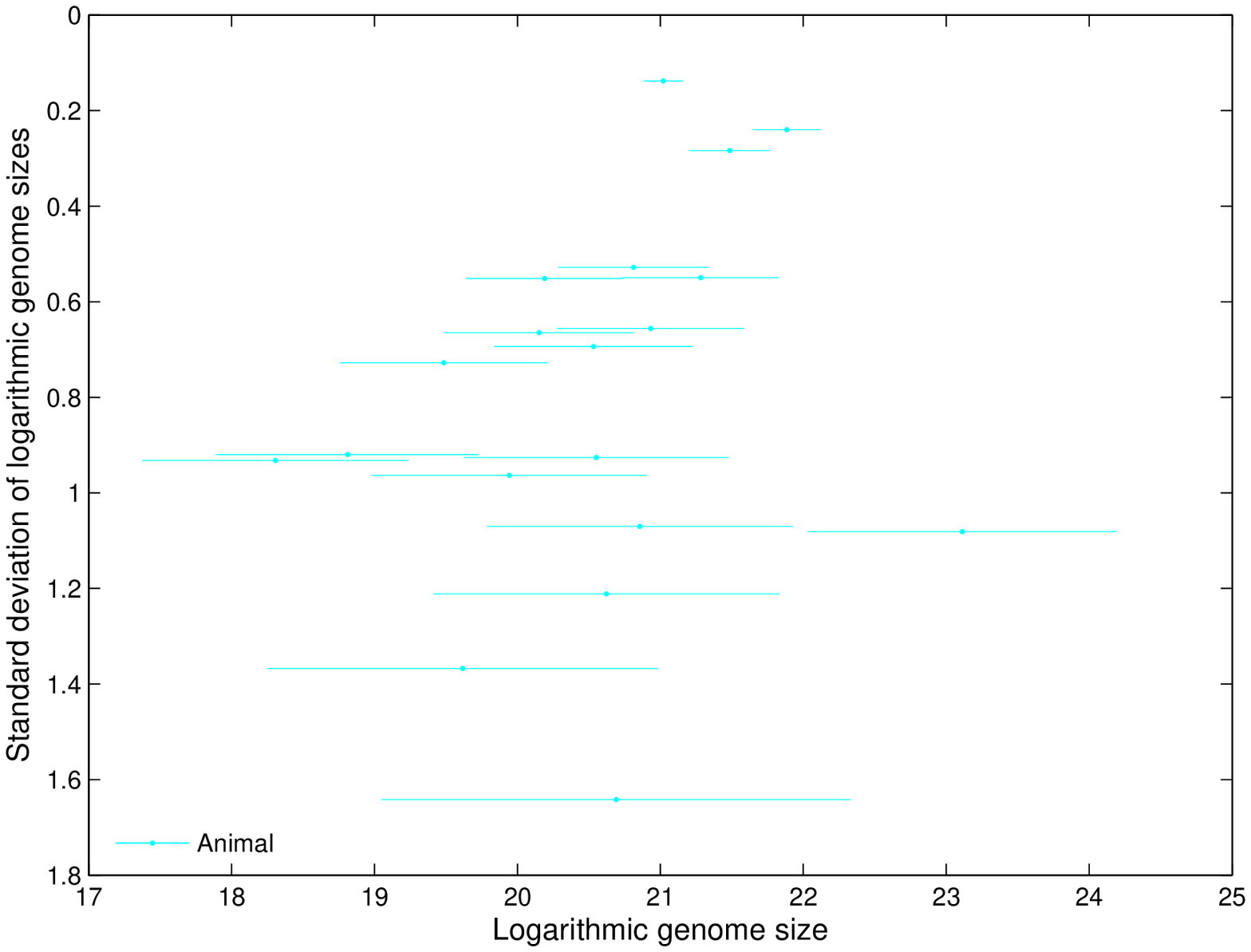}\\
  {\small \bf c} \includegraphics[width=9.5cm]{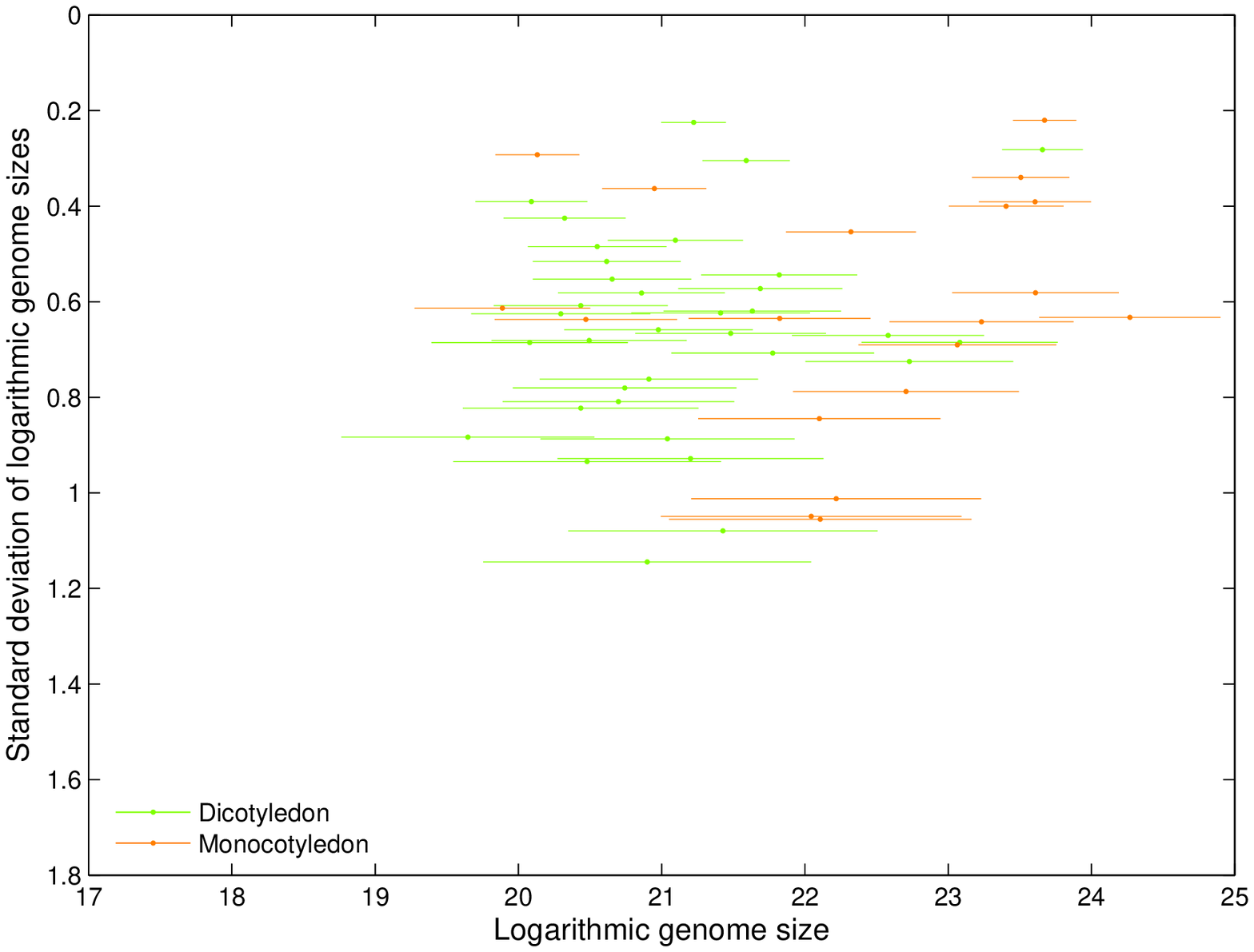}
  \caption{\small Statistical features of genome sizes among taxa in observation, based on genome sizes of contemporary species. {\bf a} Log-normal distributions of genome sizes for animal, plant, and both of them in observation. {\bf b} The $\Lambda$-shaped genome size layout for animal. {\bf c} The V-shaped genome size layout for Angiosperm, where the majority of Dicotyledon taxa are on the left and the majority of Monocotyledon taxa on the right.}
 \end{figure}

\clearpage
\begin{figure}
  \centering
  {\small \bf a} \includegraphics[width=7cm]{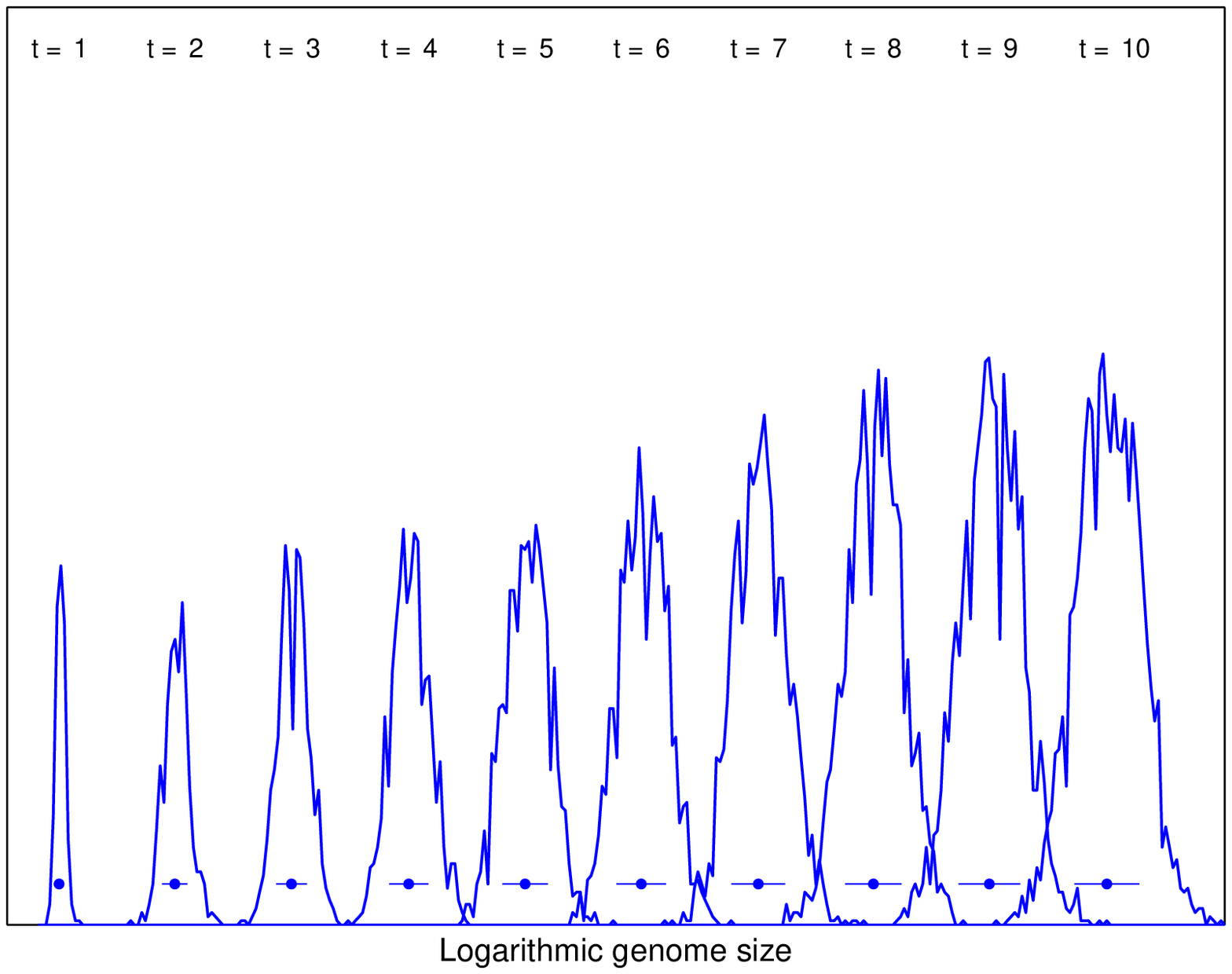}
  {\small \bf b} \includegraphics[width=7cm]{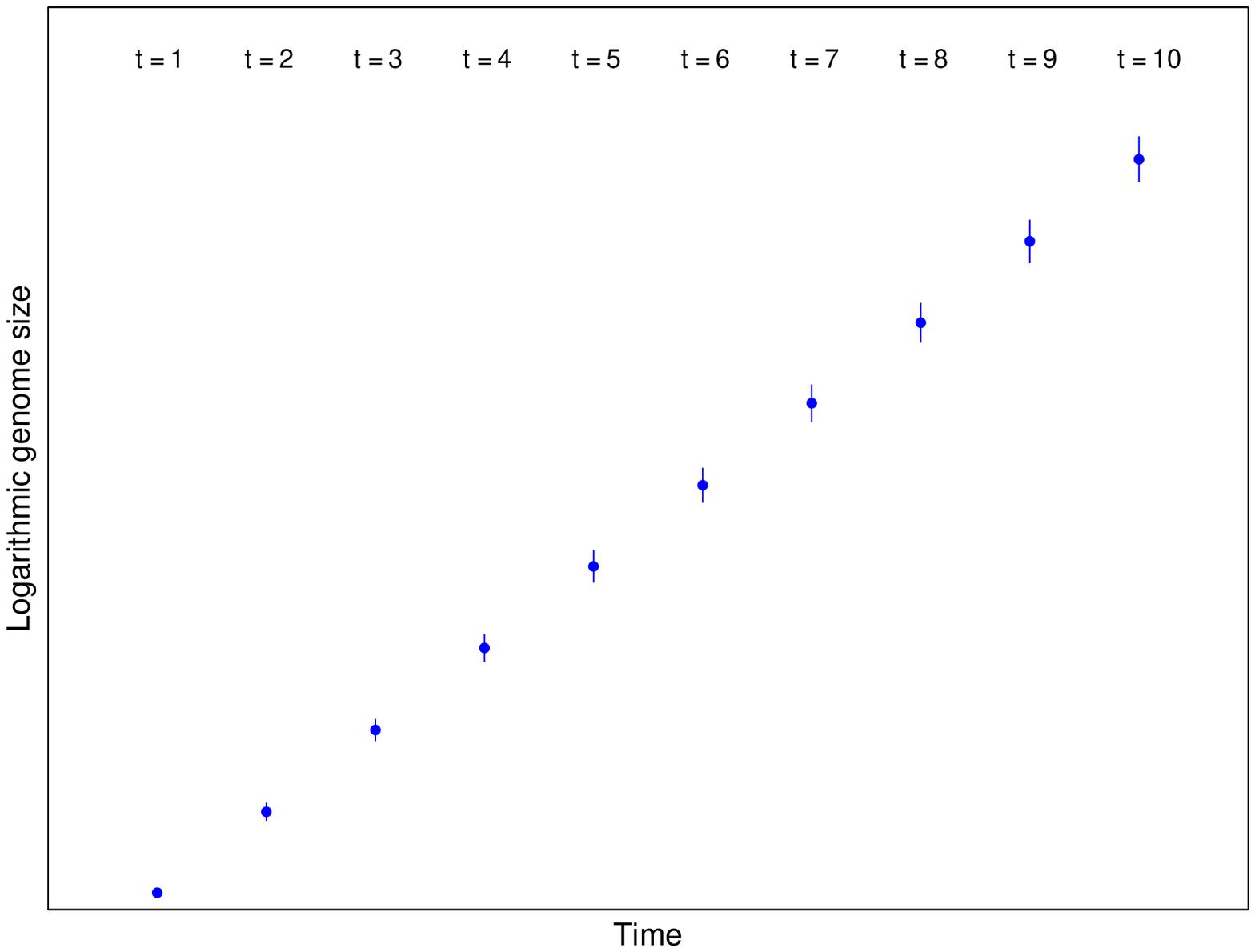}\\
  {\small \bf c} \includegraphics[width=5.5cm]{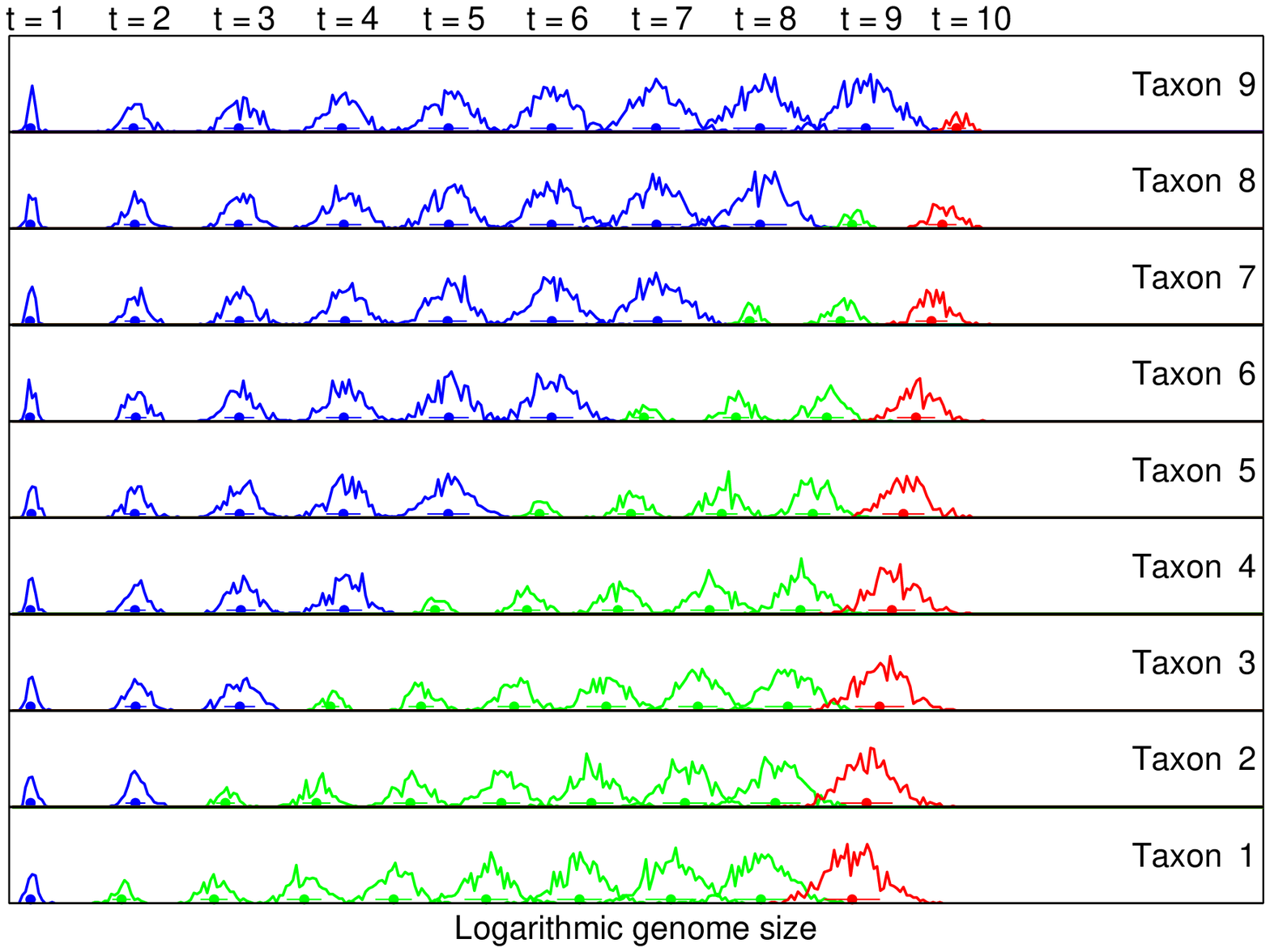}
  {\small \bf d} \includegraphics[width=5.5cm]{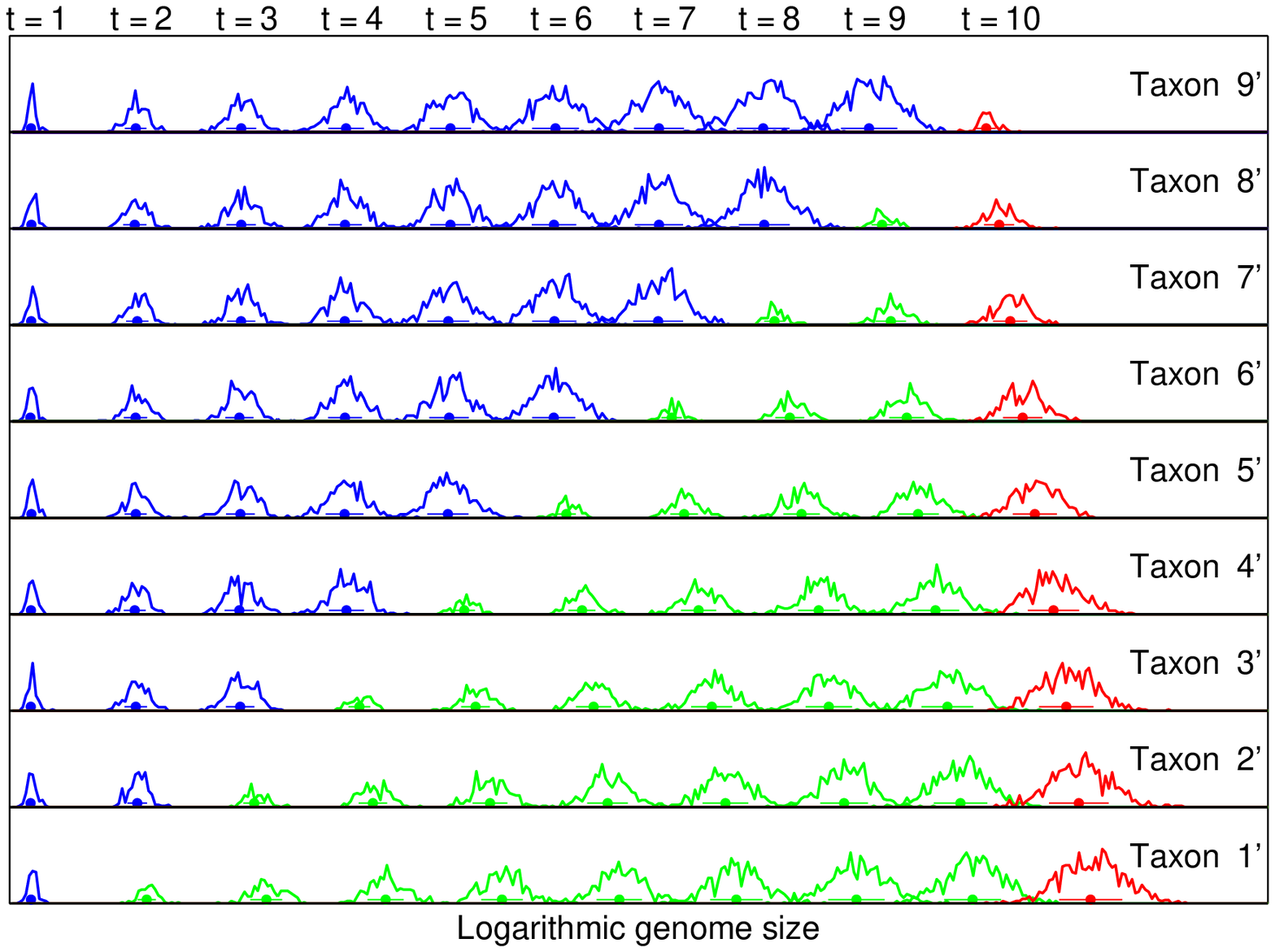}
  {\small \bf e} \includegraphics[width=5.5cm]{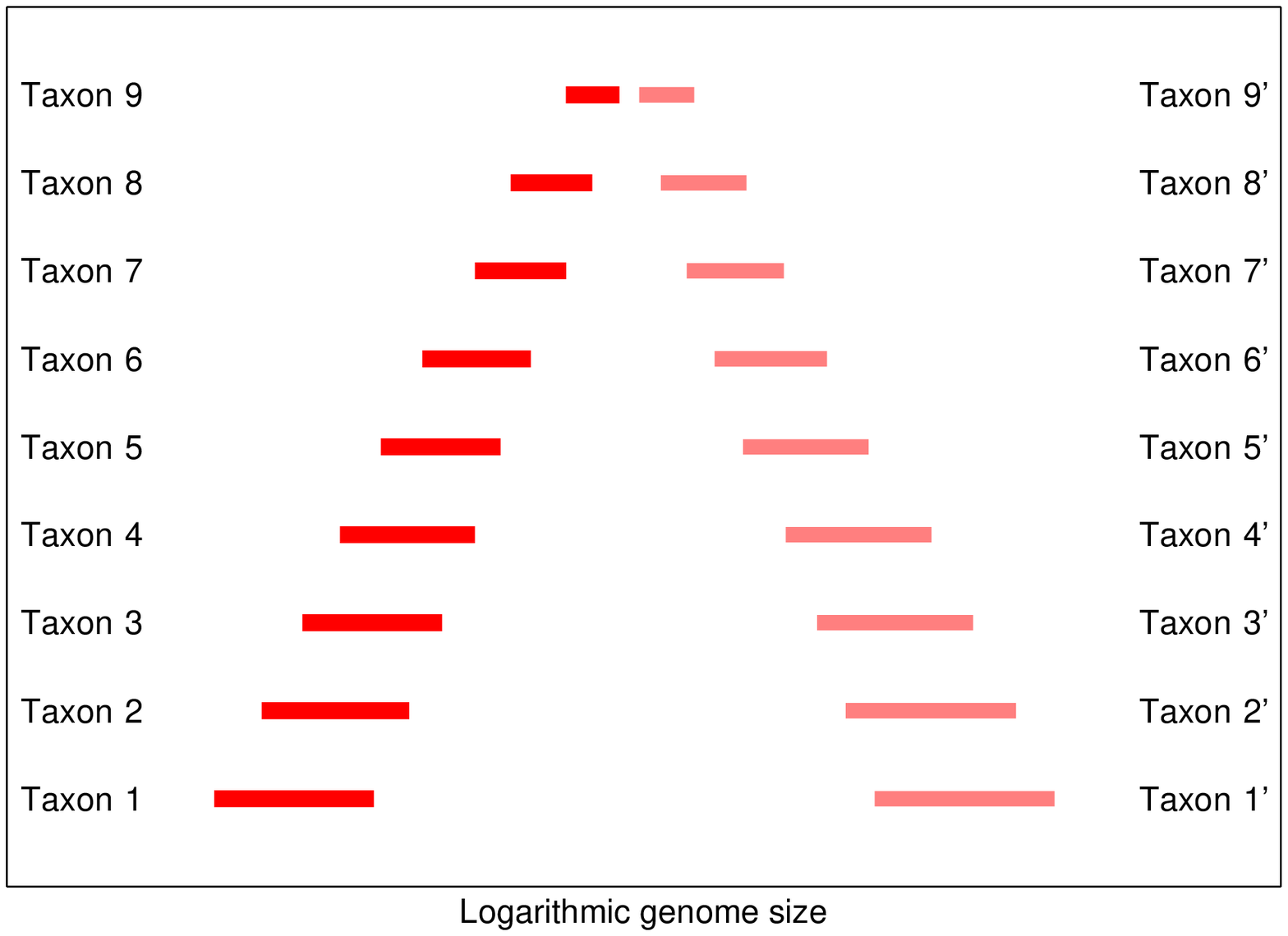}\\
  {\small \bf f} \includegraphics[width=5.5cm]{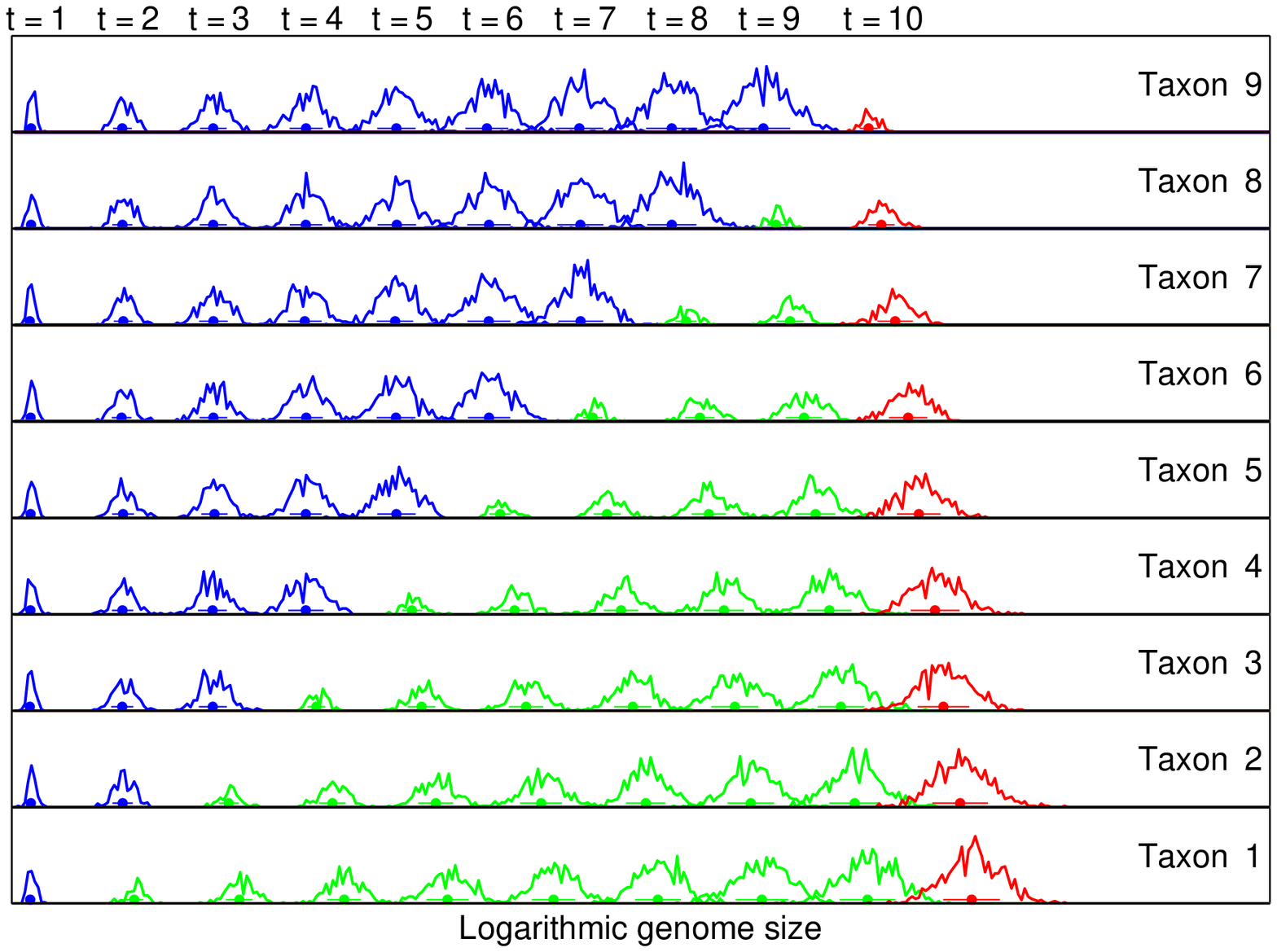}
  {\small \bf g} \includegraphics[width=5.5cm]{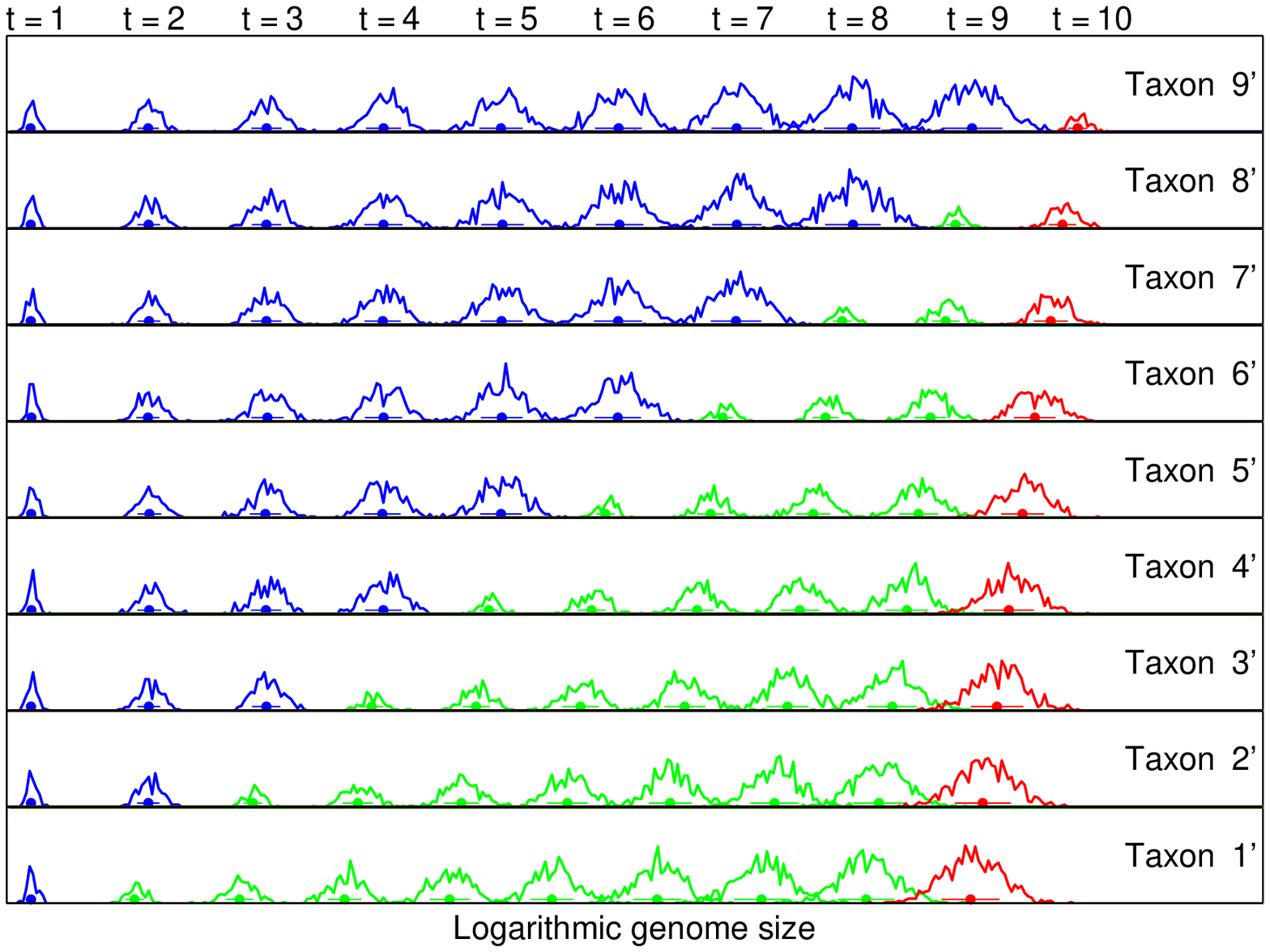}
  {\small \bf h} \includegraphics[width=5.5cm]{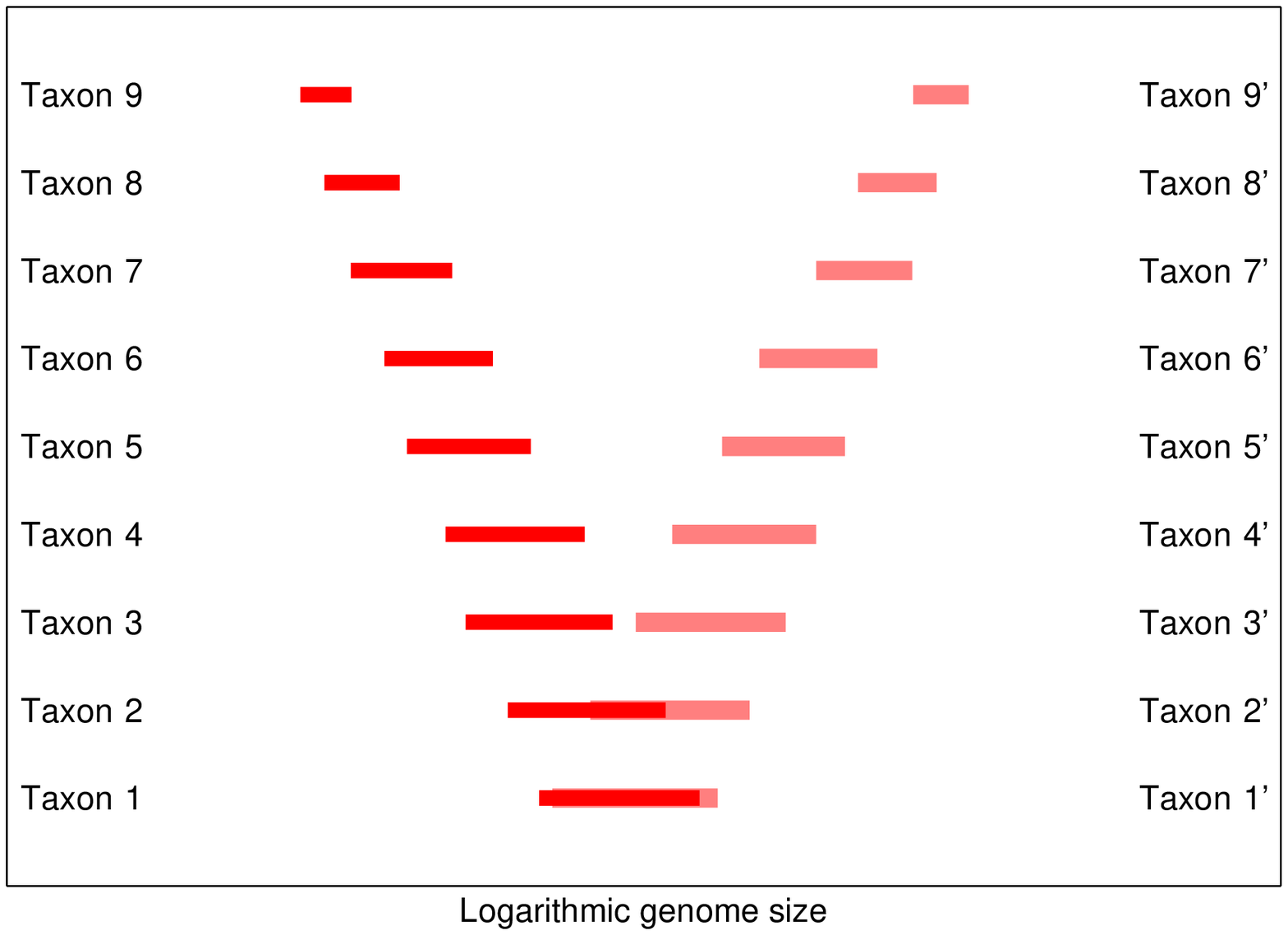}\\
  \caption{\small Simulation of the statistical features of genome sizes among taxa based on a log-normal stochastic process model. (a$\sim$b) Simulation of the genome size evolution for a taxon based on the log-normal stochastic process model. {\bf a} Simulation of the log-normal distributions for a taxon at different stages $t=1, 2, ..., 10$ in genome size evolution. {\bf b} Simulation of the exponential growth trend in genome size evolution based on the linear relationship between time $t$ and the logarithmic means of genome sizes simulated in Fig 7a, which agrees with the linear relationship between time $t$ and the logarithmic means of genome sizes in observation (Fig 1a). (c$\sim$e) Simulation of the $\Lambda$-shaped genome size layout. {\bf c} Simulation of the left part of the $\Lambda$-shaped genome size layout, where the probability for ancestor taxa is constant $P_a=0.2$, and the probability for descendant taxa is $P_d=0.175$, which corresponds to a relatively slow genome size growth rate for the descendant taxa. The ancestor taxa are in blue, while the descendant taxa are in green; especially the taxa at time $t=10$ represents the modern or the present taxa, which are in red (the same colour convention below). {\bf d} Simulation of the right part of the $\Lambda$-shaped genome size layout, where the probability for ancestor taxa is constant $P_a=0.2$, and the probability for descendant taxa is $P_d=0.225$, which corresponds to a relatively fast genome size growth rate for the descendant taxa. {\bf e} Simulation of the $\Lambda$-shaped genome size layout based on the results at time $t=10$ in Fig 7c and 7d. (f$\sim$h) Simulation of the V-shaped genome size layout. {\bf f} Simulation of the left part of the V-shaped genome size layout, where the probability for descendant taxa is constant $P_d=0.2$, and the probability for ancestor taxa is $P_a=0.175$, which corresponds to a relatively slow genome size growth rate for the ancestor taxa. {\bf g} Simulation of the right part of the V-shaped genome size layout, where the probability for descendant taxa is constant $P_d=0.2$, and the probability for ancestor taxa is $P_a=0.225$, which corresponds to a relatively fast genome size growth rate for the ancestor taxa. {\bf h} Simulation of the V-shaped genome size layout based on the results at time $t=10$ in Fig 7f and 7g.}
\end{figure}

\clearpage
\begin{figure}
  \centering
  {\small \bf a} \includegraphics[width=7.5cm]{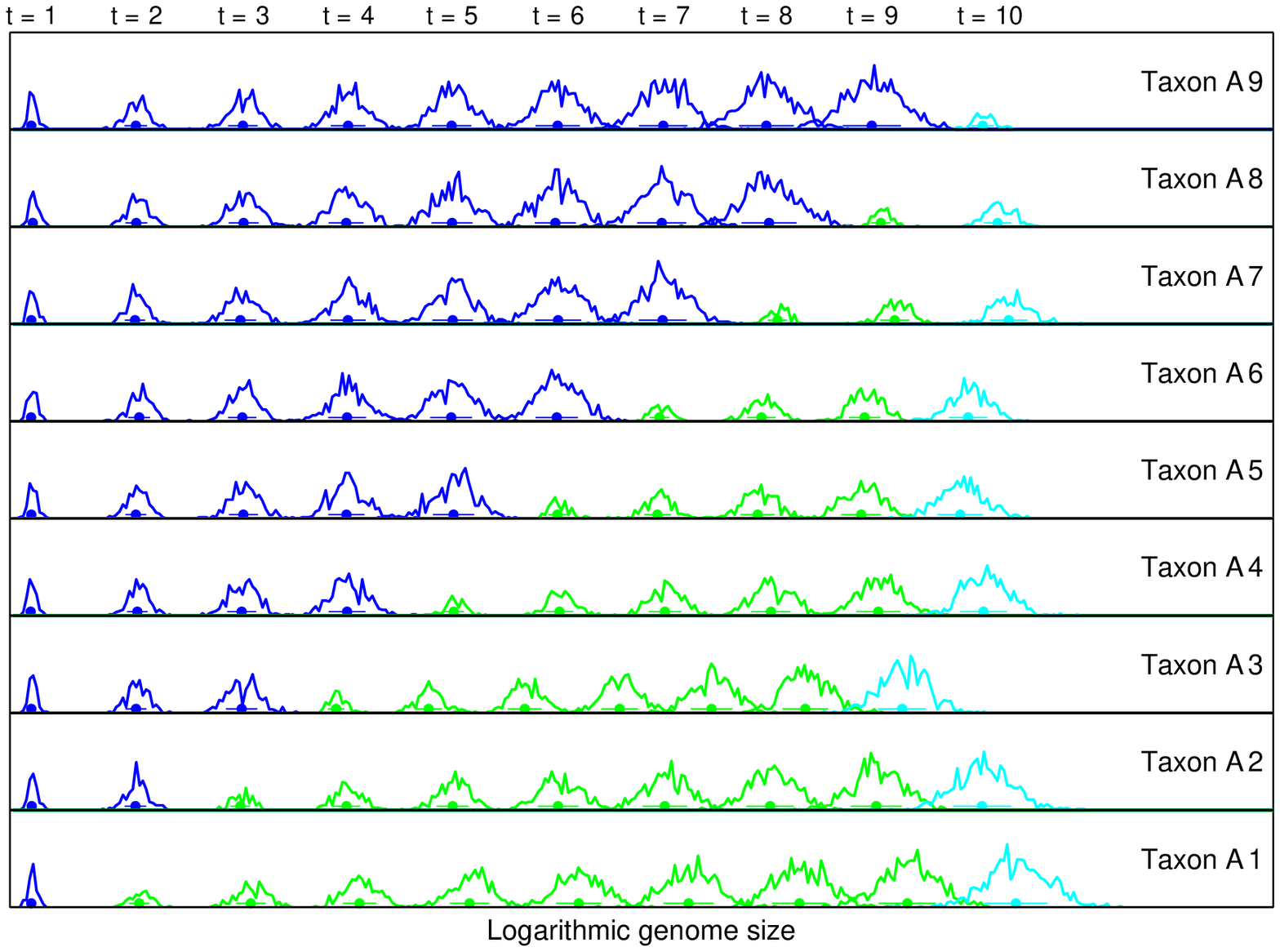}
  {\small \bf b} \includegraphics[width=7.5cm]{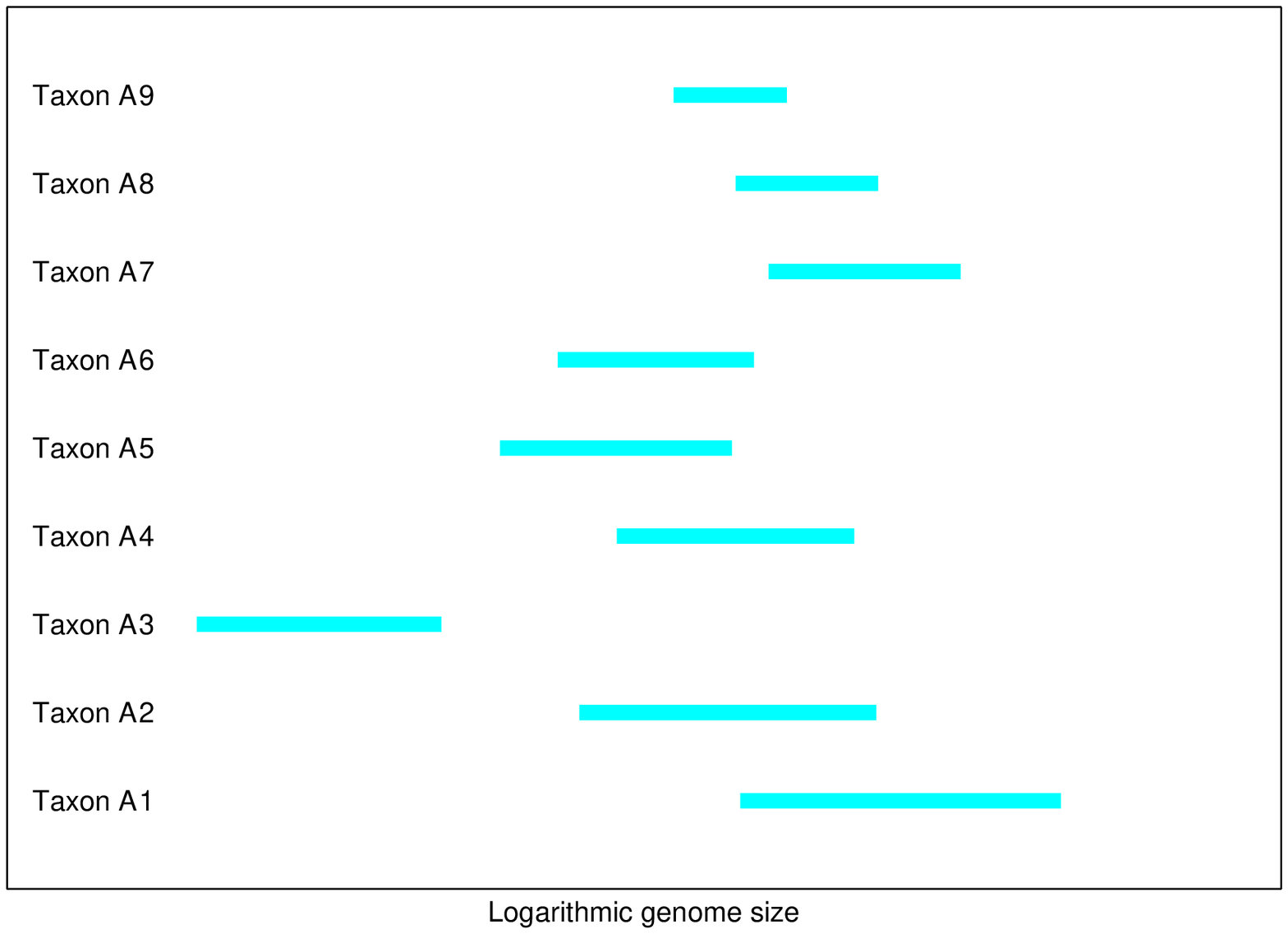}\\
  {\small \bf c} \includegraphics[width=7.5cm]{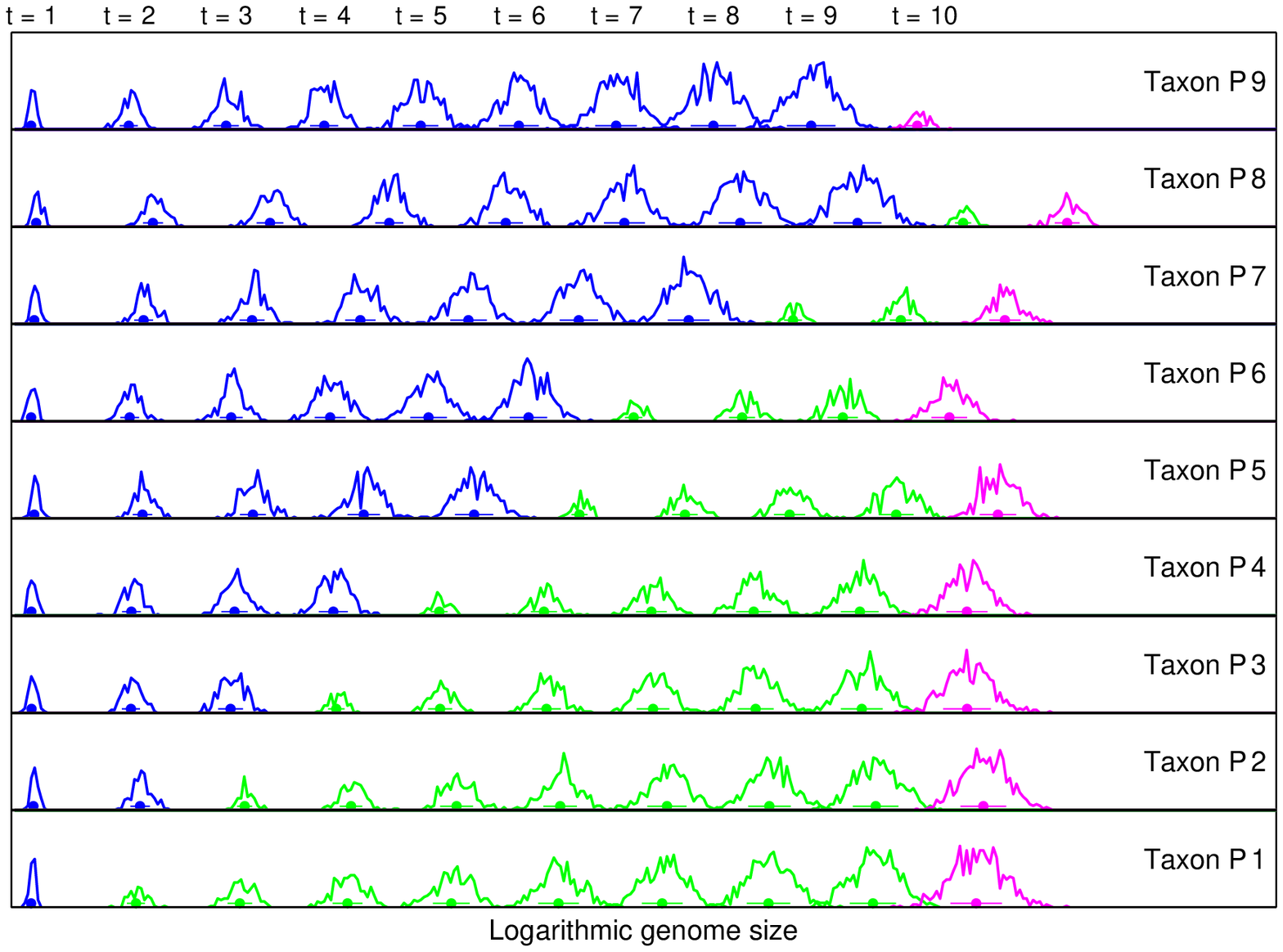}
  {\small \bf d} \includegraphics[width=7.5cm]{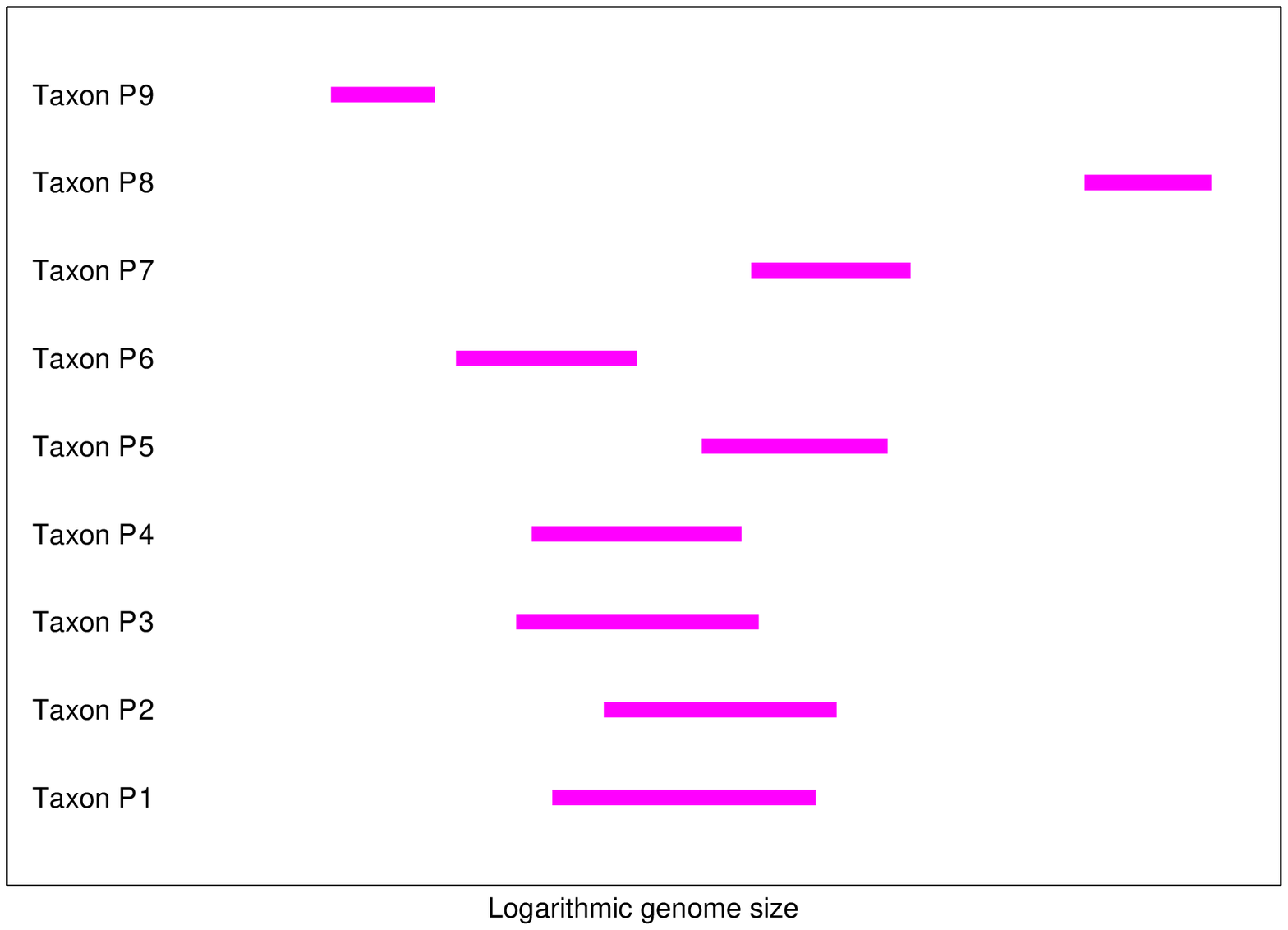}\\
  {\small \bf e} \includegraphics[width=7.5cm]{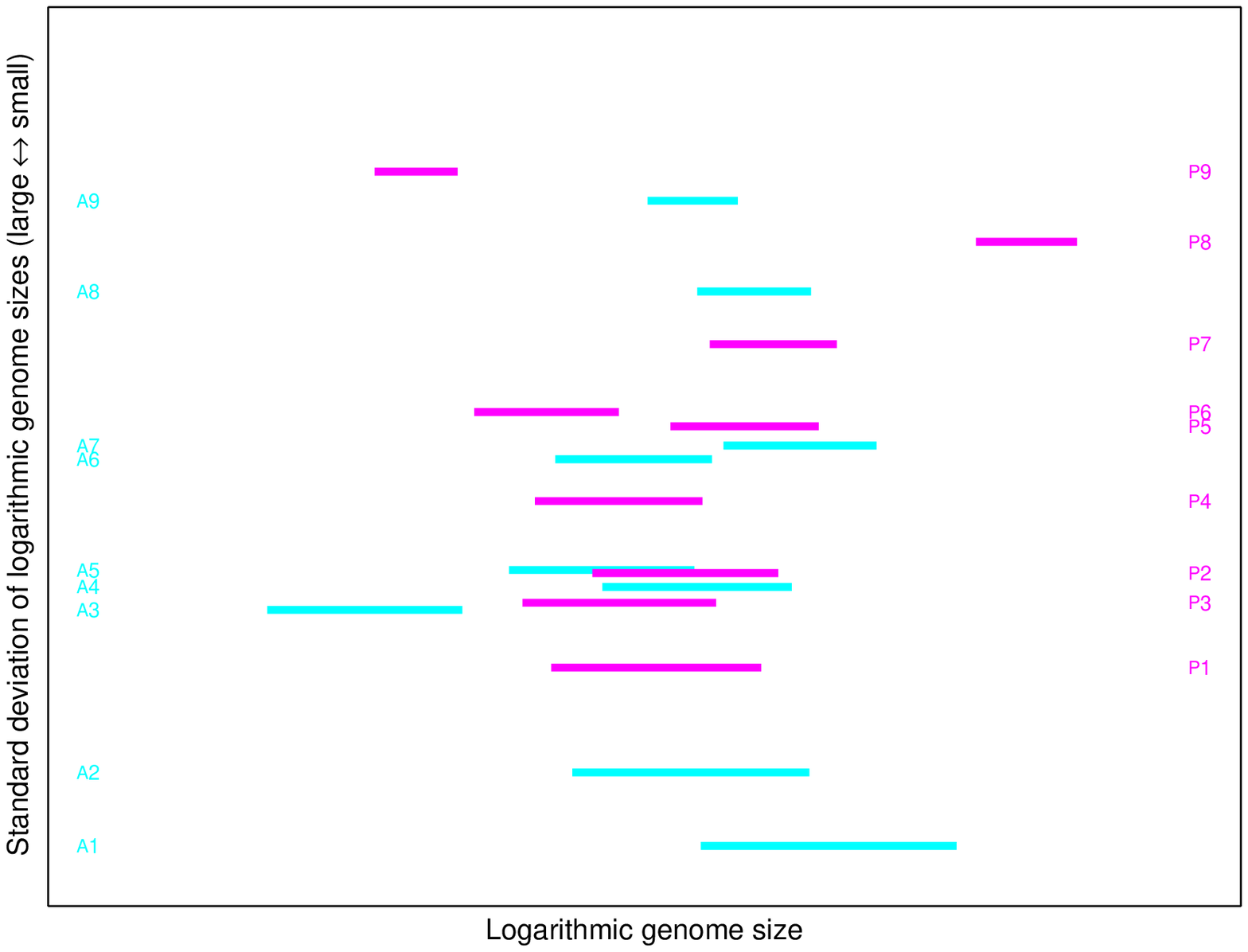}
  {\small \bf f} \includegraphics[width=7.5cm]{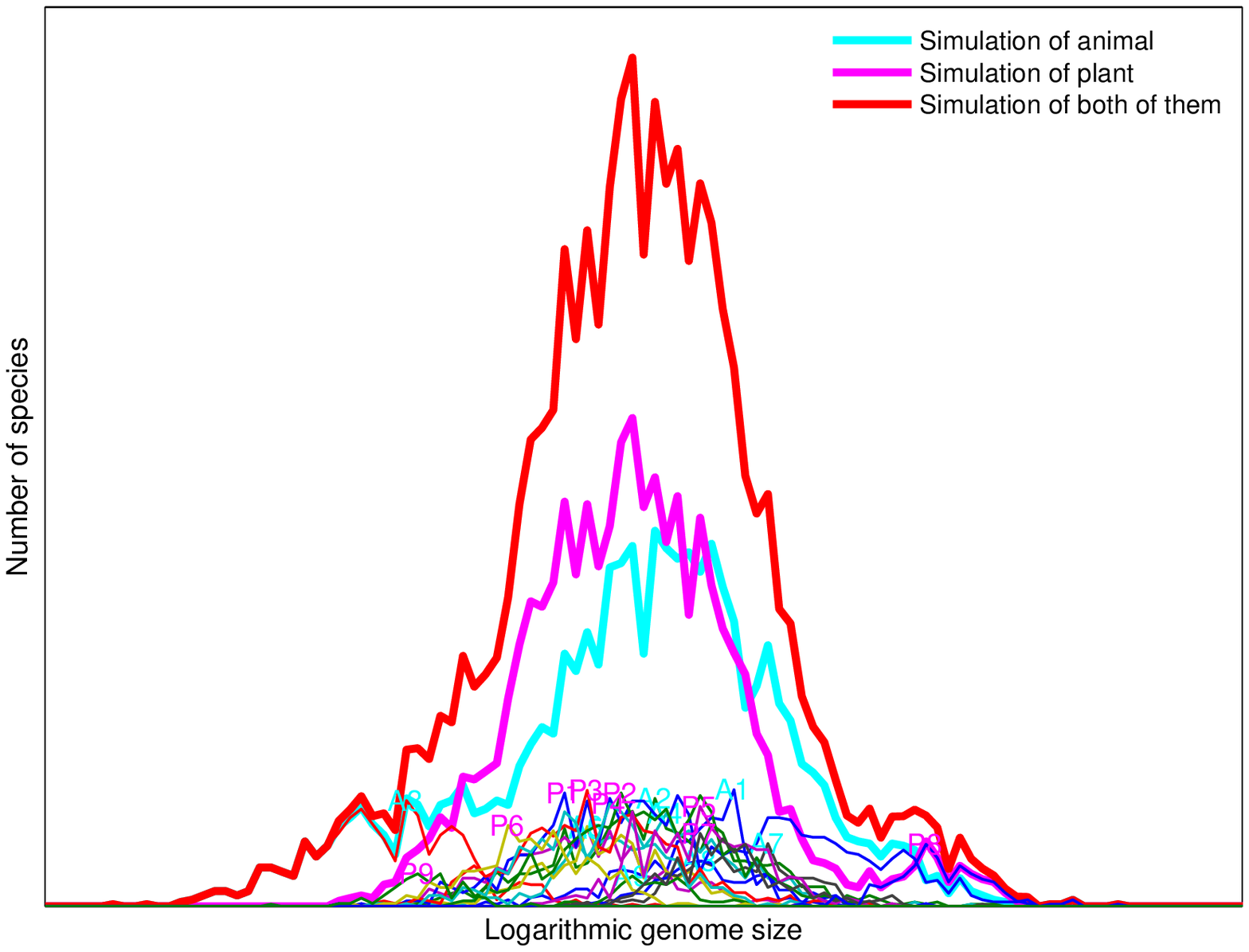}\\
  \caption{\small Explanation of the statistical features of genome sizes for animal and plant. (a$\sim$b) Explanation of the $\Lambda$-shaped genome size layout for animal. {\bf a} Simulation of the genome size evolution for animal, where the probability for ancestor taxa is constant $P_a=0.2$, and the probability for descendant taxa $P_d$ varies randomly from $0.175$ to $0.225$. {\bf b} The variation ranges of genome sizes for the simulated animal taxa $Taxon\ A1 \sim A9$, based on the results at time $t=10$ in Fig 8a. (c$\sim$d) Explanation of the V-shaped genome size layout for Angiosperms. {\bf c} Simulation of the genome size evolution for Angiosperms, where the probability for descendant taxa is constant $P_d=0.2$, and the probability for ancestor taxa $P_a$ varies randomly from $0.175$ to $0.225$. {\bf d} The variation ranges of genome sizes for the simulated taxa $Taxon\ P1 \sim P9$, based on the results at time $t=10$ in Fig 8c. {\bf e} The $\Lambda$-shaped genome size layout for animal and the V-shaped genome size layout for Angiosperms based on the results in Fig 8b and 8d. {\bf f} Simulation of the log-normal genome size distributions for animal, plant, and both of them, where the genome size distribution for animal is obtained by summing up the genome size distributions of $Taxon\ A1 \sim A9$ at time $t=10$; the genome size distribution for plant is obtained by summing up the genome size distributions of $Taxon\ P1 \sim P9$ at time $t=10$; and the genome size distribution for Eukarya is obtained by summing up the genome size distributions for animal and plant. This simulation agrees with the observation based on genome size databases (Fig 6a).}
\end{figure}

\end{document}